\documentclass[aps,prl,reprint,nofootinbib,floatfix]{revtex4-2}

\usepackage[T1]{fontenc}
\usepackage{amsmath,amssymb,bm}
\usepackage{mathtools}
\usepackage{newtxtext,newtxmath}
\usepackage{graphicx}
\usepackage[colorlinks=true,allcolors=blue]{hyperref}

\newcommand{\Tr}{\operatorname{Tr}}
\newcommand{\csch}{\operatorname{csch}}
\newcommand{\FI}{\mathcal{F}}

\begin{document}

\title{Gaussian Fisher Information Is Superadditive}

\author{Jiaxin Liu}
\author{Zuoxian Wang}
\author{Danyue Ma}
\email{madanyue0419@buaa.edu.cn}
\affiliation{School of Instrumentation and Optoelectronic Engineering, Beihang University, Haidian, Beijing 100191, China}
\date{October 2, 2026}

\begin{abstract}
Quantum Fisher information adds over independent probes, so a single parameter is best measured probe by probe. We show that Gaussian measurements, the linear optics and homodyne detection of optical and microwave experiments, break this rule: two independent modes are better measured together. The uncertainty principle leaves a linear detector half of phase space, and for two modes the choice of half is a resource. Bell homodyne, a balanced beam splitter followed by two homodyne detectors, beats the best separate readout when the modes differ in quadrature width and in how the width responds to the parameter. Heterodyne detection splits a mode on a beam splitter to read both quadratures and pays one unit of vacuum noise for the empty port. Bell homodyne fills that port with the second mode, so the noise turns into signal. We prove that the gain stays below $(\sqrt2-1)^2=17.157\%$ for every parameter carried by widths. For thermal modes at one temperature, the canonical case, a gain was conjectured impossible. We prove that any frequency difference opens a temperature window and that the gain peaks at $12.699\%$ at frequency ratio $3.318$, where Bell homodyne is the optimal Gaussian measurement. Standard hardware reaches the gain: two coupled resonators with two homodyne detectors, or a phase-preserving amplifier whose idler band is fed by the same thermal source.
\end{abstract}

\maketitle

Restricting the measurements available to an experiment can change the architecture of the optimal protocol. Quantum Fisher information is additive over independent probes, and each probe has its own optimal measurement, so one detector per mode reaches the best precision when every measurement is allowed~\cite{BraunsteinCaves1994,Mehboudi2019,Correa2015,Demkowicz2020}. That measurement is a projection fixed by the state whose parameter we want~\cite{Correa2015,Glatthard2023,Seah2019}. Real hardware fixes a smaller class, and the classical Fisher information of its outcomes can lose the additivity~\cite{Potts2019,Hovhannisyan2021}. In optics and microwave circuits that class is Gaussian, meaning linear optics, ancillary Gaussian states and homodyne detection~\cite{Weedbrook2012,BraunsteinVanLoock2005,Oh2019,Safranek2019,Nichols2018,Bressanini2024,Chang2026,Fadel2025}, as in gravitational-wave detectors and sensor networks~\cite{Ganapathy2023,Guo2020,Xia2020,Kwon2022}. A parameter carried only by quadrature widths makes the comparison sharp, because Gaussian detectors read widths directly; temperature is the canonical case~\cite{Rubio2021,Mehboudi2022,Glatthard2022,DePasquale2016,Planella2022,Brattegard2024,Tumbiolo2026} and the setting of a conjecture. So for two uncorrelated modes whose widths carry one parameter, can a joint Gaussian measurement beat the best separate readout, and by how much?

With no restriction on the detector, collective measurements gain nothing for a single parameter, and every known collective advantage involves several parameters at once~\cite{MassarPopescu1995,Albarelli2019,Das2025}. Those advantages are now routine~\cite{Hou2021,Conlon2023,XiaB2023,Mi2025,LiY2026}, the hierarchy of local, adaptive and collective strategies is mapped~\cite{Conlon2021,ChenH2022,ConlonPRA2025}, and local measurements often suffice in sensor networks~\cite{Proctor2018,Pezze2025}. For two classes of thermal models the answer is no. Cenni \emph{et al.} conjectured that no joint Gaussian measurement on equilibrium modes beats the separate-readout optimum~\cite{Cenni2022}. Navarro \emph{et al.} proved it for two families of Gaussian models in which temperature enters only through the covariance matrix~\cite{Navarro2025}. Their cases are identical copies of one state or the isothermal model $V=\nu SS^{T}$ with symplectic $S$, in which every mode carries the same symplectic eigenvalue $\nu$, the noise left once beam splitters and squeezers are undone. Call two modes symplectically isothermal when they share one $\nu$, and physically isothermal when they share one temperature. A shared temperature leaves each mode its own ratio $\hbar\omega_i/k_{\mathrm B}T$, so modes of different frequency are physically isothermal without being symplectically isothermal, and the theorem does not reach them.

Here we show that a collective advantage arises from a mismatch of widths and responses between the two modes, and that thermal equilibrium supplies it through the gap between physical and symplectic isothermality. Heterodyne detection reads both quadratures of a mode by splitting it on a beam splitter, and the vacuum that enters the empty port adds one unit of noise. Bell homodyne puts the second probe in that port, so the added noise carries signal of its own, and when the two modes respond differently the joint record beats the two separate optima. The gain is bounded, and the bound is the result. We derive a closed-form onset law showing that any nonzero frequency splitting opens a finite temperature window; for every width law the gain stays below $(\sqrt2-1)^2=17.157\%$, a supremum set by the kink where the best single-mode measurement switches from homodyne to heterodyne, and for temperature the maximum is $12.699\%$, fixed by the thermal law with nothing to tune. At that maximum and on two rectangles of nearby pairs we prove in interval arithmetic that no Gaussian measurement does better. Two coupled resonators reach the optimum with two homodynes on the bare resonators, or with one fixed unbalanced beam splitter under the exact bilinear coupling, and we show that a phase-preserving amplifier whose idler port is fed by the second mode is this receiver. The inputs stay uncorrelated and phase-insensitive throughout, so no correlation in the state can be the source. Additivity of Fisher information is therefore a property of unlimited measurement power: under Gaussian measurements, two independent probes are worth more together than apart. Full derivations are in the Supplemental Material~\cite{Supplement}.

\begin{figure}[!t]
 \includegraphics{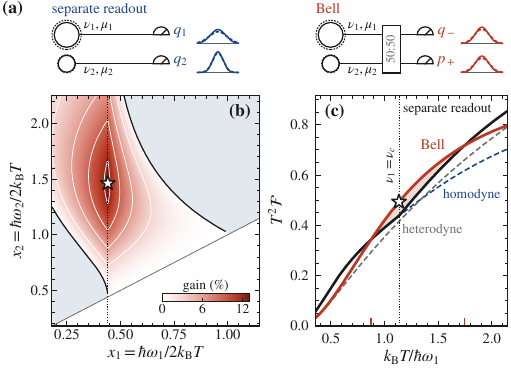}
 \caption{\label{fig:effect}
 \textbf{(a)} The two receivers at the cap pair. Discs are the two thermal modes drawn to their widths $\nu_i$ [Eq.~\eqref{eq:thermalV}], dashed at a higher temperature; each meter's record is drawn as a histogram at $T$ (solid) and $T(1+\delta)$ (dashed). Separate readout: one homodyne per mode reads $q_1$ and $q_2$, two widths and two responses. Bell: a balanced beam splitter feeds two homodynes reading $q_-=(q_1-q_2)/\sqrt2$ and $p_+=(p_1+p_2)/\sqrt2$, two records of one width $(\nu_1+\nu_2)/2$ and one response $(\mu_1+\mu_2)/2$; with vacuum in the lower port the same receiver is heterodyne detection of mode 1. \textbf{(b)} Bell gain over the separate-readout optimum for $x_2\ge x_1$, blank below the diagonal $\omega_1=\omega_2$, from white at zero to red at the cap, white contours every $3\%$. Pale grey where separate readout wins, dotted the crossover $\nu_1=\nu_c$, star the $12.699\%$ cap. \textbf{(c)} $T^2\FI$ against $k_{\mathrm B}T/\hbar\omega_1$ at $\omega_2/\omega_1=3.318$. The separate-readout optimum is the envelope of its homodyne and heterodyne branches (dashed); Bell wins on $0.8728<k_{\mathrm B}T/\hbar\omega_1<1.7455$ (shaded, red ticks).}
\end{figure}

\textit{Two numbers per mode.---}
A Gaussian detector sees two numbers per mode, a width and its response. Two uncorrelated modes that look the same along every quadrature have covariance $\nu_1I_2\oplus\nu_2I_2$, with $I_2$ the vacuum covariance and $\oplus$ placing the two blocks on the diagonal. A parameter $\Theta$ that enters only through the widths $\nu_i(\Theta)$ has responses $\mu_i=\Theta\,\partial_\Theta\nu_i$. Write $x_i=\hbar\omega_i/(2k_{\mathrm B}T)$; heating slides both modes down the $x$ axis. Two independent modes at temperature $T$ have
\begin{equation}
 V(T)=\nu_1 I_2\oplus\nu_2 I_2,\qquad
 \nu_i=\coth x_i,\qquad
 \mu_i=x_i\csch^2 x_i .
 \label{eq:thermalV}
\end{equation}
Each $\nu_i$ is a quadrature variance in vacuum units, and $\mu_i=T\partial_T\nu_i$ says how fast it moves with temperature. A zero-mean Gaussian measurement is fixed by one seed covariance $M$, the detector's own noise, and its data are Gaussian with covariance $V+M$. Scaled by $T^2$, their Fisher information is dimensionless and reads~\cite{Cenni2022}
\begin{equation}
 T^2\FI(V;M)=\frac{1}{2}\Tr\!\left[\left((V+M)^{-1}T\partial_T V\right)^2\right].
 \label{eq:gaussianFI}
\end{equation}
Here $T\partial_TV=\mu_1I_2\oplus\mu_2I_2$ is the whole signal and $V+M$ the whole noise. A separate readout is a seed with no entries linking the modes. A seed is pure when its determinant is one, and pure seeds suffice, because Williamson's theorem writes any mixed seed as a pure seed plus parameter-independent noise, which is classical post-processing (Supplemental Sec.~S10).

\textit{The separate-readout optimum.---}
A thermal mode looks the same along every quadrature and the seed has determinant one, so one number is left, the squeezing $r$ of the seed $M_i(r)=\operatorname{diag}(r,1/r)$. Maximizing Eq.~\eqref{eq:gaussianFI} over $r$ against the boundaries $r\to0,\infty$ gives (End Matter A; Supplemental Secs.~S1--S2)
\begin{equation}
 T^2\FI_i^{\mathrm{loc}}=\mu_i^2\max\left\{\frac{1}{2\nu_i^2},\frac{1}{(\nu_i+1)^2}\right\}.
 \label{eq:localmode}
\end{equation}
The boundary $r\to\infty$ is homodyne detection, which reads one quadrature with no added noise and leaves $1/(2\nu_i^2)$ in the bracket. The stationary point $r=1$ is heterodyne detection, which reads both quadratures at the cost of one extra unit of vacuum noise and leaves $1/(\nu_i+1)^2$. Equation~\eqref{eq:localmode} holds for any width and any response. The branches exchange places at $\nu_c=1+\sqrt2$, that is at $x_c=\operatorname{arccoth}(1+\sqrt2)=0.4406867935$, with homodyne winning for $x_i>x_c$. The optimum is therefore the upper envelope of two crossing curves with a kink at $\nu_c$, where one sharp quadrature is worth both with their extra vacuum unit. Summed over the two modes it is the benchmark $\FI_{\mathrm{local}}^{\max}=\FI_1^{\mathrm{loc}}+\FI_2^{\mathrm{loc}}$ that any joint Gaussian measurement has to beat, and feeding the first outcome forward to the second detector does not raise it.

\textit{Bell homodyne.---}
Mix the modes on a balanced beam splitter and homodyne $q_-=(q_1-q_2)/\sqrt2$ and $p_+=(p_1+p_2)/\sqrt2$, the continuous-variable Bell measurement~\cite{Spedalieri2013}. These quadratures commute, so one shot records both sharply. Each output carries half of each mode, so both records have variance $(\nu_1+\nu_2)/2$ and scaled derivative $(\mu_1+\mu_2)/2$. Homodyne detection is the infinite-squeezing limit of the seed, and there Eq.~\eqref{eq:gaussianFI} gives two independent records of the same Gaussian whose Fisher informations add, cancelling the factor $1/2$. A single squared ratio is left,
\begin{equation}
 T^2\FI_{\mathrm{Bell}}=\left(\frac{\mu_1+\mu_2}{\nu_1+\nu_2}\right)^{\!2}.
 \label{eq:bellFI}
\end{equation}
Figure~\ref{fig:effect}(a) draws both receivers. Every Gaussian measurement reads one commuting half of the quadratures sharply, those of the modes and of any vacuum ancilla it adds, and gives up the other half. For one thermal mode the only choice is that of Eq.~\eqref{eq:localmode}, one sharp quadrature or both with a vacuum unit. For two modes the halves differ: one homodyne per mode keeps $q_1$ and $q_2$, and Bell homodyne keeps $q_-$ and $p_+$. Equation~\eqref{eq:localmode} sums two squared ratios, and Eq.~\eqref{eq:bellFI} squares one ratio of sums. They coincide when the modes respond alike, and only a mismatch of responses separates them. At $x_1=x_c$ and $x_2=3/2$ the benchmark gives $T^2\FI=0.43325$ and Bell homodyne $0.48816$, a gain of $12.6738\%$ (Supplemental Sec.~S3).

\begin{figure}[!tb]
 \includegraphics{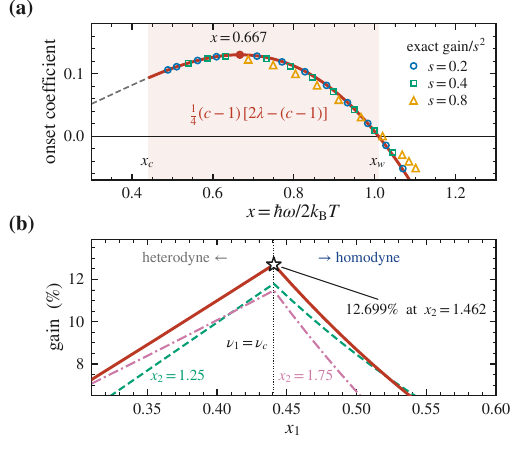}
 \caption{\label{fig:universality}
 \textbf{(a)} The onset coefficient of Eq.~\eqref{eq:onset} against the common $x$, dashed below $x_c$, off the homodyne branch. It is positive on the window $(x_c,x_w)=(0.4407,1.0096)$ and peaks at $x=0.66678$. Symbols: the exact gain divided by $s^2$ for $s=0.2$, $0.4$, $0.8$; they collapse onto the law. \textbf{(b)} Gain against $x_1$ at $x_2=1.462$ (red), $1.25$ and $1.75$. All three kink at $x_c$, where the separate-readout optimum switches branch.}
\end{figure}

\textit{Any splitting suffices.---}
Start from two modes at the same $x$ and pull them apart by $s=\ln(\omega_2/\omega_1)$, so that $x_{1,2}=xe^{\mp s/2}$. Exchanging the modes sends $s\to-s$ and leaves both Fisher informations unchanged, so the gain is even in $s$. We derive
\begin{equation}
 \frac{\FI_{\mathrm{Bell}}}{\FI_{\mathrm{local}}^{\max}}-1
 =\frac{(c-1)\left[2\lambda-(c-1)\right]}{4}\,s^2+O(s^4),
 \label{eq:onset}
\end{equation}
with $c=2x\coth 2x$ and $\lambda=2x/\sinh 2x$. Here $\lambda=\mu/\nu$, and $c$ grows as the mode freezes out. Since $\coth u>1/u$, the factor $c-1$ is positive at every temperature, so the gain is positive exactly when $2\lambda>c-1$. That inequality reads $\sinh 2x+4x-2x\cosh 2x>0$, and its left side crosses zero once, at $x_w=1.0095565499$. Equation~\eqref{eq:onset} compares against the homodyne branch of Eq.~\eqref{eq:localmode}, the better branch only for $x>x_c$, so this homodyne-branch coefficient is positive on $x\in(x_c,x_w)$ [Fig.~\ref{fig:universality}(a)]. Both $x_c$ and $x_w$ are pure numbers, so any small splitting opens a window at any frequency, and covariance-only additivity fails somewhere in temperature. In temperature the window is a band whose width approaches the fixed factor $x_w/x_c=2.2909$ as the splitting vanishes, and a larger splitting narrows it but never closes it (Supplemental Sec.~S5) [Fig.~\ref{fig:effect}(b,c)]. For every width law the homodyne-branch gain is $\psi_\lambda(2\psi_\nu-\psi_\lambda)$ at leading order, with $\psi_\nu$ and $\psi_\lambda$ the half-differences of $\ln\nu$ and $\ln\lambda$ between the modes. It is positive when $\psi_\lambda$ and $2\psi_\nu-\psi_\lambda$ have the same sign (End Matter B; Supplemental Secs.~S4 and S6).

\textit{The universal bound.---}
Label the modes so that $x_2>x_1$. Over every such pair we prove
\begin{equation}
  \frac{\FI_{\mathrm{Bell}}}{\FI_{\mathrm{local}}^{\max}}-1\;\le\;12.699322\%,
  \label{eq:cap}
\end{equation}
attained only at $x_1=x_c$ exactly and $x_2=1.46219$, that is $\omega_2/\omega_1=3.31798$ [Fig.~\ref{fig:effect}(b)] (Supplemental Sec.~S5). The first value is the crossover itself, where Eq.~\eqref{eq:localmode} changes branch and has no derivative. At the optimal $x_2$ the one-sided slopes are $+0.4214$ rising into $x_c$ and $-0.7494$ falling away, and their opposite signs put the maximum on the kink [Fig.~\ref{fig:universality}(b)]. For arbitrary widths and responses write $\Gamma(\nu)=\min\{2\nu^2,(\nu+1)^2\}$ for the reciprocal of the bracket in Eq.~\eqref{eq:localmode}; we prove
\begin{equation}
 \frac{\FI_{\mathrm{Bell}}}{\FI_{\mathrm{local}}^{\max}}
 \le\frac{\Gamma(\nu_1)+\Gamma(\nu_2)}{(\nu_1+\nu_2)^2}\le 4-2\sqrt2 .
 \label{eq:universal}
\end{equation}
The first inequality is the best response ratio, tight when $\mu_2/\mu_1=\Gamma(\nu_2)/\Gamma(\nu_1)$; the second is a case analysis over the two branches (End Matter C; Supplemental Secs.~S5--S6). The supremum, a gain of $(\sqrt2-1)^2=17.157\%$, is approached only as one width tends to the vacuum value while the other sits at $\nu_c$ and responds $\nu_c^2$ times as strongly. No smooth family attains it, since a width that touches the vacuum has stopped responding. Thermal equilibrium reaches $12.699\%$ with its response ratio fixed by the thermal law, the bound at that pair is $13.60\%$, and both are pinned to the kink. The bound does not grow with the number of modes. For $N$ modes the best passive network followed by homodynes is a disjoint pairing, proved for $N=3$ and found in every configuration tested at $N=4$ and $5$ (Supplemental Sec.~S6), so the excess over the separate-readout optimum is pairwise and stays below $(\sqrt2-1)^2$.

\textit{Gaussian optimality.---}
We now search every Gaussian receiver. Every pure two-mode seed is one beam splitter acting on two squeezed vacua, and End Matter D proves that tilting one squeezing axis against the other lowers the Fisher information. The aligned seeds that remain carry two squeezings and one mixing angle, compressed into a finite box with the infinite-squeezing limits on its faces. At the reference pair $x_1=x_c$, $x_2=3/2$ we prove in interval arithmetic that Bell homodyne beats every seed in the box, tying only at the two balanced corners where the seed is the Bell measurement itself. Figure~\ref{fig:optimality}(a) draws the same shortfall at the pair of Eq.~\eqref{eq:cap}, covered by the rectangle proof of Supplemental Sec.~S9. With the Gaussian Fisher information $\FI_{\mathrm{global}}^{\mathrm G}$, the largest Fisher information any Gaussian measurement on the two modes reaches, we have proved $\FI_{\mathrm{global}}^{\mathrm G}=\FI_{\mathrm{Bell}}>\FI_{\mathrm{local}}^{\max}$ at both pairs. With Taylor models for $\nu_i$ and $\mu_i$ (End Matter D; Supplemental Secs.~S7--S10), we prove Bell optimality for every thermal pair with $x_1\in[0.36,0.52]$ and $x_2\in[1.15,1.85]$, around the maximal-gain ridge at frequency ratios $2.2$ to $5.1$. A second rectangle, $x_1\in[0.44,0.50]$ and $x_2\in[0.82,0.94]$, sits near ratio $1.9$. Passive receivers need no rectangles: for every width and response, no passive network of the two modes and one vacuum ancilla beats the better of Bell homodyne and the separate-readout optimum, by the three-mode pairing theorem above with the ancilla as third mode (Supplemental Sec.~S6).

\begin{figure}[!t]
 \includegraphics{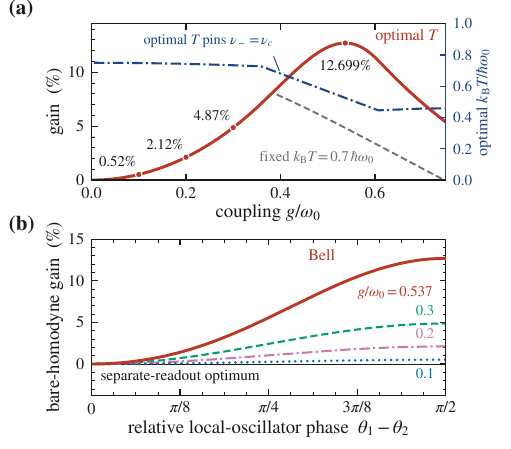}
 \caption{\label{fig:resonators}
 \textbf{(a)} Bell gain over the best separate readout of the normal modes for two identical resonators of frequency $\omega_0$ and coupling $g$. Red: temperature re-optimized at each $g$; grey dashed: $k_{\mathrm B}T=0.7\,\hbar\omega_0$. Marked points at $g/\omega_0=0.1$, $0.2$, $0.3$ and $0.5368$. Right axis, dash-dot: the optimal $k_{\mathrm B}T/\hbar\omega_0$, equal to $(1-g/\omega_0)/2x_c$ for $0.360\le g/\omega_0\le0.607$, so the lower normal mode sits on the crossover. \textbf{(b)} Gain of two bare-resonator homodynes against their relative local-oscillator phase at those couplings, from the separate-readout optimum of the normal modes at $0$ to the Bell value at $\pi/2$.}
\end{figure}

\textit{Coupled resonators.---}
Two identical resonators with a tunable coupling are a common platform~\cite{Miyanaga2021,Bothner2021}, and the second of those rectangles covers an operating point at $g/\omega_0=0.3$. A coupling $g$ that swaps excitations between them makes normal modes $(a_1\pm a_2)/\sqrt2$ at frequencies $\omega_0\pm g$, so the coupled Gibbs state is an unequal-frequency thermal pair of ratio $(\omega_0+g)/(\omega_0-g)$. With the temperature optimized at each coupling, the gain over the best separate readout of the normal modes rises from $0.52\%$ at $g/\omega_0=0.1$ to $12.699\%$ at $0.5368$, where the normal-mode ratio is $3.318$ [Fig.~\ref{fig:resonators}(a)]. The two strongest marked couplings lie inside the two rectangles, so Bell homodyne is the Gaussian optimum there. The Bell splitter applied after the normal-mode transformation is the exchange matrix, so the Bell quadratures of the normal modes are the bare-resonator quadratures $q_2$ and $p_1$. Two ordinary homodyne detectors with local oscillators set $90^\circ$ apart reproduce the Bell statistics exactly under the rotating-wave coupling. The coupled state is a thermal product only in the normal-mode basis, so the conjecture's probes are the normal modes, and a separate readout of them has to recombine the two resonators. The collective advantage is a phase setting, and no intermediate phase beats the endpoints [Fig.~\ref{fig:resonators}(b)]. The exact bilinear coupling $\varkappa q_1q_2$ moves the normal modes to $\omega_\pm=\omega_0\sqrt{1\pm\varkappa/\omega_0}$ and squeezes them relative to the bare quadratures. We show that the optimal gain stays the same function of the normal-mode ratio, delivered by one fixed unbalanced beam splitter, and that the bare homodynes still keep $8.45\%$ at the maximal-gain ratio (Supplemental Secs.~S11--S12).

\textit{Heterodyne with a live ancilla.---}
The uncertainty principle acts twice. It forbids sharp readings of $q$ and $p$ of one mode, so heterodyne carries one extra vacuum unit. It allows sharp readings of $q_1-q_2$ and $p_1+p_2$ of two modes, the commuting pair of Einstein, Podolsky and Rosen, and Bell homodyne reads that pair. Send the second mode to vacuum, a temperature-independent width $\nu_2=1$ with $\mu_2=0$, and Eq.~\eqref{eq:bellFI} returns $\mu_1^2/(\nu_1+1)^2$, the heterodyne branch of Eq.~\eqref{eq:localmode} for mode 1 alone. The Arthurs--Kelly unit of noise~\cite{ArthursKelly1965} that an empty port would add is now shared between two probes and carries signal, and the advantage appears once that signal outweighs the noise, which unequal responses decide. Identical modes have nothing to collect, and there Bell homodyne at best ties. At high temperature both widths run far above $\nu_c$, so Eq.~\eqref{eq:localmode} reads each mode by heterodyne, and each response approaches its own width, $\mu_i\to\nu_i$. Each heterodyne reading then carries a full unit, $\mu_i^2/(\nu_i+1)^2\to1$, while the two Bell records carry $[(\mu_1+\mu_2)/(\nu_1+\nu_2)]^2\to1$ between them, so separate readout wins at every frequency pair. The widths also far exceed the one unit of vacuum noise the extra port costs. One limit removes the mismatch and the other removes the cost of reading two quadratures at once. Sharing a temperature fixes $T$ alone, while the additivity theorem needs one $\nu$~\cite{Navarro2025}. Our inputs stay uncorrelated Gibbs states and our passive network adds no energy, so the resource is in the widths~\cite{Chatterjee2026}; nonequilibrium schemes work the other way, with squeezed or correlated probes~\cite{Mirkhalaf2024}.

\textit{The amplifier as receiver.---}
A phase-preserving amplifier of gain $G$ maps the signal $a_s$ to $b_s=\sqrt G\,a_s+\sqrt{G-1}\,a_i^\dagger$, where $a_i$ is the idler. At large gain its output, read in phase and in quadrature, measures $q_s+q_i$ and $p_s-p_i$, the Bell pair of signal and idler. With a vacuum idler this measurement is heterodyne, and the idler's vacuum is Caves's half quantum of added noise~\cite{Caves1982,Clerk2010}. An idler band fed by the same thermal source makes it Bell homodyne on the pair $(\omega_s,\omega_i)$. The frequencies differ by construction, with $\omega_s+\omega_i=2\omega_p$ under a four-wave pump at $\omega_p$~\cite{Macklin2015} and $\omega_s+\omega_i=\omega_p$ under a three-wave pump~\cite{Bergeal2010,RoyDevoret2016}. Feed a matched load at $218$ mK to a traveling-wave amplifier pumped at $7$ GHz, record the $4$ GHz output in phase and in quadrature, and estimate the temperature from the record's variance; a cold load on the $10$ GHz idler band gives the heterodyne benchmark and a degenerate pump the homodyne one. Transmission $L$ before the amplifier and excess noise $n_{\mathrm{ex}}$ enter as one common homodyne efficiency $\eta=L/(1+n_{\mathrm{ex}})$ of every record, the loss model of the next paragraph (Supplemental Sec.~S13). The Bell gain is $10.0\%$ ideally, $7.7\%$ at $\eta=0.8$ and $5.0\%$ at $0.65$, and twenty thousand samples resolve a five percent difference. Those numbers assume an ideal first stage; at $30$ dB the same-chain gain is $9.4\%$, and at $20$ dB, where the degenerate benchmark receives a quarter of the following amplifier's noise, it drops to $4.0\%$ (End Matter F; Supplemental Sec.~S15).

\textit{Losses and the Gaussian limit.---}
At the maximal-gain pair, Bell homodyne beats an equally lossy separate readout once the common homodyne efficiency exceeds $57.95\%$, and the ideal lossless bound once its own efficiency exceeds $90.20\%$ [Fig.~\ref{fig:optimality}(c); Supplemental Sec.~S13]. The quantum Fisher information of a thermal mode, $T^2\FI_Q=x^2\csch^2x$, is reached by photon counting, and the best separate readout keeps the fraction $\max\{1/(2\cosh^2x),e^{-2x}\}$ of it, $41.4\%$ at the crossover. At the maximal-gain pair separate readout keeps $30.2\%$ of the pair's quantum Fisher information and Bell homodyne $34.0\%$. Every Gaussian receiver shares that gap. Double heterodyne, the natural starting seed, is an exact stationary point of Eq.~\eqref{eq:gaussianFI} for every thermal product state, so a gradient search from it never finds the gain; the optimum lies at infinite squeezing, and below $5.4$ dB the best finite seed still loses (End Matter E; Supplemental Sec.~S13).

\textit{Discussion and conclusions.---}
Parametric frequency conversion reaches modes that are neither coupled nor amplified together~\cite{Abdo2013,Lecocq2016}, a thermal field is a bank of many modes at one temperature, and a four-wave pump pairs that bank band by band (Supplemental Sec.~S14). In conclusion, the Gaussian Fisher information of uncorrelated modes is superadditive. A collective measurement beats the best separate readout for a single parameter, the excess is pairwise, and we have traced it to one mechanism, the second probe filling the vacuum port of a heterodyne. We have bounded the gain by $(\sqrt2-1)^2$ for every width law and proved, in interval arithmetic, both that its thermal maximum of $12.699\%$ sits on the homodyne-heterodyne crossover and that no Gaussian measurement does better there. Every heterodyne detector and every phase-preserving amplifier has such a port. A second probe of the same parameter in that port turns the vacuum unit of added noise into signal.

\noindent\textit{Acknowledgments.---}\ This work is supported by the Fundamental
and Interdisciplinary Disciplines Breakthrough Plan of the Ministry of Education
of China under Project No.~JYB2025XDXM606 and by the National Natural Science
Foundation of China under Grant No.~42388101.

\vspace{-6pt}%
\makeatletter\let\ms@addcontentsline\addcontentsline\renewcommand{\addcontentsline}[3]{}\makeatother
\section*{Data Availability}%
\makeatletter\let\addcontentsline\ms@addcontentsline\makeatother
\vspace{-8pt}
The data that support the findings of this Letter are available within the
Letter and its Supplemental Material; no additional data were generated.

\makeatletter
\let\ms@addcontentsline\addcontentsline
\renewcommand{\addcontentsline}[3]{}
\bibliography{references}

\begin{thebibliography}{63}%
\makeatletter
\providecommand \@ifxundefined [1]{%
 \@ifx{#1\undefined}
}%
\providecommand \@ifnum [1]{%
 \ifnum #1\expandafter \@firstoftwo
 \else \expandafter \@secondoftwo
 \fi
}%
\providecommand \@ifx [1]{%
 \ifx #1\expandafter \@firstoftwo
 \else \expandafter \@secondoftwo
 \fi
}%
\providecommand \natexlab [1]{#1}%
\providecommand \enquote  [1]{``#1''}%
\providecommand \bibnamefont  [1]{#1}%
\providecommand \bibfnamefont [1]{#1}%
\providecommand \citenamefont [1]{#1}%
\providecommand \href@noop [0]{\@secondoftwo}%
\providecommand \href [0]{\begingroup \@sanitize@url \@href}%
\providecommand \@href[1]{\@@startlink{#1}\@@href}%
\providecommand \@@href[1]{\endgroup#1\@@endlink}%
\providecommand \@sanitize@url [0]{\catcode `\\12\catcode `\$12\catcode
  `\&12\catcode `\#12\catcode `\^12\catcode `\_12\catcode `\%12\relax}%
\providecommand \@@startlink[1]{}%
\providecommand \@@endlink[0]{}%
\providecommand \url  [0]{\begingroup\@sanitize@url \@url }%
\providecommand \@url [1]{\endgroup\@href {#1}{\urlprefix }}%
\providecommand \urlprefix  [0]{URL }%
\providecommand \Eprint [0]{\href }%
\providecommand \doibase [0]{https://doi.org/}%
\providecommand \selectlanguage [0]{\@gobble}%
\providecommand \bibinfo  [0]{\@secondoftwo}%
\providecommand \bibfield  [0]{\@secondoftwo}%
\providecommand \translation [1]{[#1]}%
\providecommand \BibitemOpen [0]{}%
\providecommand \bibitemStop [0]{}%
\providecommand \bibitemNoStop [0]{.\EOS\space}%
\providecommand \EOS [0]{\spacefactor3000\relax}%
\providecommand \BibitemShut  [1]{\csname bibitem#1\endcsname}%
\let\auto@bib@innerbib\@empty
\bibitem [{\citenamefont {Braunstein}\ and\ \citenamefont
  {Caves}(1994)}]{BraunsteinCaves1994}%
  \BibitemOpen
  \bibfield  {author} {\bibinfo {author} {\bibfnamefont {S.~L.}\ \bibnamefont
  {Braunstein}}\ and\ \bibinfo {author} {\bibfnamefont {C.~M.}\ \bibnamefont
  {Caves}},\ }\bibfield  {title} {\bibinfo {title} {Statistical distance and
  the geometry of quantum states},\ }\href
  {https://doi.org/10.1103/PhysRevLett.72.3439} {\bibfield  {journal} {\bibinfo
   {journal} {Phys. Rev. Lett.}\ }\textbf {\bibinfo {volume} {72}},\ \bibinfo
  {pages} {3439} (\bibinfo {year} {1994})}\BibitemShut {NoStop}%
\bibitem [{\citenamefont {Mehboudi}\ \emph {et~al.}(2019)\citenamefont
  {Mehboudi}, \citenamefont {Sanpera},\ and\ \citenamefont
  {Correa}}]{Mehboudi2019}%
  \BibitemOpen
  \bibfield  {author} {\bibinfo {author} {\bibfnamefont {M.}~\bibnamefont
  {Mehboudi}}, \bibinfo {author} {\bibfnamefont {A.}~\bibnamefont {Sanpera}},\
  and\ \bibinfo {author} {\bibfnamefont {L.~A.}\ \bibnamefont {Correa}},\
  }\bibfield  {title} {\bibinfo {title} {Thermometry in the quantum regime:
  Recent theoretical progress},\ }\href
  {https://doi.org/10.1088/1751-8121/ab2828} {\bibfield  {journal} {\bibinfo
  {journal} {J. Phys. A: Math. Theor.}\ }\textbf {\bibinfo {volume} {52}},\
  \bibinfo {pages} {303001} (\bibinfo {year} {2019})}\BibitemShut {NoStop}%
\bibitem [{\citenamefont {Correa}\ \emph {et~al.}(2015)\citenamefont {Correa},
  \citenamefont {Mehboudi}, \citenamefont {Adesso},\ and\ \citenamefont
  {Sanpera}}]{Correa2015}%
  \BibitemOpen
  \bibfield  {author} {\bibinfo {author} {\bibfnamefont {L.~A.}\ \bibnamefont
  {Correa}}, \bibinfo {author} {\bibfnamefont {M.}~\bibnamefont {Mehboudi}},
  \bibinfo {author} {\bibfnamefont {G.}~\bibnamefont {Adesso}},\ and\ \bibinfo
  {author} {\bibfnamefont {A.}~\bibnamefont {Sanpera}},\ }\bibfield  {title}
  {\bibinfo {title} {Individual quantum probes for optimal thermometry},\
  }\href {https://doi.org/10.1103/PhysRevLett.114.220405} {\bibfield  {journal}
  {\bibinfo  {journal} {Phys. Rev. Lett.}\ }\textbf {\bibinfo {volume} {114}},\
  \bibinfo {pages} {220405} (\bibinfo {year} {2015})}\BibitemShut {NoStop}%
\bibitem [{\citenamefont {Demkowicz-Dobrza{\'n}ski}\ \emph
  {et~al.}(2020)\citenamefont {Demkowicz-Dobrza{\'n}ski}, \citenamefont
  {G{\'o}recki},\ and\ \citenamefont {Gu{\c t}{\u a}}}]{Demkowicz2020}%
  \BibitemOpen
  \bibfield  {author} {\bibinfo {author} {\bibfnamefont {R.}~\bibnamefont
  {Demkowicz-Dobrza{\'n}ski}}, \bibinfo {author} {\bibfnamefont
  {W.}~\bibnamefont {G{\'o}recki}},\ and\ \bibinfo {author} {\bibfnamefont
  {M.}~\bibnamefont {Gu{\c t}{\u a}}},\ }\bibfield  {title} {\bibinfo {title}
  {Multi-parameter estimation beyond quantum {F}isher information},\ }\href
  {https://doi.org/10.1088/1751-8121/ab8ef3} {\bibfield  {journal} {\bibinfo
  {journal} {J. Phys. A: Math. Theor.}\ }\textbf {\bibinfo {volume} {53}},\
  \bibinfo {pages} {363001} (\bibinfo {year} {2020})}\BibitemShut {NoStop}%
\bibitem [{\citenamefont {Glatthard}\ \emph {et~al.}(2023)\citenamefont
  {Glatthard}, \citenamefont {Hovhannisyan}, \citenamefont {Perarnau-Llobet},
  \citenamefont {Correa},\ and\ \citenamefont {Miller}}]{Glatthard2023}%
  \BibitemOpen
  \bibfield  {author} {\bibinfo {author} {\bibfnamefont {J.}~\bibnamefont
  {Glatthard}}, \bibinfo {author} {\bibfnamefont {K.~V.}\ \bibnamefont
  {Hovhannisyan}}, \bibinfo {author} {\bibfnamefont {M.}~\bibnamefont
  {Perarnau-Llobet}}, \bibinfo {author} {\bibfnamefont {L.~A.}\ \bibnamefont
  {Correa}},\ and\ \bibinfo {author} {\bibfnamefont {H.~J.~D.}\ \bibnamefont
  {Miller}},\ }\bibfield  {title} {\bibinfo {title} {Energy measurements remain
  thermometrically optimal beyond weak coupling},\ }\href
  {https://doi.org/10.22331/q-2023-11-28-1190} {\bibfield  {journal} {\bibinfo
  {journal} {Quantum}\ }\textbf {\bibinfo {volume} {7}},\ \bibinfo {pages}
  {1190} (\bibinfo {year} {2023})}\BibitemShut {NoStop}%
\bibitem [{\citenamefont {Seah}\ \emph {et~al.}(2019)\citenamefont {Seah},
  \citenamefont {Nimmrichter}, \citenamefont {Grimmer}, \citenamefont {Santos},
  \citenamefont {Scarani},\ and\ \citenamefont {Landi}}]{Seah2019}%
  \BibitemOpen
  \bibfield  {author} {\bibinfo {author} {\bibfnamefont {S.}~\bibnamefont
  {Seah}}, \bibinfo {author} {\bibfnamefont {S.}~\bibnamefont {Nimmrichter}},
  \bibinfo {author} {\bibfnamefont {D.}~\bibnamefont {Grimmer}}, \bibinfo
  {author} {\bibfnamefont {J.~P.}\ \bibnamefont {Santos}}, \bibinfo {author}
  {\bibfnamefont {V.}~\bibnamefont {Scarani}},\ and\ \bibinfo {author}
  {\bibfnamefont {G.~T.}\ \bibnamefont {Landi}},\ }\bibfield  {title} {\bibinfo
  {title} {Collisional quantum thermometry},\ }\href
  {https://doi.org/10.1103/PhysRevLett.123.180602} {\bibfield  {journal}
  {\bibinfo  {journal} {Phys. Rev. Lett.}\ }\textbf {\bibinfo {volume} {123}},\
  \bibinfo {pages} {180602} (\bibinfo {year} {2019})}\BibitemShut {NoStop}%
\bibitem [{\citenamefont {Potts}\ \emph {et~al.}(2019)\citenamefont {Potts},
  \citenamefont {Brask},\ and\ \citenamefont {Brunner}}]{Potts2019}%
  \BibitemOpen
  \bibfield  {author} {\bibinfo {author} {\bibfnamefont {P.~P.}\ \bibnamefont
  {Potts}}, \bibinfo {author} {\bibfnamefont {J.~B.}\ \bibnamefont {Brask}},\
  and\ \bibinfo {author} {\bibfnamefont {N.}~\bibnamefont {Brunner}},\
  }\bibfield  {title} {\bibinfo {title} {Fundamental limits on low-temperature
  quantum thermometry with finite resolution},\ }\href
  {https://doi.org/10.22331/q-2019-07-09-161} {\bibfield  {journal} {\bibinfo
  {journal} {Quantum}\ }\textbf {\bibinfo {volume} {3}},\ \bibinfo {pages}
  {161} (\bibinfo {year} {2019})}\BibitemShut {NoStop}%
\bibitem [{\citenamefont {Hovhannisyan}\ \emph {et~al.}(2021)\citenamefont
  {Hovhannisyan}, \citenamefont {J{\o}rgensen}, \citenamefont {Landi},
  \citenamefont {Alhambra}, \citenamefont {Brask},\ and\ \citenamefont
  {Perarnau-Llobet}}]{Hovhannisyan2021}%
  \BibitemOpen
  \bibfield  {author} {\bibinfo {author} {\bibfnamefont {K.~V.}\ \bibnamefont
  {Hovhannisyan}}, \bibinfo {author} {\bibfnamefont {M.~R.}\ \bibnamefont
  {J{\o}rgensen}}, \bibinfo {author} {\bibfnamefont {G.~T.}\ \bibnamefont
  {Landi}}, \bibinfo {author} {\bibfnamefont {{\'A}.~M.}\ \bibnamefont
  {Alhambra}}, \bibinfo {author} {\bibfnamefont {J.~B.}\ \bibnamefont
  {Brask}},\ and\ \bibinfo {author} {\bibfnamefont {M.}~\bibnamefont
  {Perarnau-Llobet}},\ }\bibfield  {title} {\bibinfo {title} {Optimal quantum
  thermometry with coarse-grained measurements},\ }\href
  {https://doi.org/10.1103/PRXQuantum.2.020322} {\bibfield  {journal} {\bibinfo
   {journal} {PRX Quantum}\ }\textbf {\bibinfo {volume} {2}},\ \bibinfo {pages}
  {020322} (\bibinfo {year} {2021})}\BibitemShut {NoStop}%
\bibitem [{\citenamefont {Weedbrook}\ \emph {et~al.}(2012)\citenamefont
  {Weedbrook}, \citenamefont {Pirandola}, \citenamefont {Garcia-Patron},
  \citenamefont {Cerf}, \citenamefont {Ralph}, \citenamefont {Shapiro},\ and\
  \citenamefont {Lloyd}}]{Weedbrook2012}%
  \BibitemOpen
  \bibfield  {author} {\bibinfo {author} {\bibfnamefont {C.}~\bibnamefont
  {Weedbrook}}, \bibinfo {author} {\bibfnamefont {S.}~\bibnamefont
  {Pirandola}}, \bibinfo {author} {\bibfnamefont {R.}~\bibnamefont
  {Garcia-Patron}}, \bibinfo {author} {\bibfnamefont {N.~J.}\ \bibnamefont
  {Cerf}}, \bibinfo {author} {\bibfnamefont {T.~C.}\ \bibnamefont {Ralph}},
  \bibinfo {author} {\bibfnamefont {J.~H.}\ \bibnamefont {Shapiro}},\ and\
  \bibinfo {author} {\bibfnamefont {S.}~\bibnamefont {Lloyd}},\ }\bibfield
  {title} {\bibinfo {title} {{G}aussian quantum information},\ }\href
  {https://doi.org/10.1103/RevModPhys.84.621} {\bibfield  {journal} {\bibinfo
  {journal} {Rev. Mod. Phys.}\ }\textbf {\bibinfo {volume} {84}},\ \bibinfo
  {pages} {621} (\bibinfo {year} {2012})}\BibitemShut {NoStop}%
\bibitem [{\citenamefont {Braunstein}\ and\ \citenamefont {van
  Loock}(2005)}]{BraunsteinVanLoock2005}%
  \BibitemOpen
  \bibfield  {author} {\bibinfo {author} {\bibfnamefont {S.~L.}\ \bibnamefont
  {Braunstein}}\ and\ \bibinfo {author} {\bibfnamefont {P.}~\bibnamefont {van
  Loock}},\ }\bibfield  {title} {\bibinfo {title} {Quantum information with
  continuous variables},\ }\href {https://doi.org/10.1103/RevModPhys.77.513}
  {\bibfield  {journal} {\bibinfo  {journal} {Rev. Mod. Phys.}\ }\textbf
  {\bibinfo {volume} {77}},\ \bibinfo {pages} {513} (\bibinfo {year}
  {2005})}\BibitemShut {NoStop}%
\bibitem [{\citenamefont {Oh}\ \emph {et~al.}(2019)\citenamefont {Oh},
  \citenamefont {Lee}, \citenamefont {Rockstuhl}, \citenamefont {Jeong},
  \citenamefont {Kim}, \citenamefont {Nha},\ and\ \citenamefont
  {Lee}}]{Oh2019}%
  \BibitemOpen
  \bibfield  {author} {\bibinfo {author} {\bibfnamefont {C.}~\bibnamefont
  {Oh}}, \bibinfo {author} {\bibfnamefont {C.}~\bibnamefont {Lee}}, \bibinfo
  {author} {\bibfnamefont {C.}~\bibnamefont {Rockstuhl}}, \bibinfo {author}
  {\bibfnamefont {H.}~\bibnamefont {Jeong}}, \bibinfo {author} {\bibfnamefont
  {J.}~\bibnamefont {Kim}}, \bibinfo {author} {\bibfnamefont {H.}~\bibnamefont
  {Nha}},\ and\ \bibinfo {author} {\bibfnamefont {S.-Y.}\ \bibnamefont {Lee}},\
  }\bibfield  {title} {\bibinfo {title} {Optimal {G}aussian measurements for
  phase estimation in single-mode {G}aussian metrology},\ }\href
  {https://doi.org/10.1038/s41534-019-0124-4} {\bibfield  {journal} {\bibinfo
  {journal} {npj Quantum Inf.}\ }\textbf {\bibinfo {volume} {5}},\ \bibinfo
  {pages} {10} (\bibinfo {year} {2019})}\BibitemShut {NoStop}%
\bibitem [{\citenamefont {{\v S}afr{\'a}nek}(2019)}]{Safranek2019}%
  \BibitemOpen
  \bibfield  {author} {\bibinfo {author} {\bibfnamefont {D.}~\bibnamefont {{\v
  S}afr{\'a}nek}},\ }\bibfield  {title} {\bibinfo {title} {Estimation of
  {G}aussian quantum states},\ }\href
  {https://doi.org/10.1088/1751-8121/aaf068} {\bibfield  {journal} {\bibinfo
  {journal} {J. Phys. A: Math. Theor.}\ }\textbf {\bibinfo {volume} {52}},\
  \bibinfo {pages} {035304} (\bibinfo {year} {2019})}\BibitemShut {NoStop}%
\bibitem [{\citenamefont {Nichols}\ \emph {et~al.}(2018)\citenamefont
  {Nichols}, \citenamefont {Liuzzo-Scorpo}, \citenamefont {Knott},\ and\
  \citenamefont {Adesso}}]{Nichols2018}%
  \BibitemOpen
  \bibfield  {author} {\bibinfo {author} {\bibfnamefont {R.}~\bibnamefont
  {Nichols}}, \bibinfo {author} {\bibfnamefont {P.}~\bibnamefont
  {Liuzzo-Scorpo}}, \bibinfo {author} {\bibfnamefont {P.~A.}\ \bibnamefont
  {Knott}},\ and\ \bibinfo {author} {\bibfnamefont {G.}~\bibnamefont
  {Adesso}},\ }\bibfield  {title} {\bibinfo {title} {Multiparameter {G}aussian
  quantum metrology},\ }\href {https://doi.org/10.1103/PhysRevA.98.012114}
  {\bibfield  {journal} {\bibinfo  {journal} {Phys. Rev. A}\ }\textbf {\bibinfo
  {volume} {98}},\ \bibinfo {pages} {012114} (\bibinfo {year}
  {2018})}\BibitemShut {NoStop}%
\bibitem [{\citenamefont {Bressanini}\ \emph {et~al.}(2024)\citenamefont
  {Bressanini}, \citenamefont {Genoni}, \citenamefont {Kim},\ and\
  \citenamefont {Paris}}]{Bressanini2024}%
  \BibitemOpen
  \bibfield  {author} {\bibinfo {author} {\bibfnamefont {G.}~\bibnamefont
  {Bressanini}}, \bibinfo {author} {\bibfnamefont {M.~G.}\ \bibnamefont
  {Genoni}}, \bibinfo {author} {\bibfnamefont {M.~S.}\ \bibnamefont {Kim}},\
  and\ \bibinfo {author} {\bibfnamefont {M.~G.~A.}\ \bibnamefont {Paris}},\
  }\bibfield  {title} {\bibinfo {title} {Multi-parameter quantum estimation of
  single- and two-mode pure {G}aussian states},\ }\href
  {https://doi.org/10.1088/1751-8121/ad6364} {\bibfield  {journal} {\bibinfo
  {journal} {J. Phys. A: Math. Theor.}\ }\textbf {\bibinfo {volume} {57}},\
  \bibinfo {pages} {315305} (\bibinfo {year} {2024})}\BibitemShut {NoStop}%
\bibitem [{\citenamefont {Chang}\ \emph {et~al.}(2026)\citenamefont {Chang},
  \citenamefont {Genoni},\ and\ \citenamefont {Albarelli}}]{Chang2026}%
  \BibitemOpen
  \bibfield  {author} {\bibinfo {author} {\bibfnamefont {S.}~\bibnamefont
  {Chang}}, \bibinfo {author} {\bibfnamefont {M.~G.}\ \bibnamefont {Genoni}},\
  and\ \bibinfo {author} {\bibfnamefont {F.}~\bibnamefont {Albarelli}},\
  }\bibfield  {title} {\bibinfo {title} {Efficiently evaluating {H}olevo, {RLD}
  and {SLD} {C}ram{\'e}r-{R}ao bounds for multiparameter quantum estimation
  with {G}aussian states},\ }\href {https://doi.org/10.1038/s42005-026-02550-6}
  {\bibfield  {journal} {\bibinfo  {journal} {Commun. Phys.}\ }\textbf
  {\bibinfo {volume} {9}},\ \bibinfo {pages} {126} (\bibinfo {year}
  {2026})}\BibitemShut {NoStop}%
\bibitem [{\citenamefont {Fadel}\ \emph {et~al.}(2025)\citenamefont {Fadel},
  \citenamefont {Roux},\ and\ \citenamefont {Gessner}}]{Fadel2025}%
  \BibitemOpen
  \bibfield  {author} {\bibinfo {author} {\bibfnamefont {M.}~\bibnamefont
  {Fadel}}, \bibinfo {author} {\bibfnamefont {N.}~\bibnamefont {Roux}},\ and\
  \bibinfo {author} {\bibfnamefont {M.}~\bibnamefont {Gessner}},\ }\bibfield
  {title} {\bibinfo {title} {Quantum metrology with a continuous-variable
  system},\ }\href {https://doi.org/10.1088/1361-6633/ae00d8} {\bibfield
  {journal} {\bibinfo  {journal} {Rep. Prog. Phys.}\ }\textbf {\bibinfo
  {volume} {88}},\ \bibinfo {pages} {106001} (\bibinfo {year}
  {2025})}\BibitemShut {NoStop}%
\bibitem [{\citenamefont {\emph{et al.} ({LIGO} {O}4~{D}etector
  {C}ollaboration)}(2023)}]{Ganapathy2023}%
  \BibitemOpen
  \bibfield  {author} {\bibinfo {author} {\bibfnamefont {D.~G.}\ \bibnamefont
  {\emph{et al.} ({LIGO} {O}4~{D}etector {C}ollaboration)}},\ }\bibfield
  {title} {\bibinfo {title} {Broadband quantum enhancement of the {LIGO}
  detectors with frequency-dependent squeezing},\ }\href
  {https://doi.org/10.1103/PhysRevX.13.041021} {\bibfield  {journal} {\bibinfo
  {journal} {Phys. Rev. X}\ }\textbf {\bibinfo {volume} {13}},\ \bibinfo
  {pages} {041021} (\bibinfo {year} {2023})}\BibitemShut {NoStop}%
\bibitem [{\citenamefont {Guo}\ \emph {et~al.}(2020)\citenamefont {Guo},
  \citenamefont {Breum}, \citenamefont {Borregaard}, \citenamefont {Izumi},
  \citenamefont {Larsen}, \citenamefont {Gehring}, \citenamefont {Christandl},
  \citenamefont {Neergaard-Nielsen},\ and\ \citenamefont {Andersen}}]{Guo2020}%
  \BibitemOpen
  \bibfield  {author} {\bibinfo {author} {\bibfnamefont {X.}~\bibnamefont
  {Guo}}, \bibinfo {author} {\bibfnamefont {C.~R.}\ \bibnamefont {Breum}},
  \bibinfo {author} {\bibfnamefont {J.}~\bibnamefont {Borregaard}}, \bibinfo
  {author} {\bibfnamefont {S.}~\bibnamefont {Izumi}}, \bibinfo {author}
  {\bibfnamefont {M.~V.}\ \bibnamefont {Larsen}}, \bibinfo {author}
  {\bibfnamefont {T.}~\bibnamefont {Gehring}}, \bibinfo {author} {\bibfnamefont
  {M.}~\bibnamefont {Christandl}}, \bibinfo {author} {\bibfnamefont {J.~S.}\
  \bibnamefont {Neergaard-Nielsen}},\ and\ \bibinfo {author} {\bibfnamefont
  {U.~L.}\ \bibnamefont {Andersen}},\ }\bibfield  {title} {\bibinfo {title}
  {Distributed quantum sensing in a continuous-variable entangled network},\
  }\href {https://doi.org/10.1038/s41567-019-0743-x} {\bibfield  {journal}
  {\bibinfo  {journal} {Nat. Phys.}\ }\textbf {\bibinfo {volume} {16}},\
  \bibinfo {pages} {281} (\bibinfo {year} {2020})}\BibitemShut {NoStop}%
\bibitem [{\citenamefont {Xia}\ \emph {et~al.}(2020)\citenamefont {Xia},
  \citenamefont {Li}, \citenamefont {Clark}, \citenamefont {Hart},
  \citenamefont {Zhuang},\ and\ \citenamefont {Zhang}}]{Xia2020}%
  \BibitemOpen
  \bibfield  {author} {\bibinfo {author} {\bibfnamefont {Y.}~\bibnamefont
  {Xia}}, \bibinfo {author} {\bibfnamefont {W.}~\bibnamefont {Li}}, \bibinfo
  {author} {\bibfnamefont {W.}~\bibnamefont {Clark}}, \bibinfo {author}
  {\bibfnamefont {D.}~\bibnamefont {Hart}}, \bibinfo {author} {\bibfnamefont
  {Q.}~\bibnamefont {Zhuang}},\ and\ \bibinfo {author} {\bibfnamefont
  {Z.}~\bibnamefont {Zhang}},\ }\bibfield  {title} {\bibinfo {title}
  {Demonstration of a reconfigurable entangled radio-frequency photonic sensor
  network},\ }\href {https://doi.org/10.1103/PhysRevLett.124.150502} {\bibfield
   {journal} {\bibinfo  {journal} {Phys. Rev. Lett.}\ }\textbf {\bibinfo
  {volume} {124}},\ \bibinfo {pages} {150502} (\bibinfo {year}
  {2020})}\BibitemShut {NoStop}%
\bibitem [{\citenamefont {Kwon}\ \emph {et~al.}(2022)\citenamefont {Kwon},
  \citenamefont {Lim}, \citenamefont {Jiang}, \citenamefont {Jeong},\ and\
  \citenamefont {Oh}}]{Kwon2022}%
  \BibitemOpen
  \bibfield  {author} {\bibinfo {author} {\bibfnamefont {H.}~\bibnamefont
  {Kwon}}, \bibinfo {author} {\bibfnamefont {Y.}~\bibnamefont {Lim}}, \bibinfo
  {author} {\bibfnamefont {L.}~\bibnamefont {Jiang}}, \bibinfo {author}
  {\bibfnamefont {H.}~\bibnamefont {Jeong}},\ and\ \bibinfo {author}
  {\bibfnamefont {C.}~\bibnamefont {Oh}},\ }\bibfield  {title} {\bibinfo
  {title} {Quantum metrological power of continuous-variable quantum
  networks},\ }\href {https://doi.org/10.1103/PhysRevLett.128.180503}
  {\bibfield  {journal} {\bibinfo  {journal} {Phys. Rev. Lett.}\ }\textbf
  {\bibinfo {volume} {128}},\ \bibinfo {pages} {180503} (\bibinfo {year}
  {2022})}\BibitemShut {NoStop}%
\bibitem [{\citenamefont {Rubio}\ \emph {et~al.}(2021)\citenamefont {Rubio},
  \citenamefont {Anders},\ and\ \citenamefont {Correa}}]{Rubio2021}%
  \BibitemOpen
  \bibfield  {author} {\bibinfo {author} {\bibfnamefont {J.}~\bibnamefont
  {Rubio}}, \bibinfo {author} {\bibfnamefont {J.}~\bibnamefont {Anders}},\ and\
  \bibinfo {author} {\bibfnamefont {L.~A.}\ \bibnamefont {Correa}},\ }\bibfield
   {title} {\bibinfo {title} {Global quantum thermometry},\ }\href
  {https://doi.org/10.1103/PhysRevLett.127.190402} {\bibfield  {journal}
  {\bibinfo  {journal} {Phys. Rev. Lett.}\ }\textbf {\bibinfo {volume} {127}},\
  \bibinfo {pages} {190402} (\bibinfo {year} {2021})}\BibitemShut {NoStop}%
\bibitem [{\citenamefont {Mehboudi}\ \emph {et~al.}(2022)\citenamefont
  {Mehboudi}, \citenamefont {J{\o}rgensen}, \citenamefont {Seah}, \citenamefont
  {Brask}, \citenamefont {Ko{\l}ody{\'n}ski},\ and\ \citenamefont
  {Perarnau-Llobet}}]{Mehboudi2022}%
  \BibitemOpen
  \bibfield  {author} {\bibinfo {author} {\bibfnamefont {M.}~\bibnamefont
  {Mehboudi}}, \bibinfo {author} {\bibfnamefont {M.~R.}\ \bibnamefont
  {J{\o}rgensen}}, \bibinfo {author} {\bibfnamefont {S.}~\bibnamefont {Seah}},
  \bibinfo {author} {\bibfnamefont {J.~B.}\ \bibnamefont {Brask}}, \bibinfo
  {author} {\bibfnamefont {J.}~\bibnamefont {Ko{\l}ody{\'n}ski}},\ and\
  \bibinfo {author} {\bibfnamefont {M.}~\bibnamefont {Perarnau-Llobet}},\
  }\bibfield  {title} {\bibinfo {title} {Fundamental limits in {B}ayesian
  thermometry and attainability via adaptive strategies},\ }\href
  {https://doi.org/10.1103/PhysRevLett.128.130502} {\bibfield  {journal}
  {\bibinfo  {journal} {Phys. Rev. Lett.}\ }\textbf {\bibinfo {volume} {128}},\
  \bibinfo {pages} {130502} (\bibinfo {year} {2022})}\BibitemShut {NoStop}%
\bibitem [{\citenamefont {Glatthard}\ \emph {et~al.}(2022)\citenamefont
  {Glatthard}, \citenamefont {Rubio}, \citenamefont {Sawant}, \citenamefont
  {Hewitt}, \citenamefont {Barontini},\ and\ \citenamefont
  {Correa}}]{Glatthard2022}%
  \BibitemOpen
  \bibfield  {author} {\bibinfo {author} {\bibfnamefont {J.}~\bibnamefont
  {Glatthard}}, \bibinfo {author} {\bibfnamefont {J.}~\bibnamefont {Rubio}},
  \bibinfo {author} {\bibfnamefont {R.}~\bibnamefont {Sawant}}, \bibinfo
  {author} {\bibfnamefont {T.}~\bibnamefont {Hewitt}}, \bibinfo {author}
  {\bibfnamefont {G.}~\bibnamefont {Barontini}},\ and\ \bibinfo {author}
  {\bibfnamefont {L.~A.}\ \bibnamefont {Correa}},\ }\bibfield  {title}
  {\bibinfo {title} {Optimal cold atom thermometry using adaptive {B}ayesian
  strategies},\ }\href {https://doi.org/10.1103/PRXQuantum.3.040330} {\bibfield
   {journal} {\bibinfo  {journal} {PRX Quantum}\ }\textbf {\bibinfo {volume}
  {3}},\ \bibinfo {pages} {040330} (\bibinfo {year} {2022})}\BibitemShut
  {NoStop}%
\bibitem [{\citenamefont {Pasquale}\ \emph {et~al.}(2016)\citenamefont
  {Pasquale}, \citenamefont {Rossini}, \citenamefont {Fazio},\ and\
  \citenamefont {Giovannetti}}]{DePasquale2016}%
  \BibitemOpen
  \bibfield  {author} {\bibinfo {author} {\bibfnamefont {A.~D.}\ \bibnamefont
  {Pasquale}}, \bibinfo {author} {\bibfnamefont {D.}~\bibnamefont {Rossini}},
  \bibinfo {author} {\bibfnamefont {R.}~\bibnamefont {Fazio}},\ and\ \bibinfo
  {author} {\bibfnamefont {V.}~\bibnamefont {Giovannetti}},\ }\bibfield
  {title} {\bibinfo {title} {Local quantum thermal susceptibility},\ }\href
  {https://doi.org/10.1038/ncomms12782} {\bibfield  {journal} {\bibinfo
  {journal} {Nat. Commun.}\ }\textbf {\bibinfo {volume} {7}},\ \bibinfo {pages}
  {12782} (\bibinfo {year} {2016})}\BibitemShut {NoStop}%
\bibitem [{\citenamefont {Planella}\ \emph {et~al.}(2022)\citenamefont
  {Planella}, \citenamefont {Cenni}, \citenamefont {Ac{\'\i}n},\ and\
  \citenamefont {Mehboudi}}]{Planella2022}%
  \BibitemOpen
  \bibfield  {author} {\bibinfo {author} {\bibfnamefont {G.}~\bibnamefont
  {Planella}}, \bibinfo {author} {\bibfnamefont {M.~F.~B.}\ \bibnamefont
  {Cenni}}, \bibinfo {author} {\bibfnamefont {A.}~\bibnamefont {Ac{\'\i}n}},\
  and\ \bibinfo {author} {\bibfnamefont {M.}~\bibnamefont {Mehboudi}},\
  }\bibfield  {title} {\bibinfo {title} {Bath-induced correlations enhance
  thermometry precision at low temperatures},\ }\href
  {https://doi.org/10.1103/PhysRevLett.128.040502} {\bibfield  {journal}
  {\bibinfo  {journal} {Phys. Rev. Lett.}\ }\textbf {\bibinfo {volume} {128}},\
  \bibinfo {pages} {040502} (\bibinfo {year} {2022})}\BibitemShut {NoStop}%
\bibitem [{\citenamefont {Brattegard}\ and\ \citenamefont
  {Mitchison}(2024)}]{Brattegard2024}%
  \BibitemOpen
  \bibfield  {author} {\bibinfo {author} {\bibfnamefont {S.}~\bibnamefont
  {Brattegard}}\ and\ \bibinfo {author} {\bibfnamefont {M.~T.}\ \bibnamefont
  {Mitchison}},\ }\bibfield  {title} {\bibinfo {title} {Thermometry by
  correlated dephasing of impurities in a one-dimensional {F}ermi gas},\ }\href
  {https://doi.org/10.1103/PhysRevA.109.023309} {\bibfield  {journal} {\bibinfo
   {journal} {Phys. Rev. A}\ }\textbf {\bibinfo {volume} {109}},\ \bibinfo
  {pages} {023309} (\bibinfo {year} {2024})}\BibitemShut {NoStop}%
\bibitem [{\citenamefont {Tumbiolo}\ \emph {et~al.}(2026)\citenamefont
  {Tumbiolo}, \citenamefont {Maccone}, \citenamefont {Macchiavello},
  \citenamefont {Paris},\ and\ \citenamefont {Guarnieri}}]{Tumbiolo2026}%
  \BibitemOpen
  \bibfield  {author} {\bibinfo {author} {\bibfnamefont {E.}~\bibnamefont
  {Tumbiolo}}, \bibinfo {author} {\bibfnamefont {L.}~\bibnamefont {Maccone}},
  \bibinfo {author} {\bibfnamefont {C.}~\bibnamefont {Macchiavello}}, \bibinfo
  {author} {\bibfnamefont {M.~G.~A.}\ \bibnamefont {Paris}},\ and\ \bibinfo
  {author} {\bibfnamefont {G.}~\bibnamefont {Guarnieri}},\ }\bibfield  {title}
  {\bibinfo {title} {Shake before use: Universal enhancement of quantum
  thermometry by unitary driving},\ }\href {https://doi.org/10.1103/z6rz-d9sq}
  {\bibfield  {journal} {\bibinfo  {journal} {Phys. Rev. Lett.}\ }\textbf
  {\bibinfo {volume} {136}},\ \bibinfo {pages} {250401} (\bibinfo {year}
  {2026})}\BibitemShut {NoStop}%
\bibitem [{\citenamefont {Massar}\ and\ \citenamefont
  {Popescu}(1995)}]{MassarPopescu1995}%
  \BibitemOpen
  \bibfield  {author} {\bibinfo {author} {\bibfnamefont {S.}~\bibnamefont
  {Massar}}\ and\ \bibinfo {author} {\bibfnamefont {S.}~\bibnamefont
  {Popescu}},\ }\bibfield  {title} {\bibinfo {title} {Optimal extraction of
  information from finite quantum ensembles},\ }\href
  {https://doi.org/10.1103/PhysRevLett.74.1259} {\bibfield  {journal} {\bibinfo
   {journal} {Phys. Rev. Lett.}\ }\textbf {\bibinfo {volume} {74}},\ \bibinfo
  {pages} {1259} (\bibinfo {year} {1995})}\BibitemShut {NoStop}%
\bibitem [{\citenamefont {Albarelli}\ \emph {et~al.}(2019)\citenamefont
  {Albarelli}, \citenamefont {Friel},\ and\ \citenamefont
  {Datta}}]{Albarelli2019}%
  \BibitemOpen
  \bibfield  {author} {\bibinfo {author} {\bibfnamefont {F.}~\bibnamefont
  {Albarelli}}, \bibinfo {author} {\bibfnamefont {J.~F.}\ \bibnamefont
  {Friel}},\ and\ \bibinfo {author} {\bibfnamefont {A.}~\bibnamefont {Datta}},\
  }\bibfield  {title} {\bibinfo {title} {Evaluating the {H}olevo
  {C}ram{\'e}r-{R}ao bound for multiparameter quantum metrology},\ }\href
  {https://doi.org/10.1103/PhysRevLett.123.200503} {\bibfield  {journal}
  {\bibinfo  {journal} {Phys. Rev. Lett.}\ }\textbf {\bibinfo {volume} {123}},\
  \bibinfo {pages} {200503} (\bibinfo {year} {2019})}\BibitemShut {NoStop}%
\bibitem [{\citenamefont {Das}\ \emph {et~al.}(2025)\citenamefont {Das},
  \citenamefont {Conlon}, \citenamefont {Suzuki}, \citenamefont {Yung},
  \citenamefont {Lam},\ and\ \citenamefont {Assad}}]{Das2025}%
  \BibitemOpen
  \bibfield  {author} {\bibinfo {author} {\bibfnamefont {A.}~\bibnamefont
  {Das}}, \bibinfo {author} {\bibfnamefont {L.~O.}\ \bibnamefont {Conlon}},
  \bibinfo {author} {\bibfnamefont {J.}~\bibnamefont {Suzuki}}, \bibinfo
  {author} {\bibfnamefont {S.~K.}\ \bibnamefont {Yung}}, \bibinfo {author}
  {\bibfnamefont {P.~K.}\ \bibnamefont {Lam}},\ and\ \bibinfo {author}
  {\bibfnamefont {S.~M.}\ \bibnamefont {Assad}},\ }\bibfield  {title} {\bibinfo
  {title} {{H}olevo {C}ram{\'e}r-{R}ao bound: How close can we get without
  entangling measurements?},\ }\href
  {https://doi.org/10.22331/q-2025-09-30-1867} {\bibfield  {journal} {\bibinfo
  {journal} {Quantum}\ }\textbf {\bibinfo {volume} {9}},\ \bibinfo {pages}
  {1867} (\bibinfo {year} {2025})}\BibitemShut {NoStop}%
\bibitem [{\citenamefont {Hou}\ \emph {et~al.}(2021)\citenamefont {Hou},
  \citenamefont {Tang}, \citenamefont {Chen}, \citenamefont {Yuan},
  \citenamefont {Xiang}, \citenamefont {Li},\ and\ \citenamefont
  {Guo}}]{Hou2021}%
  \BibitemOpen
  \bibfield  {author} {\bibinfo {author} {\bibfnamefont {Z.}~\bibnamefont
  {Hou}}, \bibinfo {author} {\bibfnamefont {J.-F.}\ \bibnamefont {Tang}},
  \bibinfo {author} {\bibfnamefont {H.}~\bibnamefont {Chen}}, \bibinfo {author}
  {\bibfnamefont {H.}~\bibnamefont {Yuan}}, \bibinfo {author} {\bibfnamefont
  {G.-Y.}\ \bibnamefont {Xiang}}, \bibinfo {author} {\bibfnamefont {C.-F.}\
  \bibnamefont {Li}},\ and\ \bibinfo {author} {\bibfnamefont {G.-C.}\
  \bibnamefont {Guo}},\ }\bibfield  {title} {\bibinfo {title} {Zero-trade-off
  multiparameter quantum estimation via simultaneously saturating multiple
  {H}eisenberg uncertainty relations},\ }\href
  {https://doi.org/10.1126/sciadv.abd2986} {\bibfield  {journal} {\bibinfo
  {journal} {Sci. Adv.}\ }\textbf {\bibinfo {volume} {7}},\ \bibinfo {pages}
  {eabd2986} (\bibinfo {year} {2021})}\BibitemShut {NoStop}%
\bibitem [{\citenamefont {Conlon}\ \emph {et~al.}(2023)\citenamefont {Conlon},
  \citenamefont {Vogl}, \citenamefont {Marciniak}, \citenamefont {Pogorelov},
  \citenamefont {Yung}, \citenamefont {Eilenberger}, \citenamefont {Berry},
  \citenamefont {Santana}, \citenamefont {Blatt}, \citenamefont {Monz},
  \citenamefont {Lam},\ and\ \citenamefont {Assad}}]{Conlon2023}%
  \BibitemOpen
  \bibfield  {author} {\bibinfo {author} {\bibfnamefont {L.~O.}\ \bibnamefont
  {Conlon}}, \bibinfo {author} {\bibfnamefont {T.}~\bibnamefont {Vogl}},
  \bibinfo {author} {\bibfnamefont {C.~D.}\ \bibnamefont {Marciniak}}, \bibinfo
  {author} {\bibfnamefont {I.}~\bibnamefont {Pogorelov}}, \bibinfo {author}
  {\bibfnamefont {S.~K.}\ \bibnamefont {Yung}}, \bibinfo {author}
  {\bibfnamefont {F.}~\bibnamefont {Eilenberger}}, \bibinfo {author}
  {\bibfnamefont {D.~W.}\ \bibnamefont {Berry}}, \bibinfo {author}
  {\bibfnamefont {F.~S.}\ \bibnamefont {Santana}}, \bibinfo {author}
  {\bibfnamefont {R.}~\bibnamefont {Blatt}}, \bibinfo {author} {\bibfnamefont
  {T.}~\bibnamefont {Monz}}, \bibinfo {author} {\bibfnamefont {P.~K.}\
  \bibnamefont {Lam}},\ and\ \bibinfo {author} {\bibfnamefont {S.~M.}\
  \bibnamefont {Assad}},\ }\bibfield  {title} {\bibinfo {title} {Approaching
  optimal entangling collective measurements on quantum computing platforms},\
  }\href {https://doi.org/10.1038/s41567-022-01875-7} {\bibfield  {journal}
  {\bibinfo  {journal} {Nat. Phys.}\ }\textbf {\bibinfo {volume} {19}},\
  \bibinfo {pages} {351} (\bibinfo {year} {2023})}\BibitemShut {NoStop}%
\bibitem [{\citenamefont {Xia}\ \emph {et~al.}(2023)\citenamefont {Xia},
  \citenamefont {Huang}, \citenamefont {Li}, \citenamefont {Wang},\ and\
  \citenamefont {Zeng}}]{XiaB2023}%
  \BibitemOpen
  \bibfield  {author} {\bibinfo {author} {\bibfnamefont {B.}~\bibnamefont
  {Xia}}, \bibinfo {author} {\bibfnamefont {J.}~\bibnamefont {Huang}}, \bibinfo
  {author} {\bibfnamefont {H.}~\bibnamefont {Li}}, \bibinfo {author}
  {\bibfnamefont {H.}~\bibnamefont {Wang}},\ and\ \bibinfo {author}
  {\bibfnamefont {G.}~\bibnamefont {Zeng}},\ }\bibfield  {title} {\bibinfo
  {title} {Toward incompatible quantum limits on multiparameter estimation},\
  }\href {https://doi.org/10.1038/s41467-023-36661-3} {\bibfield  {journal}
  {\bibinfo  {journal} {Nat. Commun.}\ }\textbf {\bibinfo {volume} {14}},\
  \bibinfo {pages} {1021} (\bibinfo {year} {2023})}\BibitemShut {NoStop}%
\bibitem [{\citenamefont {Mi}\ \emph {et~al.}(2025)\citenamefont {Mi},
  \citenamefont {Wang},\ and\ \citenamefont {Zhang}}]{Mi2025}%
  \BibitemOpen
  \bibfield  {author} {\bibinfo {author} {\bibfnamefont {M.}~\bibnamefont
  {Mi}}, \bibinfo {author} {\bibfnamefont {B.}~\bibnamefont {Wang}},\ and\
  \bibinfo {author} {\bibfnamefont {L.}~\bibnamefont {Zhang}},\ }\bibfield
  {title} {\bibinfo {title} {Approaching the multiparameter quantum
  {C}ram{\'e}r-{R}ao bound via classical correlation and entangling
  measurements},\ }\href {https://doi.org/10.1103/37kb-qq6k} {\bibfield
  {journal} {\bibinfo  {journal} {Phys. Rev. Lett.}\ }\textbf {\bibinfo
  {volume} {135}},\ \bibinfo {pages} {110804} (\bibinfo {year}
  {2025})}\BibitemShut {NoStop}%
\bibitem [{\citenamefont {Li}\ \emph {et~al.}(2026)\citenamefont {Li},
  \citenamefont {Joosten}, \citenamefont {Baamara}, \citenamefont {Colciaghi},
  \citenamefont {Sinatra}, \citenamefont {Treutlein},\ and\ \citenamefont
  {Zibold}}]{LiY2026}%
  \BibitemOpen
  \bibfield  {author} {\bibinfo {author} {\bibfnamefont {Y.}~\bibnamefont
  {Li}}, \bibinfo {author} {\bibfnamefont {L.}~\bibnamefont {Joosten}},
  \bibinfo {author} {\bibfnamefont {Y.}~\bibnamefont {Baamara}}, \bibinfo
  {author} {\bibfnamefont {P.}~\bibnamefont {Colciaghi}}, \bibinfo {author}
  {\bibfnamefont {A.}~\bibnamefont {Sinatra}}, \bibinfo {author} {\bibfnamefont
  {P.}~\bibnamefont {Treutlein}},\ and\ \bibinfo {author} {\bibfnamefont
  {T.}~\bibnamefont {Zibold}},\ }\bibfield  {title} {\bibinfo {title}
  {Multiparameter estimation with an array of entangled atomic sensors},\
  }\href {https://doi.org/10.1126/science.adt2442} {\bibfield  {journal}
  {\bibinfo  {journal} {Science}\ }\textbf {\bibinfo {volume} {391}},\ \bibinfo
  {pages} {374} (\bibinfo {year} {2026})}\BibitemShut {NoStop}%
\bibitem [{\citenamefont {Conlon}\ \emph {et~al.}(2021)\citenamefont {Conlon},
  \citenamefont {Suzuki}, \citenamefont {Lam},\ and\ \citenamefont
  {Assad}}]{Conlon2021}%
  \BibitemOpen
  \bibfield  {author} {\bibinfo {author} {\bibfnamefont {L.~O.}\ \bibnamefont
  {Conlon}}, \bibinfo {author} {\bibfnamefont {J.}~\bibnamefont {Suzuki}},
  \bibinfo {author} {\bibfnamefont {P.~K.}\ \bibnamefont {Lam}},\ and\ \bibinfo
  {author} {\bibfnamefont {S.~M.}\ \bibnamefont {Assad}},\ }\bibfield  {title}
  {\bibinfo {title} {Efficient computation of the {N}agaoka-{H}ayashi bound for
  multiparameter estimation with separable measurements},\ }\href
  {https://doi.org/10.1038/s41534-021-00414-1} {\bibfield  {journal} {\bibinfo
  {journal} {npj Quantum Inf.}\ }\textbf {\bibinfo {volume} {7}},\ \bibinfo
  {pages} {110} (\bibinfo {year} {2021})}\BibitemShut {NoStop}%
\bibitem [{\citenamefont {Chen}\ \emph {et~al.}(2022)\citenamefont {Chen},
  \citenamefont {Chen},\ and\ \citenamefont {Yuan}}]{ChenH2022}%
  \BibitemOpen
  \bibfield  {author} {\bibinfo {author} {\bibfnamefont {H.}~\bibnamefont
  {Chen}}, \bibinfo {author} {\bibfnamefont {Y.}~\bibnamefont {Chen}},\ and\
  \bibinfo {author} {\bibfnamefont {H.}~\bibnamefont {Yuan}},\ }\bibfield
  {title} {\bibinfo {title} {Incompatibility measures in multiparameter quantum
  estimation under hierarchical quantum measurements},\ }\href
  {https://doi.org/10.1103/PhysRevA.105.062442} {\bibfield  {journal} {\bibinfo
   {journal} {Phys. Rev. A}\ }\textbf {\bibinfo {volume} {105}},\ \bibinfo
  {pages} {062442} (\bibinfo {year} {2022})}\BibitemShut {NoStop}%
\bibitem [{\citenamefont {Conlon}\ \emph {et~al.}(2025)\citenamefont {Conlon},
  \citenamefont {Koh}, \citenamefont {Shajilal}, \citenamefont {Sidhu},
  \citenamefont {Lam},\ and\ \citenamefont {Assad}}]{ConlonPRA2025}%
  \BibitemOpen
  \bibfield  {author} {\bibinfo {author} {\bibfnamefont {L.~O.}\ \bibnamefont
  {Conlon}}, \bibinfo {author} {\bibfnamefont {J.~M.}\ \bibnamefont {Koh}},
  \bibinfo {author} {\bibfnamefont {B.}~\bibnamefont {Shajilal}}, \bibinfo
  {author} {\bibfnamefont {J.}~\bibnamefont {Sidhu}}, \bibinfo {author}
  {\bibfnamefont {P.~K.}\ \bibnamefont {Lam}},\ and\ \bibinfo {author}
  {\bibfnamefont {S.~M.}\ \bibnamefont {Assad}},\ }\bibfield  {title} {\bibinfo
  {title} {Attainability of quantum state discrimination bounds with collective
  measurements on finite copies},\ }\href
  {https://doi.org/10.1103/PhysRevA.111.022438} {\bibfield  {journal} {\bibinfo
   {journal} {Phys. Rev. A}\ }\textbf {\bibinfo {volume} {111}},\ \bibinfo
  {pages} {022438} (\bibinfo {year} {2025})}\BibitemShut {NoStop}%
\bibitem [{\citenamefont {Proctor}\ \emph {et~al.}(2018)\citenamefont
  {Proctor}, \citenamefont {Knott},\ and\ \citenamefont
  {Dunningham}}]{Proctor2018}%
  \BibitemOpen
  \bibfield  {author} {\bibinfo {author} {\bibfnamefont {T.~J.}\ \bibnamefont
  {Proctor}}, \bibinfo {author} {\bibfnamefont {P.~A.}\ \bibnamefont {Knott}},\
  and\ \bibinfo {author} {\bibfnamefont {J.~A.}\ \bibnamefont {Dunningham}},\
  }\bibfield  {title} {\bibinfo {title} {Multiparameter estimation in networked
  quantum sensors},\ }\href {https://doi.org/10.1103/PhysRevLett.120.080501}
  {\bibfield  {journal} {\bibinfo  {journal} {Phys. Rev. Lett.}\ }\textbf
  {\bibinfo {volume} {120}},\ \bibinfo {pages} {080501} (\bibinfo {year}
  {2018})}\BibitemShut {NoStop}%
\bibitem [{\citenamefont {Pezz{\`e}}\ and\ \citenamefont
  {Smerzi}(2025)}]{Pezze2025}%
  \BibitemOpen
  \bibfield  {author} {\bibinfo {author} {\bibfnamefont {L.}~\bibnamefont
  {Pezz{\`e}}}\ and\ \bibinfo {author} {\bibfnamefont {A.}~\bibnamefont
  {Smerzi}},\ }\bibfield  {title} {\bibinfo {title} {Distributed quantum
  multiparameter estimation with optimal local measurements},\ }\href
  {https://doi.org/10.1103/f2jf-bg7g} {\bibfield  {journal} {\bibinfo
  {journal} {Phys. Rev. Lett.}\ }\textbf {\bibinfo {volume} {135}},\ \bibinfo
  {pages} {260805} (\bibinfo {year} {2025})}\BibitemShut {NoStop}%
\bibitem [{\citenamefont {Cenni}\ \emph {et~al.}(2022)\citenamefont {Cenni},
  \citenamefont {Lami}, \citenamefont {Acin},\ and\ \citenamefont
  {Mehboudi}}]{Cenni2022}%
  \BibitemOpen
  \bibfield  {author} {\bibinfo {author} {\bibfnamefont {M.~F.~B.}\
  \bibnamefont {Cenni}}, \bibinfo {author} {\bibfnamefont {L.}~\bibnamefont
  {Lami}}, \bibinfo {author} {\bibfnamefont {A.}~\bibnamefont {Acin}},\ and\
  \bibinfo {author} {\bibfnamefont {M.}~\bibnamefont {Mehboudi}},\ }\bibfield
  {title} {\bibinfo {title} {Thermometry of {G}aussian quantum systems using
  {G}aussian measurements},\ }\href {https://doi.org/10.22331/q-2022-06-23-743}
  {\bibfield  {journal} {\bibinfo  {journal} {Quantum}\ }\textbf {\bibinfo
  {volume} {6}},\ \bibinfo {pages} {743} (\bibinfo {year} {2022})}\BibitemShut
  {NoStop}%
\bibitem [{\citenamefont {Navarro}\ \emph {et~al.}(2025)\citenamefont
  {Navarro}, \citenamefont {Morelli}, \citenamefont {Sanz},\ and\ \citenamefont
  {Mehboudi}}]{Navarro2025}%
  \BibitemOpen
  \bibfield  {author} {\bibinfo {author} {\bibfnamefont {J.}~\bibnamefont
  {Navarro}}, \bibinfo {author} {\bibfnamefont {S.}~\bibnamefont {Morelli}},
  \bibinfo {author} {\bibfnamefont {M.}~\bibnamefont {Sanz}},\ and\ \bibinfo
  {author} {\bibfnamefont {M.}~\bibnamefont {Mehboudi}},\ }\href@noop {}
  {\bibinfo {title} {On super additivity of {F}isher information in fully
  {G}aussian metrology}} (\bibinfo {year} {2025}),\ \Eprint
  {https://arxiv.org/abs/2512.20534} {arXiv:2512.20534 [quant-ph]} \BibitemShut
  {NoStop}%
\bibitem [{Sup()}]{Supplement}%
  \BibitemOpen
  \href@noop {} {}\bibinfo {note} {See the Supplemental Material appended to
  this preprint for the conventions, the full derivations behind the
  separate-readout bound, the onset law, the thermal cap and the universal
  bound, the pure-seed reduction and the exact-arithmetic proofs, mixed seeds,
  coupled resonators with and without the rotating-wave approximation, detector
  loss and finite samples, the frequency-bank pairing and the amplifier model,
  which includes
  Refs.~\cite{Eichler2011,Kotler2021,Lu2023,Menzel2010,MercierDeLepinay2021,OckeloenKorppi2016,Scigliuzzo2020}.}\BibitemShut
  {Stop}%
\bibitem [{\citenamefont {Spedalieri}\ \emph {et~al.}(2013)\citenamefont
  {Spedalieri}, \citenamefont {Ottaviani},\ and\ \citenamefont
  {Pirandola}}]{Spedalieri2013}%
  \BibitemOpen
  \bibfield  {author} {\bibinfo {author} {\bibfnamefont {G.}~\bibnamefont
  {Spedalieri}}, \bibinfo {author} {\bibfnamefont {C.}~\bibnamefont
  {Ottaviani}},\ and\ \bibinfo {author} {\bibfnamefont {S.}~\bibnamefont
  {Pirandola}},\ }\bibfield  {title} {\bibinfo {title} {Covariance matrices
  under {B}ell-like detections},\ }\href
  {https://doi.org/10.1142/S123016121350011X} {\bibfield  {journal} {\bibinfo
  {journal} {Open Syst. Inf. Dyn.}\ }\textbf {\bibinfo {volume} {20}},\
  \bibinfo {pages} {1350011} (\bibinfo {year} {2013})}\BibitemShut {NoStop}%
\bibitem [{\citenamefont {Miyanaga}\ \emph {et~al.}(2021)\citenamefont
  {Miyanaga}, \citenamefont {Tomonaga}, \citenamefont {Ito}, \citenamefont
  {Mukai},\ and\ \citenamefont {Tsai}}]{Miyanaga2021}%
  \BibitemOpen
  \bibfield  {author} {\bibinfo {author} {\bibfnamefont {T.}~\bibnamefont
  {Miyanaga}}, \bibinfo {author} {\bibfnamefont {A.}~\bibnamefont {Tomonaga}},
  \bibinfo {author} {\bibfnamefont {H.}~\bibnamefont {Ito}}, \bibinfo {author}
  {\bibfnamefont {H.}~\bibnamefont {Mukai}},\ and\ \bibinfo {author}
  {\bibfnamefont {J.~S.}\ \bibnamefont {Tsai}},\ }\bibfield  {title} {\bibinfo
  {title} {Ultrastrong tunable coupler between superconducting {LC}
  resonators},\ }\href {https://doi.org/10.1103/PhysRevApplied.16.064041}
  {\bibfield  {journal} {\bibinfo  {journal} {Phys. Rev. Applied}\ }\textbf
  {\bibinfo {volume} {16}},\ \bibinfo {pages} {064041} (\bibinfo {year}
  {2021})}\BibitemShut {NoStop}%
\bibitem [{\citenamefont {Bothner}\ \emph {et~al.}(2021)\citenamefont
  {Bothner}, \citenamefont {Rodrigues},\ and\ \citenamefont
  {Steele}}]{Bothner2021}%
  \BibitemOpen
  \bibfield  {author} {\bibinfo {author} {\bibfnamefont {D.}~\bibnamefont
  {Bothner}}, \bibinfo {author} {\bibfnamefont {I.~C.}\ \bibnamefont
  {Rodrigues}},\ and\ \bibinfo {author} {\bibfnamefont {G.~A.}\ \bibnamefont
  {Steele}},\ }\bibfield  {title} {\bibinfo {title} {Photon-pressure strong
  coupling between two superconducting circuits},\ }\href
  {https://doi.org/10.1038/s41567-020-0987-5} {\bibfield  {journal} {\bibinfo
  {journal} {Nat. Phys.}\ }\textbf {\bibinfo {volume} {17}},\ \bibinfo {pages}
  {85} (\bibinfo {year} {2021})}\BibitemShut {NoStop}%
\bibitem [{\citenamefont {Arthurs}\ and\ \citenamefont
  {Kelly}(1965)}]{ArthursKelly1965}%
  \BibitemOpen
  \bibfield  {author} {\bibinfo {author} {\bibfnamefont {E.}~\bibnamefont
  {Arthurs}}\ and\ \bibinfo {author} {\bibfnamefont {J.~L.}\ \bibnamefont
  {Kelly}, \bibfnamefont {Jr.}},\ }\bibfield  {title} {\bibinfo {title} {On the
  simultaneous measurement of a pair of conjugate observables},\ }\href
  {https://doi.org/10.1002/j.1538-7305.1965.tb01684.x} {\bibfield  {journal}
  {\bibinfo  {journal} {Bell Syst. Tech. J.}\ }\textbf {\bibinfo {volume}
  {44}},\ \bibinfo {pages} {725} (\bibinfo {year} {1965})}\BibitemShut
  {NoStop}%
\bibitem [{\citenamefont {Chatterjee}\ \emph {et~al.}(2026)\citenamefont
  {Chatterjee}, \citenamefont {Pandit}, \citenamefont {Singh}, \citenamefont
  {Chattopadhyay},\ and\ \citenamefont {Andersen}}]{Chatterjee2026}%
  \BibitemOpen
  \bibfield  {author} {\bibinfo {author} {\bibfnamefont {K.}~\bibnamefont
  {Chatterjee}}, \bibinfo {author} {\bibfnamefont {T.}~\bibnamefont {Pandit}},
  \bibinfo {author} {\bibfnamefont {V.}~\bibnamefont {Singh}}, \bibinfo
  {author} {\bibfnamefont {P.}~\bibnamefont {Chattopadhyay}},\ and\ \bibinfo
  {author} {\bibfnamefont {U.~L.}\ \bibnamefont {Andersen}},\ }\href@noop {}
  {\bibinfo {title} {Even-odd splitting of the {G}aussian quantum {F}isher
  information: From symplectic geometry to metrology}} (\bibinfo {year}
  {2026}),\ \Eprint {https://arxiv.org/abs/2601.06513} {arXiv:2601.06513
  [quant-ph]} \BibitemShut {NoStop}%
\bibitem [{\citenamefont {Mirkhalaf}\ \emph {et~al.}(2024)\citenamefont
  {Mirkhalaf}, \citenamefont {Mehboudi}, \citenamefont {Nafari~Qaleh},\ and\
  \citenamefont {Rahimi-Keshari}}]{Mirkhalaf2024}%
  \BibitemOpen
  \bibfield  {author} {\bibinfo {author} {\bibfnamefont {S.}~\bibnamefont
  {Mirkhalaf}}, \bibinfo {author} {\bibfnamefont {M.}~\bibnamefont {Mehboudi}},
  \bibinfo {author} {\bibfnamefont {Z.}~\bibnamefont {Nafari~Qaleh}},\ and\
  \bibinfo {author} {\bibfnamefont {S.}~\bibnamefont {Rahimi-Keshari}},\
  }\bibfield  {title} {\bibinfo {title} {Operational significance of
  nonclassicality in nonequilibrium {G}aussian quantum thermometry},\ }\href
  {https://doi.org/10.1088/1367-2630/ad23a1} {\bibfield  {journal} {\bibinfo
  {journal} {New J. Phys.}\ }\textbf {\bibinfo {volume} {26}},\ \bibinfo
  {pages} {023046} (\bibinfo {year} {2024})}\BibitemShut {NoStop}%
\bibitem [{\citenamefont {Caves}(1982)}]{Caves1982}%
  \BibitemOpen
  \bibfield  {author} {\bibinfo {author} {\bibfnamefont {C.~M.}\ \bibnamefont
  {Caves}},\ }\bibfield  {title} {\bibinfo {title} {Quantum limits on noise in
  linear amplifiers},\ }\href {https://doi.org/10.1103/PhysRevD.26.1817}
  {\bibfield  {journal} {\bibinfo  {journal} {Phys. Rev. D}\ }\textbf {\bibinfo
  {volume} {26}},\ \bibinfo {pages} {1817} (\bibinfo {year}
  {1982})}\BibitemShut {NoStop}%
\bibitem [{\citenamefont {Clerk}\ \emph {et~al.}(2010)\citenamefont {Clerk},
  \citenamefont {Devoret}, \citenamefont {Girvin}, \citenamefont {Marquardt},\
  and\ \citenamefont {Schoelkopf}}]{Clerk2010}%
  \BibitemOpen
  \bibfield  {author} {\bibinfo {author} {\bibfnamefont {A.~A.}\ \bibnamefont
  {Clerk}}, \bibinfo {author} {\bibfnamefont {M.~H.}\ \bibnamefont {Devoret}},
  \bibinfo {author} {\bibfnamefont {S.~M.}\ \bibnamefont {Girvin}}, \bibinfo
  {author} {\bibfnamefont {F.}~\bibnamefont {Marquardt}},\ and\ \bibinfo
  {author} {\bibfnamefont {R.~J.}\ \bibnamefont {Schoelkopf}},\ }\bibfield
  {title} {\bibinfo {title} {Introduction to quantum noise, measurement, and
  amplification},\ }\href {https://doi.org/10.1103/RevModPhys.82.1155}
  {\bibfield  {journal} {\bibinfo  {journal} {Rev. Mod. Phys.}\ }\textbf
  {\bibinfo {volume} {82}},\ \bibinfo {pages} {1155} (\bibinfo {year}
  {2010})}\BibitemShut {NoStop}%
\bibitem [{\citenamefont {Macklin}\ \emph {et~al.}(2015)\citenamefont
  {Macklin}, \citenamefont {O'Brien}, \citenamefont {Hover}, \citenamefont
  {Schwartz}, \citenamefont {Bolkhovsky}, \citenamefont {Zhang}, \citenamefont
  {Oliver},\ and\ \citenamefont {Siddiqi}}]{Macklin2015}%
  \BibitemOpen
  \bibfield  {author} {\bibinfo {author} {\bibfnamefont {C.}~\bibnamefont
  {Macklin}}, \bibinfo {author} {\bibfnamefont {K.}~\bibnamefont {O'Brien}},
  \bibinfo {author} {\bibfnamefont {D.}~\bibnamefont {Hover}}, \bibinfo
  {author} {\bibfnamefont {M.~E.}\ \bibnamefont {Schwartz}}, \bibinfo {author}
  {\bibfnamefont {V.}~\bibnamefont {Bolkhovsky}}, \bibinfo {author}
  {\bibfnamefont {X.}~\bibnamefont {Zhang}}, \bibinfo {author} {\bibfnamefont
  {W.~D.}\ \bibnamefont {Oliver}},\ and\ \bibinfo {author} {\bibfnamefont
  {I.}~\bibnamefont {Siddiqi}},\ }\bibfield  {title} {\bibinfo {title} {A
  near-quantum-limited {J}osephson traveling-wave parametric amplifier},\
  }\href {https://doi.org/10.1126/science.aaa8525} {\bibfield  {journal}
  {\bibinfo  {journal} {Science}\ }\textbf {\bibinfo {volume} {350}},\ \bibinfo
  {pages} {307} (\bibinfo {year} {2015})}\BibitemShut {NoStop}%
\bibitem [{\citenamefont {Bergeal}\ \emph {et~al.}(2010)\citenamefont
  {Bergeal}, \citenamefont {Schackert}, \citenamefont {Metcalfe}, \citenamefont
  {Vijay}, \citenamefont {Manucharyan}, \citenamefont {Frunzio}, \citenamefont
  {Prober}, \citenamefont {Schoelkopf}, \citenamefont {Girvin},\ and\
  \citenamefont {Devoret}}]{Bergeal2010}%
  \BibitemOpen
  \bibfield  {author} {\bibinfo {author} {\bibfnamefont {N.}~\bibnamefont
  {Bergeal}}, \bibinfo {author} {\bibfnamefont {F.}~\bibnamefont {Schackert}},
  \bibinfo {author} {\bibfnamefont {M.}~\bibnamefont {Metcalfe}}, \bibinfo
  {author} {\bibfnamefont {R.}~\bibnamefont {Vijay}}, \bibinfo {author}
  {\bibfnamefont {V.~E.}\ \bibnamefont {Manucharyan}}, \bibinfo {author}
  {\bibfnamefont {L.}~\bibnamefont {Frunzio}}, \bibinfo {author} {\bibfnamefont
  {D.~E.}\ \bibnamefont {Prober}}, \bibinfo {author} {\bibfnamefont {R.~J.}\
  \bibnamefont {Schoelkopf}}, \bibinfo {author} {\bibfnamefont {S.~M.}\
  \bibnamefont {Girvin}},\ and\ \bibinfo {author} {\bibfnamefont {M.~H.}\
  \bibnamefont {Devoret}},\ }\bibfield  {title} {\bibinfo {title}
  {Phase-preserving amplification near the quantum limit with a {J}osephson
  ring modulator},\ }\href {https://doi.org/10.1038/nature09035} {\bibfield
  {journal} {\bibinfo  {journal} {Nature}\ }\textbf {\bibinfo {volume} {465}},\
  \bibinfo {pages} {64} (\bibinfo {year} {2010})}\BibitemShut {NoStop}%
\bibitem [{\citenamefont {Roy}\ and\ \citenamefont
  {Devoret}(2016)}]{RoyDevoret2016}%
  \BibitemOpen
  \bibfield  {author} {\bibinfo {author} {\bibfnamefont {A.}~\bibnamefont
  {Roy}}\ and\ \bibinfo {author} {\bibfnamefont {M.}~\bibnamefont {Devoret}},\
  }\bibfield  {title} {\bibinfo {title} {Introduction to parametric
  amplification of quantum signals with {J}osephson circuits},\ }\href
  {https://doi.org/10.1016/j.crhy.2016.07.012} {\bibfield  {journal} {\bibinfo
  {journal} {C. R. Phys.}\ }\textbf {\bibinfo {volume} {17}},\ \bibinfo {pages}
  {740} (\bibinfo {year} {2016})}\BibitemShut {NoStop}%
\bibitem [{\citenamefont {Abdo}\ \emph {et~al.}(2013)\citenamefont {Abdo},
  \citenamefont {Sliwa}, \citenamefont {Schackert}, \citenamefont {Bergeal},
  \citenamefont {Hatridge}, \citenamefont {Frunzio}, \citenamefont {Stone},\
  and\ \citenamefont {Devoret}}]{Abdo2013}%
  \BibitemOpen
  \bibfield  {author} {\bibinfo {author} {\bibfnamefont {B.}~\bibnamefont
  {Abdo}}, \bibinfo {author} {\bibfnamefont {K.}~\bibnamefont {Sliwa}},
  \bibinfo {author} {\bibfnamefont {F.}~\bibnamefont {Schackert}}, \bibinfo
  {author} {\bibfnamefont {N.}~\bibnamefont {Bergeal}}, \bibinfo {author}
  {\bibfnamefont {M.}~\bibnamefont {Hatridge}}, \bibinfo {author}
  {\bibfnamefont {L.}~\bibnamefont {Frunzio}}, \bibinfo {author} {\bibfnamefont
  {A.~D.}\ \bibnamefont {Stone}},\ and\ \bibinfo {author} {\bibfnamefont
  {M.}~\bibnamefont {Devoret}},\ }\bibfield  {title} {\bibinfo {title} {Full
  coherent frequency conversion between two propagating microwave modes},\
  }\href {https://doi.org/10.1103/PhysRevLett.110.173902} {\bibfield  {journal}
  {\bibinfo  {journal} {Phys. Rev. Lett.}\ }\textbf {\bibinfo {volume} {110}},\
  \bibinfo {pages} {173902} (\bibinfo {year} {2013})}\BibitemShut {NoStop}%
\bibitem [{\citenamefont {Lecocq}\ \emph {et~al.}(2016)\citenamefont {Lecocq},
  \citenamefont {Clark}, \citenamefont {Simmonds}, \citenamefont {Aumentado},\
  and\ \citenamefont {Teufel}}]{Lecocq2016}%
  \BibitemOpen
  \bibfield  {author} {\bibinfo {author} {\bibfnamefont {F.}~\bibnamefont
  {Lecocq}}, \bibinfo {author} {\bibfnamefont {J.~B.}\ \bibnamefont {Clark}},
  \bibinfo {author} {\bibfnamefont {R.~W.}\ \bibnamefont {Simmonds}}, \bibinfo
  {author} {\bibfnamefont {J.}~\bibnamefont {Aumentado}},\ and\ \bibinfo
  {author} {\bibfnamefont {J.~D.}\ \bibnamefont {Teufel}},\ }\bibfield  {title}
  {\bibinfo {title} {Mechanically mediated microwave frequency conversion in
  the quantum regime},\ }\href {https://doi.org/10.1103/PhysRevLett.116.043601}
  {\bibfield  {journal} {\bibinfo  {journal} {Phys. Rev. Lett.}\ }\textbf
  {\bibinfo {volume} {116}},\ \bibinfo {pages} {043601} (\bibinfo {year}
  {2016})}\BibitemShut {NoStop}%
\bibitem [{\citenamefont {Eichler}\ \emph {et~al.}(2011)\citenamefont
  {Eichler}, \citenamefont {Bozyigit}, \citenamefont {Lang}, \citenamefont
  {Baur}, \citenamefont {Steffen}, \citenamefont {Fink}, \citenamefont
  {Filipp},\ and\ \citenamefont {Wallraff}}]{Eichler2011}%
  \BibitemOpen
  \bibfield  {author} {\bibinfo {author} {\bibfnamefont {C.}~\bibnamefont
  {Eichler}}, \bibinfo {author} {\bibfnamefont {D.}~\bibnamefont {Bozyigit}},
  \bibinfo {author} {\bibfnamefont {C.}~\bibnamefont {Lang}}, \bibinfo {author}
  {\bibfnamefont {M.}~\bibnamefont {Baur}}, \bibinfo {author} {\bibfnamefont
  {L.}~\bibnamefont {Steffen}}, \bibinfo {author} {\bibfnamefont {J.~M.}\
  \bibnamefont {Fink}}, \bibinfo {author} {\bibfnamefont {S.}~\bibnamefont
  {Filipp}},\ and\ \bibinfo {author} {\bibfnamefont {A.}~\bibnamefont
  {Wallraff}},\ }\bibfield  {title} {\bibinfo {title} {Photon state tomography
  for two-mode correlated itinerant microwave fields},\ }\href
  {https://doi.org/10.1103/PhysRevLett.107.113601} {\bibfield  {journal}
  {\bibinfo  {journal} {Phys. Rev. Lett.}\ }\textbf {\bibinfo {volume} {107}},\
  \bibinfo {pages} {113601} (\bibinfo {year} {2011})}\BibitemShut {NoStop}%
\bibitem [{\citenamefont {Kotler}\ \emph {et~al.}(2021)\citenamefont {Kotler},
  \citenamefont {Peterson}, \citenamefont {Shojaee}, \citenamefont {Lecocq},
  \citenamefont {Cicak}, \citenamefont {Kwiatkowski}, \citenamefont {Geller},
  \citenamefont {Glancy}, \citenamefont {Knill}, \citenamefont {Simmonds},
  \citenamefont {Aumentado},\ and\ \citenamefont {Teufel}}]{Kotler2021}%
  \BibitemOpen
  \bibfield  {author} {\bibinfo {author} {\bibfnamefont {S.}~\bibnamefont
  {Kotler}}, \bibinfo {author} {\bibfnamefont {G.~A.}\ \bibnamefont
  {Peterson}}, \bibinfo {author} {\bibfnamefont {E.}~\bibnamefont {Shojaee}},
  \bibinfo {author} {\bibfnamefont {F.}~\bibnamefont {Lecocq}}, \bibinfo
  {author} {\bibfnamefont {K.}~\bibnamefont {Cicak}}, \bibinfo {author}
  {\bibfnamefont {A.}~\bibnamefont {Kwiatkowski}}, \bibinfo {author}
  {\bibfnamefont {S.}~\bibnamefont {Geller}}, \bibinfo {author} {\bibfnamefont
  {S.}~\bibnamefont {Glancy}}, \bibinfo {author} {\bibfnamefont
  {E.}~\bibnamefont {Knill}}, \bibinfo {author} {\bibfnamefont {R.~W.}\
  \bibnamefont {Simmonds}}, \bibinfo {author} {\bibfnamefont {J.}~\bibnamefont
  {Aumentado}},\ and\ \bibinfo {author} {\bibfnamefont {J.~D.}\ \bibnamefont
  {Teufel}},\ }\bibfield  {title} {\bibinfo {title} {Direct observation of
  deterministic macroscopic entanglement},\ }\href
  {https://doi.org/10.1126/science.abf2998} {\bibfield  {journal} {\bibinfo
  {journal} {Science}\ }\textbf {\bibinfo {volume} {372}},\ \bibinfo {pages}
  {622} (\bibinfo {year} {2021})}\BibitemShut {NoStop}%
\bibitem [{\citenamefont {Lu}\ \emph {et~al.}(2023)\citenamefont {Lu},
  \citenamefont {Maiti}, \citenamefont {Garmon}, \citenamefont {Ganjam},
  \citenamefont {Zhang}, \citenamefont {Claes}, \citenamefont {Frunzio},
  \citenamefont {Girvin},\ and\ \citenamefont {Schoelkopf}}]{Lu2023}%
  \BibitemOpen
  \bibfield  {author} {\bibinfo {author} {\bibfnamefont {Y.}~\bibnamefont
  {Lu}}, \bibinfo {author} {\bibfnamefont {A.}~\bibnamefont {Maiti}}, \bibinfo
  {author} {\bibfnamefont {J.~W.~O.}\ \bibnamefont {Garmon}}, \bibinfo {author}
  {\bibfnamefont {S.}~\bibnamefont {Ganjam}}, \bibinfo {author} {\bibfnamefont
  {Y.}~\bibnamefont {Zhang}}, \bibinfo {author} {\bibfnamefont
  {J.}~\bibnamefont {Claes}}, \bibinfo {author} {\bibfnamefont
  {L.}~\bibnamefont {Frunzio}}, \bibinfo {author} {\bibfnamefont {S.~M.}\
  \bibnamefont {Girvin}},\ and\ \bibinfo {author} {\bibfnamefont {R.~J.}\
  \bibnamefont {Schoelkopf}},\ }\bibfield  {title} {\bibinfo {title}
  {High-fidelity parametric beamsplitting with a parity-protected converter},\
  }\href {https://doi.org/10.1038/s41467-023-41104-0} {\bibfield  {journal}
  {\bibinfo  {journal} {Nat. Commun.}\ }\textbf {\bibinfo {volume} {14}},\
  \bibinfo {pages} {5767} (\bibinfo {year} {2023})}\BibitemShut {NoStop}%
\bibitem [{\citenamefont {Menzel}\ \emph {et~al.}(2010)\citenamefont {Menzel},
  \citenamefont {Deppe}, \citenamefont {Mariantoni}, \citenamefont {Caballero},
  \citenamefont {Baust}, \citenamefont {Niemczyk}, \citenamefont {Hoffmann},
  \citenamefont {Marx}, \citenamefont {Solano},\ and\ \citenamefont
  {Gross}}]{Menzel2010}%
  \BibitemOpen
  \bibfield  {author} {\bibinfo {author} {\bibfnamefont {E.~P.}\ \bibnamefont
  {Menzel}}, \bibinfo {author} {\bibfnamefont {F.}~\bibnamefont {Deppe}},
  \bibinfo {author} {\bibfnamefont {M.}~\bibnamefont {Mariantoni}}, \bibinfo
  {author} {\bibfnamefont {M.~{\'A}.~A.}\ \bibnamefont {Caballero}}, \bibinfo
  {author} {\bibfnamefont {A.}~\bibnamefont {Baust}}, \bibinfo {author}
  {\bibfnamefont {T.}~\bibnamefont {Niemczyk}}, \bibinfo {author}
  {\bibfnamefont {E.}~\bibnamefont {Hoffmann}}, \bibinfo {author}
  {\bibfnamefont {A.}~\bibnamefont {Marx}}, \bibinfo {author} {\bibfnamefont
  {E.}~\bibnamefont {Solano}},\ and\ \bibinfo {author} {\bibfnamefont
  {R.}~\bibnamefont {Gross}},\ }\bibfield  {title} {\bibinfo {title} {Dual-path
  state reconstruction scheme for propagating quantum microwaves and detector
  noise tomography},\ }\href {https://doi.org/10.1103/PhysRevLett.105.100401}
  {\bibfield  {journal} {\bibinfo  {journal} {Phys. Rev. Lett.}\ }\textbf
  {\bibinfo {volume} {105}},\ \bibinfo {pages} {100401} (\bibinfo {year}
  {2010})}\BibitemShut {NoStop}%
\bibitem [{\citenamefont {de~L{\'e}pinay}\ \emph {et~al.}(2021)\citenamefont
  {de~L{\'e}pinay}, \citenamefont {Ockeloen-Korppi}, \citenamefont {Woolley},\
  and\ \citenamefont {Sillanp{\"a}{\"a}}}]{MercierDeLepinay2021}%
  \BibitemOpen
  \bibfield  {author} {\bibinfo {author} {\bibfnamefont {L.~M.}\ \bibnamefont
  {de~L{\'e}pinay}}, \bibinfo {author} {\bibfnamefont {C.~F.}\ \bibnamefont
  {Ockeloen-Korppi}}, \bibinfo {author} {\bibfnamefont {M.~J.}\ \bibnamefont
  {Woolley}},\ and\ \bibinfo {author} {\bibfnamefont {M.~A.}\ \bibnamefont
  {Sillanp{\"a}{\"a}}},\ }\bibfield  {title} {\bibinfo {title} {Quantum
  mechanics-free subsystem with mechanical oscillators},\ }\href
  {https://doi.org/10.1126/science.abf5389} {\bibfield  {journal} {\bibinfo
  {journal} {Science}\ }\textbf {\bibinfo {volume} {372}},\ \bibinfo {pages}
  {625} (\bibinfo {year} {2021})}\BibitemShut {NoStop}%
\bibitem [{\citenamefont {Ockeloen-Korppi}\ \emph {et~al.}(2016)\citenamefont
  {Ockeloen-Korppi}, \citenamefont {Damsk{\"a}gg}, \citenamefont
  {Pirkkalainen}, \citenamefont {Clerk}, \citenamefont {Woolley},\ and\
  \citenamefont {Sillanp{\"a}{\"a}}}]{OckeloenKorppi2016}%
  \BibitemOpen
  \bibfield  {author} {\bibinfo {author} {\bibfnamefont {C.~F.}\ \bibnamefont
  {Ockeloen-Korppi}}, \bibinfo {author} {\bibfnamefont {E.}~\bibnamefont
  {Damsk{\"a}gg}}, \bibinfo {author} {\bibfnamefont {J.-M.}\ \bibnamefont
  {Pirkkalainen}}, \bibinfo {author} {\bibfnamefont {A.~A.}\ \bibnamefont
  {Clerk}}, \bibinfo {author} {\bibfnamefont {M.~J.}\ \bibnamefont {Woolley}},\
  and\ \bibinfo {author} {\bibfnamefont {M.~A.}\ \bibnamefont
  {Sillanp{\"a}{\"a}}},\ }\bibfield  {title} {\bibinfo {title} {Quantum
  backaction evading measurement of collective mechanical modes},\ }\href
  {https://doi.org/10.1103/PhysRevLett.117.140401} {\bibfield  {journal}
  {\bibinfo  {journal} {Phys. Rev. Lett.}\ }\textbf {\bibinfo {volume} {117}},\
  \bibinfo {pages} {140401} (\bibinfo {year} {2016})}\BibitemShut {NoStop}%
\bibitem [{\citenamefont {Scigliuzzo}\ \emph {et~al.}(2020)\citenamefont
  {Scigliuzzo}, \citenamefont {Bengtsson}, \citenamefont {Besse}, \citenamefont
  {Wallraff}, \citenamefont {Delsing},\ and\ \citenamefont
  {Gasparinetti}}]{Scigliuzzo2020}%
  \BibitemOpen
  \bibfield  {author} {\bibinfo {author} {\bibfnamefont {M.}~\bibnamefont
  {Scigliuzzo}}, \bibinfo {author} {\bibfnamefont {A.}~\bibnamefont
  {Bengtsson}}, \bibinfo {author} {\bibfnamefont {J.-C.}\ \bibnamefont
  {Besse}}, \bibinfo {author} {\bibfnamefont {A.}~\bibnamefont {Wallraff}},
  \bibinfo {author} {\bibfnamefont {P.}~\bibnamefont {Delsing}},\ and\ \bibinfo
  {author} {\bibfnamefont {S.}~\bibnamefont {Gasparinetti}},\ }\bibfield
  {title} {\bibinfo {title} {Primary thermometry of propagating microwaves in
  the quantum regime},\ }\href {https://doi.org/10.1103/PhysRevX.10.041054}
  {\bibfield  {journal} {\bibinfo  {journal} {Phys. Rev. X}\ }\textbf {\bibinfo
  {volume} {10}},\ \bibinfo {pages} {041054} (\bibinfo {year}
  {2020})}\BibitemShut {NoStop}%
\end{thebibliography}%
\let\addcontentsline\ms@addcontentsline
\makeatother

\clearpage
\onecolumngrid
\vspace{6pt}
\begin{center}\large\textbf{End Matter}\end{center}
\vspace{4pt}
\twocolumngrid

\begin{figure*}[!tp]
 \includegraphics{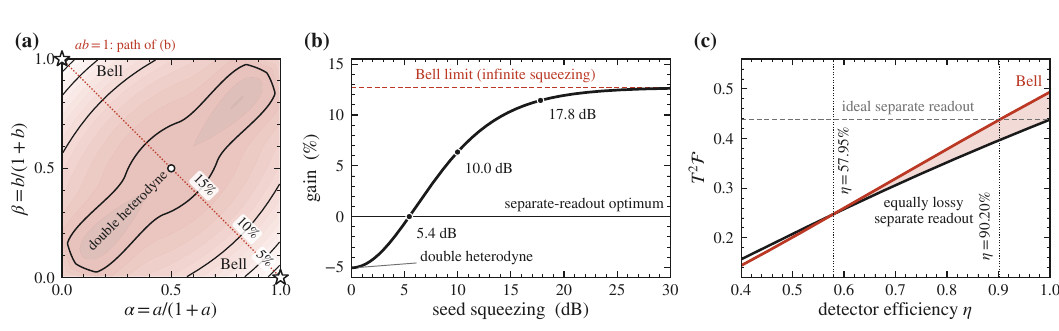}
 \caption{\label{fig:optimality}
 All panels at the cap $x_1=x_c$, $x_2=1.462$. \textbf{(a)} Shortfall $1-\FI(M)/\FI_{\mathrm{Bell}}$ over pure seeds $M$, written as a beam splitter of angle $\theta$ acting on squeezed vacua of strengths $a$ and $b$ (End Matter D) and minimized over $\tau=\cos^2\theta$ at each point. The axes $\alpha=a/(1+a)$ and $\beta=b/(1+b)$ put the infinite-squeezing limits on the edges. The shortfall vanishes only at the two starred Bell corners; double heterodyne is at $15.7\%$, a stationary point of Eq.~\eqref{eq:gaussianFI}. The dotted anti-diagonal carries the balanced seeds $ab=1$ of (b). \textbf{(b)} Best finite-seed gain against squeezing: negative at $0$ dB, zero at $5.4$ dB, half the Bell gain at $10.0$ dB, nine tenths at $17.8$ dB. \textbf{(c)} $T^2\FI$ against the efficiency $\eta$ shared by all homodynes. Bell overtakes an equally lossy separate readout at $\eta=57.95\%$ and the ideal lossless bound at $\eta=90.20\%$.}
\end{figure*}

\textit{A. Separate-readout bound.---}
For one mode the seed is $M(r)=\operatorname{diag}(r,1/r)$, so $V+M=\operatorname{diag}(\nu+r,\nu+1/r)$ and Eq.~\eqref{eq:gaussianFI} reads
\begin{equation}
 T^2\FI(r)=\frac{\mu^2}{2}\left[\frac{1}{(\nu+r)^2}+\frac{r^2}{(1+\nu r)^2}\right].
 \label{eq:emA}
\end{equation}
Differentiating and clearing the positive denominators leaves the stationarity condition $r(\nu+r)^{3}=(1+\nu r)^{3}$. Both sides are cubes, so with $y=r^{1/3}$ this reads $y^{4}-\nu y^{3}+\nu y-1=0$, which factors as $(y-1)(y+1)(y^2-\nu y+1)=0$. The root $y=1$ is the vacuum seed $r=1$, heterodyne detection, and $y=-1$ is unphysical. The quadratic has real roots only for $\nu\ge2$, and its two roots multiply to one, so the corresponding $r=y^{3}$ form a reciprocal pair exchanged by swapping the quadratures. Evaluating Eq.~\eqref{eq:emA} there and comparing with $r=1$ and with the limits $r\to0,\infty$ shows both to be minima. The maximum is therefore always $r=1$ or the infinite-squeezing boundary, which is Eq.~\eqref{eq:localmode} (Supplemental Secs.~S1--S2).

An adaptive separate readout measures mode~1, then sets the second detector from that outcome. The two modes are uncorrelated, so the joint outcome density factorizes for every setting, and the Fisher chain rule gives $\FI=\FI_1(M_1)+\mathbb{E}_{o_1}[\FI_2(M_2(o_1))]$. Each term is bounded by its own optimum in Eq.~\eqref{eq:localmode}, so the sum is bounded by $\FI_{\mathrm{local}}^{\max}$. Randomized settings are the same argument with a fixed averaging distribution.

\textit{B. Onset law.---}
On the homodyne branch of Eq.~\eqref{eq:localmode} the two competitors are
$T^2\FI_{\mathrm{local}}^{\max}=\tfrac12(\lambda_1^2+\lambda_2^2)$ and $T^2\FI_{\mathrm{Bell}}=\bar\lambda^2$, with
$\lambda_i=\mu_i/\nu_i$ and $\bar\lambda=(\nu_1\lambda_1+\nu_2\lambda_2)/(\nu_1+\nu_2)$. The benchmark is a mean of squares and the Bell value is the square of a $\nu$-weighted mean, so the whole comparison is between those two ways of averaging one quantity.

Two logarithmic derivatives generate the expansion. With $\lambda=2x/\sinh2x$ and $\nu=\coth x$,
\begin{equation}
 \frac{d\ln\lambda}{d\ln x}=1-2x\coth2x=1-c,\qquad
 \frac{d\ln\nu}{d\ln x}=-\lambda .
 \label{eq:emB}
\end{equation}
Set $x_{1,2}=xe^{\mp s/2}$, write $f=\ln\lambda$ and $w=\ln\nu$ as functions of $\ln x$, and expand both competitors to second order in $s$. The mean of squares gives $\lambda^2[1+\tfrac14f''s^2+\tfrac12f'^2s^2]$, while $\bar\lambda=\lambda[1+\tfrac18(f''+f'^2+2f'w')s^2]$, so the Bell value is $\lambda^2[1+\tfrac14(f''+f'^2+2f'w')s^2]$. Every second derivative cancels in the ratio, which leaves
$\FI_{\mathrm{Bell}}/\FI_{\mathrm{local}}^{\max}-1=\tfrac14(2f'w'-f'^2)s^2+O(s^4)$. Equation~\eqref{eq:emB} gives $f'=1-c$ and $w'=-\lambda$, so $2f'w'-f'^{2}=2\lambda(c-1)-(c-1)^{2}=(c-1)[2\lambda-(c-1)]$, the bracket of Eq.~\eqref{eq:onset}.

The bracket $2\lambda-(c-1)$ is positive exactly when $h(x)=\sinh2x+4x-2x\cosh2x>0$. That function vanishes at $x=0$ and has derivative $h'(x)=4-4x\sinh2x$, which changes sign once, so $h$ rises and then falls and has a single positive root, $x_w=1.0095565499$.

Nothing in that comparison of averages is thermal. For any width law write $\psi_\nu=\tfrac12(\ln\nu_1-\ln\nu_2)$ and $\psi_\lambda=\tfrac12(\ln\lambda_1-\ln\lambda_2)$, so that $\nu_{1,2}=\tilde\nu e^{\pm\psi_\nu}$ and $\lambda_{1,2}=\tilde\lambda e^{\pm\psi_\lambda}$ about the geometric means $\tilde\nu$ and $\tilde\lambda$. The mean of squares is $\tilde\lambda^2\cosh2\psi_\lambda$, the $\nu$-weighted mean is $\bar\lambda=\tilde\lambda(\cosh\psi_\lambda+\tanh\psi_\nu\sinh\psi_\lambda)$, and second order gives $\FI_{\mathrm{Bell}}/\FI_{\mathrm{local}}^{\max}-1=\psi_\lambda(2\psi_\nu-\psi_\lambda)$. The thermal splitting with Eq.~\eqref{eq:emB} gives $\psi_\nu=\lambda s/2$ and $\psi_\lambda=(c-1)s/2$, and $\psi_\lambda(2\psi_\nu-\psi_\lambda)=\tfrac14(c-1)[2\lambda-(c-1)]s^2$ is Eq.~\eqref{eq:onset} (Supplemental Secs.~S4 and S6).

\textit{C. Universal bound.---}
With $m=\mu_2/\mu_1$ and $\ell_i=1/\Gamma(\nu_i)$, Eqs.~\eqref{eq:localmode} and \eqref{eq:bellFI} give $\FI_{\mathrm{Bell}}/\FI_{\mathrm{local}}^{\max}=(1+m)^2/[(\nu_1+\nu_2)^2(\ell_1+m^2\ell_2)]$. By the Cauchy--Schwarz inequality the ratio is largest at $\ell_1=m\ell_2$, that is $m=\Gamma(\nu_2)/\Gamma(\nu_1)$, where it equals $[\Gamma(\nu_1)+\Gamma(\nu_2)]/(\nu_1+\nu_2)^2$, the first inequality of Eq.~\eqref{eq:universal}. For the second, order $\nu_1\le\nu_2$. With both widths below $\nu_c$ the bound is $2(\nu_1^2+\nu_2^2)/(\nu_1+\nu_2)^2$, decreasing in $\nu_1$ and, at $\nu_1=1$, increasing in $\nu_2$, so it is largest as $\nu_1\to1$ and $\nu_2\to\nu_c$. With $\nu_1\le\nu_c\le\nu_2$ the bound is $[2\nu_1^2+(\nu_2+1)^2]/(\nu_1+\nu_2)^2$, decreasing in $\nu_1$ and then in $\nu_2$, with the same corner. With both widths above $\nu_c$ the derivative in $\nu_1$ is negative, which reduces it to the previous case. The corner value is $4-2\sqrt2$, and no smooth family attains it, because a width equal to the vacuum value has zero response, $\mu_1=0$. The thermal specialization is in Supplemental Sec.~S5 and the full width-law proof in Sec.~S6.

\textit{D. Proof of Gaussian optimality.---}
The search runs over pure two-mode seeds $M$. The phase of each output leaves the outcome statistics unchanged, so we fix it and lose nothing. The Bloch--Messiah form remains, in which every pure two-mode seed is one beam splitter of mixing angle $\theta$ acting on two squeezed vacua, $M=B_\theta[D(a)\oplus R_\phi D(b)R_\phi^{T}]B_\theta^{T}$ with $D(a)=\operatorname{diag}(a,a^{-1})$. The second squeezer is tilted by $R_\phi$, the rotation by $\phi$ in the quadrature plane. Write $u=\cos2\phi$, so that $u=\pm1$ are the seeds whose squeezing axes line up and the interior holds the tilted ones. In this variable Eq.~\eqref{eq:gaussianFI} is a quadratic in $u$ divided by the square of a function that is itself linear in $u$. For $x_2>x_1$ we prove that the numerator of $d\FI/du$ is a straight line in $u$ with nonnegative slope, so it changes sign at most once, from negative to positive. The Fisher information therefore falls and then rises, which makes any interior stationary point a minimum, so the correlations between $q$ and $p$ that a tilt creates never help.

The candidates left are the aligned seeds, which in the quadrature ordering $(q_1,q_2,p_1,p_2)$ read $M=X\oplus X^{-1}$ with $X=R_\theta\operatorname{diag}(a,b)R_\theta^{T}$; the beam-splitter angle $\theta$ now sits inside $X$. The compactification $\alpha=a/(1+a)$, $\beta=b/(1+b)$, $\tau=\cos^2\theta$ maps that family onto the unit cube, with the infinite-squeezing limits on its faces. Clearing the positive denominator of $\FI_{\mathrm{Bell}}-\FI(M)$ turns it into a polynomial of degree four in each variable. We split the cube into subboxes and expand that polynomial in the Bernstein basis of each, where a list of positive coefficients proves positivity. The arithmetic is interval arithmetic throughout, at $120$ digits for a single thermal pair and $100$ digits for a region of them, so each coefficient is an enclosure of the exact value.

\begin{table}[!t]
\caption{\label{tab:cert}Bernstein proofs of Bell optimality. On every tile listed Bell homodyne is proved strictly optimal, and the only zeros are structural. A tile is a region of thermal pairs; each is proved on Bernstein subboxes of the seed cube.}
\begin{ruledtabular}
\begin{tabular}{lcc}
Domain & Tiles & Bernstein coefficients\\
\hline
$x_1=x_c$, $x_2=3/2$ & 1 & 742 positive, 8 zeros\\
$x_1\in[0.36,0.52]$, $x_2\in[1.15,1.85]$ & 22 & positive, 8 zeros per tile\\
$x_1\in[0.44,0.50]$, $x_2\in[0.82,0.94]$ & 4 & positive, 8 zeros per tile\\
\end{tabular}
\end{ruledtabular}
\end{table}

At the reference pair $x_1=x_c$, $x_2=3/2$, six subboxes carry $750$ Bernstein coefficients, $742$ strictly positive and $8$ zero. Those eight are structural, sitting at the two balanced infinite-squeezing corners, which are the Bell measurement itself. To prove optimality on a region of thermal pairs, we replace $\nu_i$ and $\mu_i$ by second-order Taylor polynomials in $x_i$ and bound each remainder over the tile. The bounded remainder then becomes a formal parameter, carried through the same Bernstein expansion. Table~\ref{tab:cert} lists the outcome, with $22$ tiles covering a neighborhood of the maximal-gain ridge and $4$ tiles covering the coupled-resonator operating point. The maximal-gain pair of Eq.~\eqref{eq:cap} lies inside the first region. The reduction and both proofs are in Supplemental Secs.~S7--S10.

\textit{E. Double heterodyne.---}
A search for the best Gaussian receiver naturally starts at double heterodyne, one vacuum-seeded detector on each mode, $M=I_4$. That point is an exact stationary point of Eq.~\eqref{eq:gaussianFI} for every thermal product state, since the gradient in the seed parameters vanishes identically. The maximum lies elsewhere, so a gradient search that starts there stays there and reports the separate-readout answer. The optimum sits where no stationarity condition points, at infinite squeezing on the boundary of the seed family, and finite seeds approach it slowly. At the maximal-gain pair of Eq.~\eqref{eq:cap} the best finite seed is still worse than the separate-readout benchmark below $5.4$ dB of quadrature squeezing. It recovers half the Bell gain at $10.0$ dB and nine tenths of it at $17.8$ dB [Fig.~\ref{fig:optimality}(b); Supplemental Sec.~S13]. Bell homodyne escapes because a balanced beam splitter followed by two homodyne detectors realizes the infinite-squeezing limit directly, with no squeezed light anywhere in the network.

\textit{F. Amplifier as receiver.---}
A phase-preserving amplifier of gain $G$ acts as $b_s=\sqrt G\,a_s+\sqrt{G-1}\,a_i^\dagger$ with signal $a_s$ and idler $a_i$. Heterodyne detection of $b_s$ records the commuting pair $q_s+q_i$ and $p_s-p_i$, up to a common scale, as $G\to\infty$. Flipping both idler quadratures, a phase rotation that leaves a thermal state unchanged, maps that pair onto $q_-$ and $p_+$ of Eq.~\eqref{eq:bellFI}. A transmission $L$ before the amplifier and $n_{\mathrm{ex}}$ quanta of excess added noise referred to its input enter the loss model of Fig.~\ref{fig:optimality}(c) as one homodyne efficiency $\eta=L/(1+n_{\mathrm{ex}})$. A following amplifier of noise $n_H$ quanta adds $n_H/G$. At $G=100$ ($20$ dB) with $n_H=10$, and $n_H/G$ replacing the record's vacuum unit, the readout keeps $0.90$ of the ideal Bell Fisher information at the $4$ and $10$ GHz pair. The efficiency $\eta=1/(1+n_H/G)$ reproduces that to $0.2\%$. At $G=1000$ it is $0.99$. Pumped degenerately, the amplifier gives the recorded quadrature the amplitude gain $\sqrt G+\sqrt{G-1}$, so the following amplifier's noise referred to the input is a quarter of the phase-preserving value, $n_H/(4G)$ quanta. It enters the homodyne width as $n_H/(2G)$ against the Bell record's $n_H/G$, so the homodyne benchmark runs at $\eta_h=L/(1+n_H/(2G))$. A same-chain comparison at $G=100$ and $n_H=10$ then gives $4.0\%$ at the $4$ and $10$ GHz pair against $9.1\%$ at one common efficiency. At $G=1000$ they differ by $0.5$ percentage points, $9.4\%$ against $9.9\%$ (Supplemental Sec.~S15).

\clearpage
\onecolumngrid

\setcounter{secnumdepth}{3}
\setcounter{section}{0}
\setcounter{equation}{0}
\setcounter{figure}{0}
\setcounter{table}{0}
\renewcommand{\thesection}{S\arabic{section}}
\renewcommand{\theequation}{S\arabic{equation}}
\renewcommand{\thetable}{S\arabic{table}}
\renewcommand{\thefigure}{S\arabic{figure}}
\renewcommand{\theHsection}{S\arabic{section}}
\renewcommand{\theHequation}{S\arabic{equation}}
\renewcommand{\theHtable}{S\arabic{table}}
\renewcommand{\theHfigure}{S\arabic{figure}}

\vspace{6pt}
\begin{center}
 {\large\textbf{Supplemental Material}}\\[3pt]
 {\large\textbf{Gaussian Fisher Information Is Superadditive}}
\end{center}
\vspace{4pt}

\tableofcontents
\clearpage

\noindent\textit{Roadmap.---} The End Matter of the Letter proves each result in short form, and this Supplement carries the full derivations. Sections~S1--S3 set the conventions, the separate-readout bound and the beam-splitter family. Sections~S4--S6 derive the onset law, the thermal cap, the universal bound and the three-mode pairing theorem. Sections~S7--S10 prove Gaussian optimality at the reference pair and on two rectangles. Sections~S11--S15 treat coupled resonators, detector loss, the frequency-bank pairing and the amplifier, and Sec.~S16 states data availability.

\section{Conventions}
\label{S:sec:conventions}

The Letter states the Gaussian Fisher information of a seed measurement and the two thermal numbers per mode without deriving either~\cite{Weedbrook2012,BraunsteinVanLoock2005}. We derive both here, and the derivation shows that a Gaussian detector sees exactly two numbers per mode, one width and one response.

A centered two-mode Gaussian state is fixed by its first moments and a covariance matrix $V$ obeying $V+i\Omega\ge0$, where $\Omega=\bigoplus_i\left(\begin{smallmatrix}0&1\\-1&0\end{smallmatrix}\right)$ collects the canonical commutators $[q_i,p_i]=i$. We order quadratures as $(q_1,p_1,q_2,p_2)$ unless a section says otherwise, and Sec.~\ref{S:sec:certpoint} uses $(q_1,q_2,p_1,p_2)$ so that the seeds treated there are block diagonal. A Gaussian measurement is fixed by one seed Gaussian state of covariance $M$ (the noise the detector itself adds) together with all displacements of that seed in phase space. Those displacements label the outcomes. Displacing the seed by a fixed vector only shifts the outcome distribution, and that shift does not depend on temperature, so it leaves every Fisher information below unchanged. The outcome density is Gaussian with covariance $V+M$, so for a temperature-dependent covariance and a temperature-independent first moment,
\begin{equation}
 T^2\FI(V;M)=\frac12\Tr\left[\left((V+M)^{-1}T\partial_TV\right)^2\right].
 \label{S:eq:FI}
\end{equation}
The factor $T^2$ makes the Fisher information dimensionless.

For two independent thermal modes,
\begin{equation}
 V=\nu_1I_2\oplus\nu_2I_2,\quad
 \nu_i=\coth x_i,\quad
 \mu_i=x_i\csch^2x_i,
 \label{S:eq:thermal}
\end{equation}
with $x_i=\hbar\omega_i/(2k_{\mathrm B}T)$ and $\mu_i=T\partial_T\nu_i$. The derivative in Eq.~\eqref{S:eq:FI} is then $T\partial_TV=\mu_1I_2\oplus\mu_2I_2$, so a Gaussian detector is sensitive to each mode only through one width and one response. The ratio
\begin{equation}
 \lambda_i=\frac{\mu_i}{\nu_i}=\frac{2x_i}{\sinh 2x_i}
 \label{S:eq:lambda}
\end{equation}
says how strongly a mode's width responds to temperature. It equals $1$ at $x_i=0$ and decreases from there, and it is the number the rest of this Supplement compares between the two modes.

\section{Separate-readout bound}

\begin{figure*}[!tp]
 \includegraphics{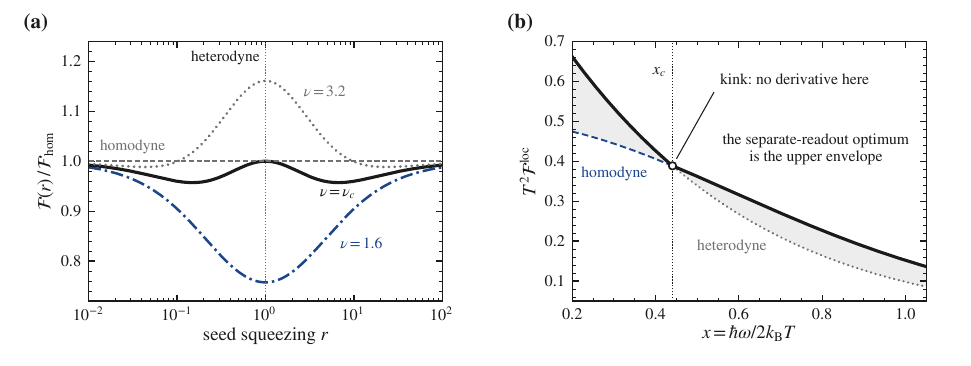}
 \caption{\label{S:fig:localbound}
 The single-mode optimum and the kink it creates. \textbf{(a)} Fisher information of one thermal
 mode against the squeezing $r$ of its seed, divided by the homodyne value, for three widths.
 Both boundaries $r\to0$ and $r\to\infty$ are homodyne detection and share one value, and the
 centre $r=1$ is heterodyne detection. Heterodyne is the better reading only for
 $\nu>\nu_c=1+\sqrt2$, and at $\nu=\nu_c$ the two tie. \textbf{(b)} The two branch values against
 temperature. The separate-readout optimum is their upper envelope, drawn solid, and the envelope has
 no derivative where the branches cross at $x_c$. The universal cap of the collective gain sits
 exactly on that corner.}
\end{figure*}
\label{S:sec:localbound}

End Matter A takes the one-mode stationarity condition to the factorization $(y-1)(y+1)(y^2-\nu_i y+1)=0$ in $y=r^{1/3}$, and checks at $\nu_i=3$ that the reciprocal pair of roots falls below the boundary value. We prove that for every $\nu_i$ at once. The one-mode seed is $M(r)=\operatorname{diag}(r,r^{-1})$ up to a phase rotation, and Eq.~\eqref{S:eq:FI} reads
\begin{equation}
 T^2\FI_i(r)=\frac{\mu_i^2}{2}\left[(\nu_i+r)^{-2}+(\nu_i+r^{-1})^{-2}\right].
 \label{S:eq:FIone}
\end{equation}
This is invariant under $r\mapsto1/r$, which only swaps the two quadratures, and its two boundary limits $r\to0$ and $r\to\infty$ are both equal to $\mu_i^2/(2\nu_i^2)$. The reciprocal stationary points are the roots of the quadratic factor $y^2-\nu_i y+1$, that is the values of $y$ with $y+y^{-1}=\nu_i$, and they are real only when $\nu_i\ge2$. Take such a $y$ and set $r=y^3$, so that $r$ is one of those stationary points. Then $\nu_i+r=(1+y^2+y^4)/y$, and $\nu_i+r^{-1}=(1+\nu_i r)/r=(1+y^2+y^4)/y^3$. The two terms of Eq.~\eqref{S:eq:FIone} therefore share one denominator, and
\begin{equation}
 T^2\FI_i(y^3)=\frac{\mu_i^2}{2}\,\frac{y^2(1+y^4)}{(1+y^2+y^4)^2}.
\end{equation}
The boundary value $\mu_i^2/(2\nu_i^2)$ becomes $\mu_i^2y^2/[2(1+y^2)^2]$, because $\nu_i=y+y^{-1}=(1+y^2)/y$. Subtracting the stationary value from it and putting the two over a common denominator leaves
\begin{equation}
 \frac{\mu_i^2}{2}\,\frac{y^6}{(1+y^2)^2(1+y^2+y^4)^2}>0,
\end{equation}
because the numerator is $y^2$ times $(1+y^2+y^4)^2-(1+y^4)(1+y^2)^2$, and that difference equals $y^4$ identically in $y$. The reciprocal stationary points therefore sit strictly below the boundary value at every $\nu_i\ge2$, so they are never the maximum. The remaining comparison is $r=1$ against the boundary, which is
\begin{equation}
 T^2\FI_i^{\mathrm{loc}}=\mu_i^2\max\left\{(2\nu_i^2)^{-1},(\nu_i+1)^{-2}\right\}.
 \label{S:eq:local}
\end{equation}

The two entries of that maximum cross where $2\nu_i^2=(\nu_i+1)^2$. Taking positive square roots gives $\sqrt2\,\nu_i=\nu_i+1$, so $\nu_i=1/(\sqrt2-1)=1+\sqrt2\equiv\nu_c$, that is $x_c=\operatorname{arccoth}(1+\sqrt2)=0.4406867935$. The homodyne entry $1/(2\nu_i^2)$ is the larger one for $\nu_i<\nu_c$, and the heterodyne entry $1/(\nu_i+1)^2$ for $\nu_i>\nu_c$. Both entries are smooth in $\nu_i$ and cross transversally, so the envelope carries a corner at $\nu_c$, where its two one-sided slopes differ, and the gain inherits that corner. Summing Eq.~\eqref{S:eq:local} over the two modes gives the benchmark $\FI_{\mathrm{local}}^{\max}=\FI_1^{\mathrm{loc}}+\FI_2^{\mathrm{loc}}$ that a joint measurement has to beat.

That benchmark holds against more than one round of detectors, and End Matter A shows it for two rounds. We prove the same bound for any number of rounds on any number of modes. Measure the modes in some order, each setting chosen from all earlier outcomes. The input is a product state, so the joint outcome density factorizes for every choice of settings, and the chain rule turns the total Fisher information into one term per mode. Each term is an average over earlier outcomes of the Fisher information of a single-mode Gaussian measurement, and Eq.~\eqref{S:eq:local} bounds every one of them. Randomized settings are the same argument with a fixed averaging distribution in place of the conditional one. Anything done to the collected outcomes afterwards is a temperature-independent map of the data, which cannot raise the Fisher information. We therefore find the separate-readout benchmark surviving adaptivity, randomization and classical post-processing together.

\section{Bell homodyne and the beam-splitter family}
\label{S:sec:passivenet}

The Letter takes the balanced beam splitter as its joint receiver and shows that it beats the separate-readout benchmark~\cite{Spedalieri2013}. Within the beam-splitter family the maximum lies at the balanced splitter or at no splitter at all, and Sec.~\ref{S:sec:ceiling} reduces every passive network to this family. A balanced splitter is one point of a one-parameter family, and the best member of that family is either the balanced splitter or no splitter at all. Take a beam splitter of transmissivity $t$. Homodyne the $q$ quadrature of one output, and the $p$ quadrature of the other. At $t=1/2$ these two records are $q_-=(q_1-q_2)/\sqrt2$ and $p_+=(p_1+p_2)/\sqrt2$, the continuous-variable Bell pair. A beam splitter acts as the same rotation on the two $q$'s and on the two $p$'s, so a thermal product state, which has no $q$--$p$ correlation, still has none after the splitter. The two records are therefore independent, and their Fisher informations add:
\begin{align}
T^2\FI_{\mathrm{hom}}(t)=&\frac12\left[\frac{t\mu_1+(1-t)\mu_2}{t\nu_1+(1-t)\nu_2}\right]^2\nonumber\\
&+\frac12\left[\frac{(1-t)\mu_1+t\mu_2}{(1-t)\nu_1+t\nu_2}\right]^2.
\label{S:eq:passive}
\end{align}
Let $z=(2t-1)^2$, so that $z$ runs over $[0,1]$, with $z=0$ the balanced splitter $t=1/2$ and $z=1$ the two ends $t=0$ and $t=1$, where the splitter mixes nothing. Write $\Sigma_\mu=\mu_1+\mu_2$, $\Delta_\mu=\mu_1-\mu_2$, $\Sigma_\nu=\nu_1+\nu_2$, and $\Delta_\nu=\nu_1-\nu_2$. Then
\begin{equation}
T^2\FI_{\mathrm{hom}}(z)=
\frac{\Delta_\mu^2\Delta_\nu^2z^2
+(\Delta_\mu^2\Sigma_\nu^2-4\Delta_\mu\Delta_\nu\Sigma_\mu\Sigma_\nu
+\Delta_\nu^2\Sigma_\mu^2)z+\Sigma_\mu^2\Sigma_\nu^2}
{(\Sigma_\nu^2-\Delta_\nu^2z)^2}.
\end{equation}
Its derivative is
\begin{equation}
\frac{d\,T^2\FI_{\mathrm{hom}}}{dz}=
\frac{(\Delta_\mu\Sigma_\nu-\Delta_\nu\Sigma_\mu)
\left[\Delta_\nu^2z(3\Delta_\mu\Sigma_\nu-\Delta_\nu\Sigma_\mu)
+\Sigma_\nu^2(\Delta_\mu\Sigma_\nu-3\Delta_\nu\Sigma_\mu)\right]}
{(\Sigma_\nu^2-\Delta_\nu^2z)^3}.
\end{equation}
Order the modes by $x_2>x_1$. Both $\nu=\coth x$ and $\mu=x\csch^2x$ decrease in $x$. For the second, $d\mu/dx=\csch^2x\,(1-2x\coth x)$, and $2x\coth x>1$ at every $x>0$. Hence $\Delta_\nu>0$ and $\Delta_\mu>0$. The combination that controls the sign is $\Delta_\mu\Sigma_\nu-\Delta_\nu\Sigma_\mu=2(\mu_1\nu_2-\mu_2\nu_1)=2\nu_1\nu_2(\lambda_1-\lambda_2)$, which is positive because $\lambda$ decreases in $x$ as well. In the bracket, $z$ multiplies $\Delta_\nu^2(3\Delta_\mu\Sigma_\nu-\Delta_\nu\Sigma_\mu)$, and splitting $3\Delta_\mu\Sigma_\nu-\Delta_\nu\Sigma_\mu$ as $2\Delta_\mu\Sigma_\nu+(\Delta_\mu\Sigma_\nu-\Delta_\nu\Sigma_\mu)$ makes it positive term by term. The bracket is therefore a line in $z$ with positive slope, so the derivative crosses zero at most once, and only from negative to positive. Since $\Sigma_\nu>\Delta_\nu>0$, the denominator $(\Sigma_\nu^2-\Delta_\nu^2z)^3$ stays positive on $0\le z\le1$. The Fisher information therefore has no interior maximum across the family, so its largest value is at one of the two ends. We therefore find the maximum at $z=0$ or $z=1$: the balanced Bell homodyne, or two separate homodynes on conjugate quadratures, which is the homodyne branch of Eq.~\eqref{S:eq:local} on both modes.

A relative phase between two local oscillators traces the same family with no beam splitter turned at all. Suppose the two probes have already been mixed by a balanced coupling. Homodyning each output at local-oscillator phases $\varphi_1$ and $\varphi_2$ then reproduces Eq.~\eqref{S:eq:passive} with $2t-1=\cos(\varphi_1-\varphi_2)$. The best phase setting is again one of the two ends. Section~\ref{S:sec:resonators} derives that setting for two coupled resonators and identifies which end is larger there. Proposition~3 of Sec.~\ref{S:sec:ceiling} extends the endpoint statement to passive networks of the two modes and one vacuum ancilla, which admit heterodyne detection.

\section{Universality}
\label{S:sec:universality}

End Matter B derives the onset law for a small splitting, with both modes on the homodyne branch. We show here how the advantage decays as $2\mu_2/\mu_1$ along the ray $x_1=x_c$, Sec.~\ref{S:sec:cap} proves that it stays open at every splitting, and we begin where End Matter B stopped.

End Matter B reduces the comparison to two ways of averaging one number. On the homodyne branch the benchmark is the mean of $\lambda_1^2$ and $\lambda_2^2$, and the Bell value is the square of the width-weighted mean of $\lambda_1$ and $\lambda_2$,
\begin{equation}
T^2\FI_{\mathrm{local}}^{\max}=\frac{\lambda_1^2+\lambda_2^2}{2},
\qquad
T^2\FI_{\mathrm{Bell}}=\left(\frac{\mu_1+\mu_2}{\nu_1+\nu_2}\right)^{\!2}
=\left(\frac{\nu_1\lambda_1+\nu_2\lambda_2}{\nu_1+\nu_2}\right)^{\!2},
\label{S:eq:twoaverages}
\end{equation}
with $\lambda_i=\mu_i/\nu_i$ of Eq.~\eqref{S:eq:lambda}. Homodyne is the better single-mode reading above the crossover $x_c$ and heterodyne below it, by the branch comparison that follows Eq.~\eqref{S:eq:local}. Since $x_i=\hbar\omega_i/(2k_{\mathrm B}T)$, a mode with large $x_i$ is cold and one with small $x_i$ is warm. The two entries of Eq.~\eqref{S:eq:twoaverages} agree when $\lambda_1=\lambda_2$, so only a difference between the two $\lambda_i$ can separate them. Expanding about a common $x$ with $x_{1,2}=xe^{\mp s/2}$ gives the onset coefficient, with $s=\ln(\omega_2/\omega_1)$ at fixed temperature. Label exchange sends $s$ to $-s$ and leaves both entries fixed, an identity in $s$ before any expansion is taken, so every odd power of $s$ vanishes at every order.

One identity generates every coefficient. The identity $\coth x=\coth 2x+\csch 2x$, multiplied by $2x$, splits $2x\coth x$ into $c+\lambda$ with $c=2x\coth 2x$, so $d\ln\mu/d\ln x=1-c-\lambda$. Write $f=\ln\lambda$ and $w=\ln\nu$ as functions of $\ln x$, with a prime for $d/d\ln x$, so that $f'=1-c$ and $w'=-\lambda$. Two closure relations, $d\lambda/d\ln x=\lambda(1-c)$ and $dc/d\ln x=c-\lambda^2$, express every higher derivative in $c$ and $\lambda$ alone, so every coefficient of the expansion is a rational function of those two numbers.

The same cancellation recurs two orders later. We expand both entries of Eq.~\eqref{S:eq:twoaverages} through $s^4$ at the common midpoint $x$ and divide. Every second derivative again cancels between the two entries, and the quartic coefficient is left carrying only $f'$, $w'$ and their third derivatives. The expansion that follows holds while both modes stay on the homodyne branch, that is for $x\ge x_ce^{s/2}$:
\begin{align}
\frac{\FI_{\mathrm{Bell}}}{\FI_{\mathrm{local}}^{\max}}-1
&=\frac{f'(2w'-f')}{4}\,s^2\nonumber\\
&\quad+\frac{5f'^4-8f'^3w'+3f'^2w'^2-2f'w'^3-(f'-w')f'''+f'w'''}{48}\,s^4+O(s^6).
\label{S:eq:fourthorder}
\end{align}
Differentiating $f'$ and $w'$ with the two closure relations expresses $f'''$ and $w'''$ in $c$ and $\lambda$. Writing $\bar c=c-1$ for the excess of $c$ over one, the quartic coefficient becomes
\begin{equation}
\frac{5\bar c^4-7\lambda\bar c^3+(\lambda^2-\lambda-1)\bar c^2+(\lambda^3+\lambda^2-1)\bar c-\lambda(\lambda^2-1)}{48},
\label{S:eq:quartic}
\end{equation}
which sits beside $\tfrac14\bar c(2\lambda-\bar c)$ in the same two variables. At the peak of the leading coefficient, $x=0.66678$, Eq.~\eqref{S:eq:quartic} equals $-0.008488$, so the law overshoots there and the quartic term pulls it back. The leading coefficient is positive exactly on $x_c<x<x_w$, with $x_w=1.0095565499$ the single positive root of $2\lambda=c-1$. We call that interval the window, and every temperature quoted below as inside the window lies in it.

Both truncations track the exact gain across the splittings an experiment would use, as Table~\ref{S:tab:onset} shows at three temperatures inside the window. The leading coefficient alone reproduces the exact gain to one part in a hundred at $s=0.2$ and four parts in a hundred at $s=0.4$. At $s=0.8$, a frequency ratio of $2.23$, it is fourteen parts in a hundred high. Adding the quartic term cuts each of those deviations: to one part in ten thousand at $s=0.2$, two parts in a thousand at $s=0.4$, and three parts in a hundred at $s=0.8$. The Letter's test of the law shows the exact gains collapsing onto the $s^2$ law, and the residual at the largest splitting is the quartic term Eq.~\eqref{S:eq:quartic} supplies.

\begin{table}[b]
\caption{\label{S:tab:onset}The onset law against exact evaluation of both Fisher informations. Each row fixes a common $x=\hbar\omega/(2k_{\mathrm B}T)$ and a splitting $s=\ln(\omega_2/\omega_1)$, so that $x_{1,2}=xe^{\mp s/2}$. Every pair listed keeps both modes above the crossover, where homodyne is the better single-mode reading. Gains are relative and quoted in percent.}
\begin{ruledtabular}
\begin{tabular}{ccccc}
$x$ & $s$ & exact gain & law through $s^2$ & law through $s^4$\\
\hline
$0.70$ & $0.2$ & $0.5150$ & $0.5168$ & $0.5150$\\
$0.70$ & $0.4$ & $2.038$  & $2.067$  & $2.038$\\
$0.70$ & $0.8$ & $7.775$  & $8.269$  & $7.806$\\
$0.80$ & $0.2$ & $0.4468$ & $0.4497$ & $0.4468$\\
$0.80$ & $0.4$ & $1.752$  & $1.799$  & $1.752$\\
$0.80$ & $0.8$ & $6.475$  & $7.196$  & $6.443$\\
$0.90$ & $0.2$ & $0.2879$ & $0.2906$ & $0.2879$\\
$0.90$ & $0.4$ & $1.121$  & $1.162$  & $1.119$\\
$0.90$ & $0.8$ & $4.065$  & $4.649$  & $3.953$\\
\end{tabular}
\end{ruledtabular}
\end{table}

Along the ray that holds the warm mode at the crossover, the advantage shrinks and never vanishes. Here we leave the symmetric parametrization: fix the warm mode at $x_1=x_c$ and send the cold mode to high frequency. Its width falls to the vacuum value and its response $\mu_2=x_2\csch^2x_2$ falls like $4x_2e^{-2x_2}$. The cold mode's own term in the benchmark is $\lambda_2^2/2$, the square of an exponentially small number, so it leaves the comparison. Mode~1 sits at the crossover, where its two readings tie, and a Bell receiver whose second port carries vacuum returns exactly that tied value. The gain therefore vanishes with $\mu_2$, and expanding the Bell ratio to first order in $\mu_2$ and squaring, we find it approaches $2\mu_2/\mu_1$. The exact gain is $1.60\%$ at a frequency ratio of $7.39$ and $6.5\times10^{-5}\%$ at a ratio of $20.1$, and Sec.~\ref{S:sec:cap} proves it positive at every finite splitting. A mode too cold to be worth its own detector still contributes once it is mixed into a joint record.

At finite splitting the window narrows and stays open. Its edges are the two temperatures where the exact gain changes sign, taken against the full separate-readout envelope. We return to the symmetric parametrization, so both edges are values of the common midpoint $x$ from which the pair is built as $x_{1,2}=xe^{\mp s/2}$. They sit at $x=0.4797$ and $x=1.0100$ for $s=0.2$, at $x=0.5067$ and $x=1.0116$ for $s=0.4$, and at $x=0.5283$ and $x=1.0200$ for $s=0.8$. The band therefore spans a factor $2.106$, $1.996$ and $1.931$. Each lower edge lies below $x_ce^{s/2}$, so the advantage extends below the branch condition under which Eq.~\eqref{S:eq:fourthorder} was taken. As $s\to0$ the edges return to $x_c$ and to the sign change $x_w$ of the leading coefficient, and the factor to $x_w/x_c=2.2909$.

\section{Thermal cap}
\label{S:sec:cap}

End Matter C proves the universal bound, and this section supplies the thermal specialization used in the Letter. We locate the thermal maximum by showing how the warm-mode maximum is pinned to the crossover, where the separate-readout benchmark changes branch, and then prove that no other pair of frequencies at any temperature does better.

The benchmark is an upper envelope of two smooth curves, homodyne worth $\mu^2/(2\nu^2)$ and heterodyne worth $\mu^2/(\nu+1)^2$, which cross at $\nu_c=1+\sqrt2$ by the branch comparison that follows Eq.~\eqref{S:eq:local}. In the temperature variable that crossing sits at $x_c=\tfrac12\ln(1+\sqrt2)$, where $\sinh 2x_c=1$ and $\cosh 2x_c=\sqrt2$, so every thermal function collapses to closed form,
\begin{equation}
\nu(x_c)=1+\sqrt2,\quad
\lambda(x_c)=2x_c,\quad
\mu(x_c)=2(1+\sqrt2)x_c,\quad
c(x_c)=2\sqrt2\,x_c,
\label{S:eq:crossvalues}
\end{equation}
and the two branches take the common value $T^2\FI^{\mathrm{loc}}(x_c)=2x_c^2$.

The corner of an upper envelope always bends upward. Above a crossing the branch with the larger slope is the one that takes over, so the slope of the envelope $E$ jumps upward as $x_1$ passes $x_c$. Write $E=T^2\FI_{\mathrm{local}}^{\max}$ for the benchmark and $G=T^2\FI_{\mathrm{Bell}}/E$ on ordered pairs $x_2>x_1$, so that $G-1$ is the relative gain. Its numerator is smooth in $x_1$ and $E$ is continuous at the corner, so the only jump in $\partial_{x_1}G$ comes from $\partial_{x_1}(1/E)=-E'/E^2$. That term reproduces the jump in $E'$ with its sign reversed and its size divided by $E^2$. At a concave corner the slope can jump from positive to negative without passing through zero, so a maximum can sit exactly on the corner and no stationarity condition finds it. The signs of the two one-sided slopes find it instead.

Because the two one-sided slopes have opposite signs, the maximum has nowhere to move off the kink. Differentiate $\ln G$ in $\ln x_1$ using $d\ln\nu/d\ln x=-\lambda$ and $d\ln\mu/d\ln x=1-c-\lambda$. Every term is common to the two sides except the one carrying mode~1's own contribution $E_1=T^2\FI_1^{\mathrm{loc}}$. The benchmark is the sum $E=E_1+E_2$ and mode~2 does not move with $x_1$, so $d\ln E/d\ln x_1=(E_1/E)\,d\ln E_1/d\ln x_1$. That single term is where the two branches differ, and it reads
\begin{equation}
2(1-c_1)\quad\text{above the crossover},
\qquad
2(1-c_1-\lambda_1)+\frac{2\mu_1}{\nu_1+1}\quad\text{below it}.
\label{S:eq:branchslopes}
\end{equation}
At $x_1=x_c$ the value above the crossover exceeds the value below it by $2x_c(2-\sqrt2)$, by Eq.~\eqref{S:eq:crossvalues}, and there $E_1=2x_c^2$. Multiplying by $E_1/E$ gives the jump in $d\ln G/d\ln x_1$. Multiplying by $G$ and dividing by $x_c$ turns that logarithmic derivative into an ordinary one. The slope $\partial_{x_1}G$ therefore drops at the crossover by $2(2-\sqrt2)\,G\,(2x_c^2)/E$. Hold the cold mode at $x_2=1.4621888$, the value fixed below. Both one-sided derivatives then evaluate to
\begin{equation}
\partial_{x_1}G\big|_{x_c^-}=+0.4214,
\qquad
\partial_{x_1}G\big|_{x_c^+}=-0.7494,
\label{S:eq:onesided}
\end{equation}
in fractional gain per unit $x_1$. The gain rises into the crossover and falls away from it, so the maximum in $x_1$ sits exactly on the kink.

In the cold mode nothing is degenerate, so the maximum is an ordinary interior root and we locate it by differentiating. Fix $x_1=x_c$, where both modes lie on the homodyne branch and $T^2\FI_{\mathrm{local}}^{\max}=(\lambda_1^2+\lambda_2^2)/2$ against $T^2\FI_{\mathrm{Bell}}=[(\mu_1+\mu_2)/(\nu_1+\nu_2)]^2$. Both are positive, so the stationarity condition can be written $d\ln G/d\ln x_2=0$. The Bell value is a square, so its logarithmic derivative is twice that of the ratio inside. The benchmark carries the same factor two from $\lambda_2^2$, and dividing it out leaves
\begin{equation}
\frac{\mu_2(1-c_2-\lambda_2)}{\mu_1+\mu_2}
+\frac{\nu_2\lambda_2}{\nu_1+\nu_2}
=\frac{\lambda_2^2\,(1-c_2)}{\lambda_1^2+\lambda_2^2},
\label{S:eq:capstat}
\end{equation}
with $c_2=2x_2\coth2x_2$ and the warm mode entering through Eq.~\eqref{S:eq:crossvalues}. The left side collects how the two sums in the Bell ratio move with $x_2$, the responses above and the widths below, and the right side how the benchmark moves. The width sum falls, since $d\ln\nu/d\ln x=-\lambda$, which is why its term enters with a plus sign. Equation~\eqref{S:eq:capstat} has one root above the crossover, at $x_2=x^*=1.4621888$, giving the frequency ratio $\omega_2/\omega_1=x^*/x_c=3.317977$ and the gain $12.699322\%$; the next paragraphs prove that this root is the global maximum.

The maximum is global, and we prove it in interval arithmetic with the tools of Secs.~\ref{S:sec:certpoint}--\ref{S:sec:region}. Write $g(x_2)=G(x_c,x_2)$ for the gain along the line $x_1=x_c$. Every quantity below is evaluated in interval arithmetic at $90$ bits of precision, so each inequality is established on a whole box together with a bound on the evaluation error, and a box on which the enclosure is too wide to decide is bisected. Three steps cover the region $x_1<x_c<x_2$ and the region $x_c<x_1<x_2$ inside the box $x_1\le3$, $x_2\le6$.

First, the line. Bisection on the sign of $g'$ encloses the root $x^*$ of Eq.~\eqref{S:eq:capstat} to $22$ digits, $x^*=1.462188764357$. On $x_c<x_2\le x_c+0.04$ the gain stays below $5\%$; on $x_c+0.04\le x_2\le x^*-10^{-5}$ the derivative $g'$ is positive, on $x^*+10^{-5}\le x_2\le3.5$ it is negative, and on $3.5\le x_2\le6$ the gain lies below $g(x^*)$. On the window $|x_2-x^*|\le10^{-5}$ the derivative obeys $|g'|\le6.9\times10^{-5}$, so the mean-value theorem bounds the gain there by $g(x^*)+7\times10^{-10}$. The gain on the line therefore has its maximum at $x^*$, and that maximum lies between $12.69932180\%$ and $12.69932187\%$.

Second, the strips on either side of the line. Equation~\eqref{S:eq:branchslopes} gives $\partial_{x_1}G>0$ on $x_c-0.02\le x_1<x_c$, $x_c\le x_2\le3.5$, and $\partial_{x_1}G<0$ on $x_c<x_1\le x_c+0.02$, $x_c+0.04\le x_2\le3.5$, established on $1.2\times10^{6}$ and $1.2\times10^{4}$ boxes. The sign pattern of Eq.~\eqref{S:eq:onesided} therefore holds for every cold mode in that range, not only at the maximum, and each warm mode within $0.02$ of the crossover is driven onto it from both sides. On the strips $G(x_1,x_2)<g(x_2)\le g(x^*)$, and the corner $x_c<x_1<x_2\le x_c+0.04$ has gain below $5\%$ directly.

Third, the rest of the box, where the gain lies below $g(x^*)$ directly: $0.05\le x_1\le x_c-0.02$ with $x_c\le x_2\le3.5$, then $x_c+0.02\le x_1\le3$ with $x_1<x_2\le3.5$, and then $3.5\le x_2\le6$ with $0.05\le x_1\le3$, on $8.5\times10^{5}$ boxes altogether. Outside the box three elementary bounds finish the argument. Since $\mu$ decreases and $\nu_2\ge1$, the Bell value obeys $T^2\FI_{\mathrm{Bell}}\le[\mu_1/(\nu_1+1)]^2(1+\mu_2/\mu_1)^2$. For $x_2\ge6$ this gives $G\le[1+\mu(6)/\mu(x_c)]^2<1.0002$ when $x_1<x_c$, and $G\le[1+\mu(6)/\mu(3)]^2<1.0099$ when $x_c\le x_1\le3$, because $2\nu_1^2\le(\nu_1+1)^2$ on that side. When both $x_i\ge3$, the benchmark $(\lambda_1^2+\lambda_2^2)/2$ is at least $(\lambda_1+\lambda_2)^2/4$ while the Bell ratio $(\lambda_1\nu_1+\lambda_2\nu_2)/(\nu_1+\nu_2)$ is at most $\nu_{\max}(\lambda_1+\lambda_2)/2$, so $G\le\coth^2 3<1.0100$. When $x_1\le0.05$, write $E_1=\mu_1^2/(\nu_1+1)^2=[2x_1/(e^{2x_1}-1)]^2<1$ and $E_2=\lambda_2^2/2\le2x_c^2$; then $T^2\FI_{\mathrm{Bell}}\le E_1(1+\nu_c\lambda_2/\mu_1)^2$ and $G\le[1+\nu_c\sqrt{2E_2}/\mu(0.05)]^2/(1+E_2)<1.030$ for every admissible $E_2$. The whole verification takes $2.0\times10^{6}$ boxes.

The same tools prove that the gain along the line never closes, so that any splitting opens a temperature window, as the Letter states. On the line, $g(x_c)=1$ and $g'(x_c)=0$, because the both-homodyne expression for $G$ is symmetric in its two arguments and equals one on the diagonal. The second derivative is $g''(x_c)=0.962$ and stays positive on $[x_c,x_c+0.04]$, established by a mean-value enclosure built from Taylor coefficients of degree three, so $g$ rises from $1$ throughout that interval; on $[x_c+0.04,3.5]$ the inequality $g>1$ holds directly, on $522$ boxes. For $x_2\ge3.5$ write $e=e^{-2x_2}\le e^{-7}$, so that $\mu_2\ge4x_2e$, $\nu_2-1=2e/(1-e)$ and $\lambda_2^2=16x_2^2e^2/(1-e^2)^2$. With $E_1=\mu_c^2/(\nu_c+1)^2$ for the warm mode's own term, the inequality $T^2\FI_{\mathrm{Bell}}>T^2\FI_{\mathrm{local}}^{\max}$ reads $(1+\mu_2/\mu_c)^2>[1+(\nu_2-1)/(\nu_c+1)]^2(1+\lambda_2^2/2E_1)$, and since $\sqrt{1+u}\le1+u/2$ it follows from
\begin{equation}
\frac{4x_2}{\mu_c}>\frac{2}{(1-e)(\nu_c+1)}+\frac{4x_2^2e\,[1+(\nu_2-1)/(\nu_c+1)]}{(1-e^2)^2E_1},
\label{S:eq:tailpositive}
\end{equation}
whose left side exceeds $6.57$ and whose right side stays below $0.71$ for every $x_2\ge3.5$. Holding the warm mode on the crossover therefore gives a positive gain at every frequency ratio, and by continuity an open temperature window around that temperature.

That argument covers two of the three sectors, and in the third the gain is negative. With $x_2>x_1$ the pairs split three ways: $x_1<x_c<x_2$, then $x_c<x_1$, and then both modes below the crossover. Equations~\eqref{S:eq:branchslopes} and \eqref{S:eq:capstat} between them cover the first two and leave the third. There the benchmark reads both modes by heterodyne, so it equals $\sum_i\mu_i^2/(\nu_i+1)^2$. Write the response sum as $\mu_1+\mu_2=\sum_i[\mu_i/(\nu_i+1)](\nu_i+1)$. The Cauchy--Schwarz inequality bounds its square by the benchmark times $(\nu_1+1)^2+(\nu_2+1)^2$. The Bell value is that square divided by $(\nu_1+\nu_2)^2$, so Bell stays below the benchmark whenever $(\nu_1+1)^2+(\nu_2+1)^2\le(\nu_1+\nu_2)^2$. After expanding and cancelling, the condition reads $(\nu_1-1)(\nu_2-1)\ge2$. Both modes lie on the warm side of the crossover, $x<x_c$, and $\nu=\coth x$ falls as $x$ rises, so both widths exceed $\nu_c$. Since $\nu_c-1=\sqrt2$, the product $(\nu_1-1)(\nu_2-1)$ equals $2$ only at $\nu_1=\nu_2=\nu_c$ and exceeds it everywhere else in the sector, so the gain there is negative.

The same inequality favours separate detectors again at high temperature, where both widths run far above $\nu_c$ and the benchmark reads each mode by heterodyne. That is the high-temperature limit of the Letter's mechanism. A matched pair is the case $\lambda_1=\lambda_2$ of Eq.~\eqref{S:eq:twoaverages}, where the two entries agree exactly above the crossover, and the same bound settles it below, so identical modes never gain. The cap therefore sits at the one temperature where the two single-mode readings are worth the same. Together with the proof above this settles the thermal maximum: $12.699322\%$ is the largest gain of Bell homodyne over the separate-readout optimum at any pair of frequencies and any temperature, and it is reached only at $x_1=x_c$ and $x_2=1.46219$. Far outside the box both responses are exponentially small and the gain falls toward zero, as the cold-mode limit of Sec.~\ref{S:sec:universality} shows.

\section{Universal bound}
\label{S:sec:ceiling}

The Letter's cap of $12.699\%$ is a thermal number, and we separate what depends on the thermal law from what does not. End Matter C of the Letter sketches the answer, and we give the full proof here. Take two uncorrelated modes with covariances $\nu_1I_2$ and $\nu_2I_2$, each width at least the vacuum value $1$, and one parameter $\Theta$ that enters the state only through those widths. Write $\mu_i=\Theta\,\partial_\Theta\nu_i$ for the response of each width, so that $\Theta=T$ with Eq.~\eqref{S:eq:thermal} returns the thermal pair of the Letter. The same two numbers per mode describe an unknown noise power, the occupation of a thermal-like source, and thermal noise that has been amplified or attenuated. The derivation of Eq.~\eqref{S:eq:local} in Sec.~\ref{S:sec:localbound} and the Bell value at $t=1/2$ in Eq.~\eqref{S:eq:passive} used the widths and responses as free symbols and no thermal law. Both therefore hold verbatim,
\begin{equation}
 \Theta^2\FI_i^{\mathrm{loc}}=\frac{\mu_i^2}{\Gamma(\nu_i)},\quad
 \Gamma(\nu)=\min\{2\nu^2,(\nu+1)^2\},\qquad
 \Theta^2\FI_{\mathrm{Bell}}=\left(\frac{\mu_1+\mu_2}{\nu_1+\nu_2}\right)^{\!2}.
 \label{S:eq:widthpair}
\end{equation}
Here $\Gamma(\nu)$ is the reciprocal of the bracket in Eq.~\eqref{S:eq:local}, its homodyne entry $2\nu^2$ below the crossover $\nu_c=1+\sqrt2$ and its heterodyne entry $(\nu+1)^2$ above it. A common factor in the responses cancels from every ratio below, and we prove that the Bell-to-local ratio has a closed-form supremum.

\textit{Proposition~1 (universal bound).} For any widths $\nu_1,\nu_2\ge1$ and any responses $\mu_1,\mu_2$,
\begin{equation}
 \frac{\FI_{\mathrm{Bell}}}{\FI_{\mathrm{local}}^{\max}}
 \;\le\;\Xi(\nu_1,\nu_2)\equiv\frac{\Gamma(\nu_1)+\Gamma(\nu_2)}{(\nu_1+\nu_2)^2}
 \;\le\;4-2\sqrt2=1.171573 .
 \label{S:eq:ceiling}
\end{equation}
The first inequality is an equality exactly when $\mu_2/\mu_1=\Gamma(\nu_2)/\Gamma(\nu_1)$, and the second is reached only at $(\nu_1,\nu_2)=(1,\nu_c)$, up to mode exchange. We call $\Xi$ the width bound: it is the largest Bell-to-local ratio that any family of responses can reach at the given widths.

\begin{figure*}[!tp]
 \includegraphics{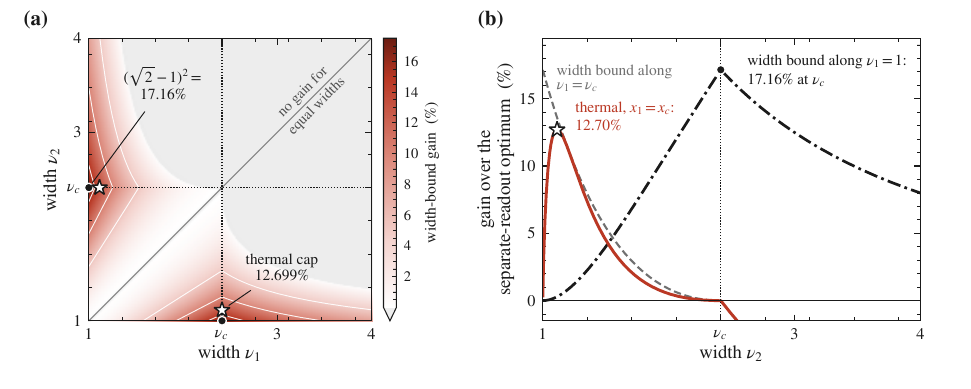}
 \caption{\label{S:fig:ceiling}
 The universal bound and its thermal instance. \textbf{(a)} The width bound as a gain, $\Xi(\nu_1,\nu_2)-1$
 of Eq.~\eqref{S:eq:ceiling}, over widths $\nu_1,\nu_2\in[1,4]$. Grey marks where it is not positive,
 and dotted lines mark the crossover $\nu=\nu_c$ on both axes. The supremum $(\sqrt2-1)^2=17.157\%$
 sits at $(1,\nu_c)$, one mode at vacuum width and the other on the crossover. The stars mark the
 thermal maximal-gain pair $(\nu_c,1.113)$. Both points recur as mirror images under mode exchange.
 \textbf{(b)} Cuts against $\nu_2$. The width bound along $\nu_1=1$ rises to $17.16\%$ at
 $\nu_2=\nu_c$ and falls beyond it, with a corner on the crossover. The thermal gain along
 $x_1=x_c$, drawn against $\nu_2=\coth x_2$, peaks at $12.699\%$ at $\nu_2=1.113$ with mode~1 held
 on the crossover. It has its own corner at $\nu_2=\nu_c$, where the two modes coincide and the gain
 vanishes. The width bound along $\nu_1=\nu_c$, dashed, bounds that thermal cut from above and meets
 zero at the same corner.}
\end{figure*}

\textit{Proof.} At fixed widths, Eq.~\eqref{S:eq:widthpair} writes the ratio on the left as $(\mu_1+\mu_2)^2$ over $(\nu_1+\nu_2)^2[\mu_1^2/\Gamma(\nu_1)+\mu_2^2/\Gamma(\nu_2)]$. Its derivative in $\mu_2$ vanishes when $\mu_1^2/\Gamma(\nu_1)+\mu_2^2/\Gamma(\nu_2)=\mu_2(\mu_1+\mu_2)/\Gamma(\nu_2)$, that is when $\mu_2/\mu_1=\Gamma(\nu_2)/\Gamma(\nu_1)$. The Cauchy--Schwarz inequality $(\mu_1+\mu_2)^2\le[\Gamma(\nu_1)+\Gamma(\nu_2)][\mu_1^2/\Gamma(\nu_1)+\mu_2^2/\Gamma(\nu_2)]$ makes that stationary point the maximum, with equality exactly there, at the value $\Xi$ in the middle of Eq.~\eqref{S:eq:ceiling}.

We then maximize $\Xi$ over the widths, ordered $\nu_1\le\nu_2$ by its exchange symmetry, in three cases. With both widths at most $\nu_c$, $\Xi=2(\nu_1^2+\nu_2^2)/(\nu_1+\nu_2)^2$, and $\partial\Xi/\partial\nu_1$ has the sign of $4\nu_2(\nu_1-\nu_2)\le0$, so $\nu_1$ falls to $1$. There $\Xi=2(1+\nu_2^2)/(1+\nu_2)^2$ has a derivative of the sign of $4(\nu_2-1)\ge0$, so $\nu_2$ rises to $\nu_c$, where $\Xi=4-2\sqrt2$. With $\nu_1\le\nu_c\le\nu_2$, $\Xi=[2\nu_1^2+(\nu_2+1)^2]/(\nu_1+\nu_2)^2$, and $\partial\Xi/\partial\nu_1$ has the sign of $4\nu_1\nu_2-2(\nu_2+1)^2\le4\nu_c\nu_2-2(\nu_2+1)^2=-2[(\nu_2-\sqrt2)^2-1]\le0$ for $\nu_2\ge\nu_c$, so $\nu_1$ falls to $1$ again. The remaining factor, $\Xi=1+2/(1+\nu_2)^2$, decreases in $\nu_2$, so $\nu_2$ falls to $\nu_c$ and to the same value. With both widths at least $\nu_c$, $\Xi=[(\nu_1+1)^2+(\nu_2+1)^2]/(\nu_1+\nu_2)^2$, and $\partial\Xi/\partial\nu_1$ has the sign of $2[\nu_2(\nu_1-\nu_2)-\nu_1-\nu_2-2]<0$, so $\nu_1$ falls to $\nu_c$, the previous case. Equal widths give $\Xi\le1$ on both branches, so identical modes never gain, whatever their responses. The supremum is therefore $4-2\sqrt2$, at $\nu_1=1$, $\nu_2=\nu_c$ and $\mu_2/\mu_1=\Gamma(\nu_c)/\Gamma(1)=\nu_c^2=3+2\sqrt2=5.828$. $\square$

No smooth family reaches that supremum. A width never falls below $1$, so a width that touches $1$ has a minimum there and $\mu_1=0$, exactly where equality asks for $\mu_2/\mu_1=\nu_c^2$. With $\mu_1=0$ the ratio is $\Gamma(\nu_c)/(1+\nu_c)^2=1$. Families therefore approach the bound $(\sqrt2-1)^2=17.157\%$ only as $\nu_1\to1$, and none attains it. Thermal equilibrium stops earlier, because the thermal law fixes the response ratio and the first inequality of Eq.~\eqref{S:eq:ceiling} is then strict. Section~\ref{S:sec:cap} finds the thermal maximum $12.699322\%$ at $\nu_1=\nu_c$ exactly and $\nu_2=\coth(1.4621888)=1.113$, with mode~1 the warm one in the Letter's labelling. The width bound $\Xi-1$ at that thermal pair is $13.60\%$. Both maxima place one mode on the kink $\nu=\nu_c$ of the separate-readout envelope and the other near vacuum, and Fig.~\ref{S:fig:ceiling} draws both.

The figure also shows the two gains vanishing together where the modes become alike, and near equal modes the comparison has a universal shape as well. Write $\lambda_i=\mu_i/\nu_i$ as in Eq.~\eqref{S:eq:lambda}, and define the two half-differences
\begin{equation}
 \psi_\nu=\tfrac12\ln\frac{\nu_1}{\nu_2},\qquad
 \psi_\lambda=\tfrac12\ln\frac{\lambda_1}{\lambda_2},
 \label{S:eq:halfdiff}
\end{equation}
of the log width and of the log response ratio. On the homodyne branch Eq.~\eqref{S:eq:twoaverages} holds in any family, with the benchmark the mean of $\lambda_1^2$ and $\lambda_2^2$ and the Bell value the square of the width-weighted mean of $\lambda_1$ and $\lambda_2$. Their ratio is unchanged by a common factor in the widths and by another in the $\lambda_i$, so with $\nu_1/\nu_2=e^{2\psi_\nu}$ and $\lambda_1/\lambda_2=e^{2\psi_\lambda}$ it is exactly $\cosh^2(\psi_\nu+\psi_\lambda)/[\cosh^2\psi_\nu\,\cosh2\psi_\lambda]$. Its logarithm, which equals the gain to second order, expands as
\begin{align}
 \ln\frac{\FI_{\mathrm{Bell}}}{\FI_{\mathrm{local}}^{\max}}
 &=2\ln\cosh(\psi_\nu+\psi_\lambda)-2\ln\cosh\psi_\nu-\ln\cosh2\psi_\lambda\nonumber\\
 &=(\psi_\nu+\psi_\lambda)^2-\psi_\nu^2-2\psi_\lambda^2+\dots
 =\psi_\lambda(2\psi_\nu-\psi_\lambda)+\dots,
 \label{S:eq:universalonset}
\end{align}
so the gain is $\psi_\lambda(2\psi_\nu-\psi_\lambda)$ to leading order, in every width family. It is positive exactly when $\psi_\lambda$ and $2\psi_\nu-\psi_\lambda$ share a sign. The two mismatches must point the same way, and the width mismatch must be at least half the mismatch of the response ratio. Above the crossover the comparison closes instead. For any responses, the third case of the proof bounds the Bell value against two heterodynes by $[(\nu_1+1)^2+(\nu_2+1)^2]/(\nu_1+\nu_2)^2$, which lies below $1$ exactly when $(\nu_1-1)(\nu_2-1)>2$. That region contains every pair near equal widths $\nu>\nu_c$, so for near-equal modes with $\nu>\nu_c$, on the warm side of the crossover, Bell homodyne never beats two heterodynes, in any family. For the thermal pair $x_{1,2}=xe^{\mp s/2}$ of Sec.~\ref{S:sec:universality}, the derivatives $d\ln\nu/d\ln x=-\lambda$ and $d\ln\lambda/d\ln x=1-c$ give $\psi_\nu=\lambda s/2$ and $\psi_\lambda=(c-1)s/2$ to leading order, with $c=2x\coth2x$. Their product in Eq.~\eqref{S:eq:universalonset} is $\tfrac14(c-1)[2\lambda-(c-1)]s^2$, the onset coefficient of the Letter. The sign condition $2\psi_\nu>\psi_\lambda$ becomes $2\lambda>c-1$, the window of Sec.~\ref{S:sec:universality}.

Every comparison so far pits one receiver, Bell homodyne, against separate detectors. The Bell receiver is one member of a class, every passive two-mode network followed by two homodynes, and we reduce that class to the family of Eq.~\eqref{S:eq:passive}. Let the network act on the two mode amplitudes by a unitary matrix $\mathcal U$. The local-oscillator phases are absorbed in $\mathcal U$ as output phase shifters, so each homodyne reads the $q$ quadrature of its output. The two records $Y_i=\sum_j[\operatorname{Re}\mathcal U_{ij}\,q_j-\operatorname{Im}\mathcal U_{ij}\,p_j]$ sit on different outputs, so one shot fixes both. Each mode enters with $\langle q_j^2\rangle=\langle p_j^2\rangle=\nu_j$ and no correlation between $q_j$ and $p_j$. The records therefore have covariance $\nu_1\Pi_1+\nu_2\Pi_2$ and response $\mu_1\Pi_1+\mu_2\Pi_2$, where $(\Pi_j)_{ik}=\operatorname{Re}(\mathcal U_{ij}\mathcal U_{kj}^*)$ is the real part of the projector onto the $j$th column of $\mathcal U$. Orthonormal rows give $\Pi_1+\Pi_2=I_2$, so both matrices are functions of $\Pi_1$ alone and commute. $\Pi_1$ is real symmetric with trace $1$ and determinant $[\operatorname{Im}(\mathcal U_{11}\mathcal U_{21}^*)]^2\ge0$, so its eigenvalues are $t$ and $1-t$ with $0\le t\le1$. The Fisher information of a Gaussian record with fixed mean is the trace rule of Eq.~\eqref{S:eq:FI}, and in the eigenbasis of $\Pi_1$ that trace is Eq.~\eqref{S:eq:passive} with $t$ in place of the transmissivity. The whole class therefore collapses to one parameter, with $t=1/2$ the Bell receiver and $t=0$ or $t=1$ two separate homodynes.

\textit{Proposition~2 (passive networks without ancillas).} For any widths and any responses, Eq.~\eqref{S:eq:passive} attains its maximum over $0\le t\le1$ at $t=1/2$ or at an end.

\textit{Proof.} In the variable $z=(2t-1)^2$ of Sec.~\ref{S:sec:passivenet}, the derivative displayed there is a positive factor, since $\Sigma_\nu^2-\Delta_\nu^2z\ge4\nu_1\nu_2$, times $\Delta_\mu\Sigma_\nu-\Delta_\nu\Sigma_\mu$ times a linear function of $z$, in the sums $\Sigma$ and differences $\Delta$ defined there. Label the modes so that $\Delta_\nu\ge0$, and reverse the sign of both responses if needed so that $\Sigma_\mu\ge0$, which leaves Eq.~\eqref{S:eq:passive} unchanged. When $\Delta_\mu\Sigma_\nu=\Delta_\nu\Sigma_\mu$ the derivative vanishes identically, the family is constant in $t$, and the maximum is attained everywhere. Otherwise the linear factor changes sign at most once, so an interior maximum needs the derivative positive at $z=0$ and negative at $z=1$. At $z=0$ it is $\Sigma_\nu^2(\Delta_\mu\Sigma_\nu-\Delta_\nu\Sigma_\mu)(\Delta_\mu\Sigma_\nu-3\Delta_\nu\Sigma_\mu)$, nonpositive whenever $\Delta_\mu\Sigma_\nu$ lies between $\Delta_\nu\Sigma_\mu$ and $3\Delta_\nu\Sigma_\mu$. At $z=1$ it is $(\Delta_\mu\Sigma_\nu-\Delta_\nu\Sigma_\mu)[\Delta_\nu^2(3\Delta_\mu\Sigma_\nu-\Delta_\nu\Sigma_\mu)+\Sigma_\nu^2(\Delta_\mu\Sigma_\nu-3\Delta_\nu\Sigma_\mu)]$. For $\Delta_\mu\Sigma_\nu>3\Delta_\nu\Sigma_\mu$ every factor is positive. For $\Delta_\mu\Sigma_\nu<\Delta_\nu\Sigma_\mu$ the term carrying $\Sigma_\nu^2>\Delta_\nu^2$ is negative and pulls the bracket below $4\Delta_\nu^2(\Delta_\mu\Sigma_\nu-\Delta_\nu\Sigma_\mu)\le0$, so the product is again positive. No case allows both signs. $\square$ Proposition~3 below settles the class with one vacuum ancilla, which admits heterodyne detection. Squeezed ancillas widen the class to every pure two-mode seed, which the following search covers.

\textit{Numerical check.} Let the seed run over the pure two-mode family of Sec.~\ref{S:sec:pureseed}, two squeezers of strength at most $e^{5}$, a tilt and a beam splitter. The search covered four hundred random width pairs, with widths between $1$ and about $6$ and responses spanning two decades. The best seed never exceeded the larger of the Bell value and the separate-readout optimum by more than one part in a million.

Mixed seeds add nothing, by the argument of Sec.~\ref{S:sec:mixed}. In interval arithmetic, optimality over all Gaussian measurements is therefore a theorem on the two rectangles of Sec.~\ref{S:sec:region} that the Letter quotes.

\textit{$N$ modes.---} The same structure governs any number of uncorrelated modes. Take $N$ modes with widths $\nu_i$ and responses $\mu_i$, a passive $N$-port network $U$ followed by one homodyne per output, and record the real quadrature of each output. The records have covariance $C_{kl}=\sum_i\nu_i\,\mathrm{Re}(U_{ki}U_{li}^{*})$, their response is the same sum with $\mu_i$ in place of $\nu_i$, and the Fisher information is $\tfrac12\Tr[(C^{-1}\partial C)^2]$; the balanced two-port with one output read in phase and the other in quadrature is Bell homodyne. Three statements are exact. A disjoint pairing of the modes, Bell homodyne on the pairs that gain and separate detectors elsewhere, belongs to the class, so the $N$-mode optimum is at least the best pairing and exceeds the separate-readout optimum whenever one pair does. The gain of a pairing is a weighted mean of the gains of its blocks, so it stays below $(\sqrt2-1)^2$ by Proposition~1. For identical modes the product state is invariant under every passive network, the $N$ records are independent with width $\nu$, and the class returns $N$ times the homodyne value, so identical modes gain nothing at any $N$. For three modes the pairing is the optimum for every width and response, by Proposition~3 below. Beyond three modes it is the optimum in every configuration we tested. We maximized the Fisher information over $U(N)$ from twelve random starting points for $320$ configurations at $N=3$ and $180$ at $N=4$, half thermal and half with independent widths between $1$ and $8$ and response ratios spanning two decades, for $60$ at $N=5$, for $750$ three-mode configurations built around the two-mode supremum, with one mode at $\nu_c$ and near-vacuum partners whose response ratios lie near $1/\nu_c^2$, and for $27$ four-mode ones with two modes at $\nu_c$. In all $1337$ cases the best network agrees with the best pairing to within $10^{-15}$ in relative terms, the largest gain found, $16.99\%$, approaches the two-mode supremum from below, and in the $187$ configurations with no gaining pair the network optimum is the separate-readout optimum. Squeezed ancillas did not change this in the eight three-mode configurations where we searched every pure seed with generator norm at most $6$: the best finite seed stayed below the network optimum each time.

\textit{Three modes.---} For three modes the pairing is a theorem.

\textit{Proposition~3 (three modes, passive networks without ancillas).} Let $\nu_i>0$ and $\mu_i$ be arbitrary, with $\lambda_i=\mu_i/\nu_i$, and let a network $\mathcal U\in U(3)$ act on the three mode amplitudes before one homodyne per output. Then
\begin{equation}
 \sup_{\mathcal U}F=\max\Big\{\tfrac12\sum_{i}\lambda_i^2,\;\tfrac12\lambda_k^2+\Big(\frac{\mu_i+\mu_j}{\nu_i+\nu_j}\Big)^{\!2}\Big\},
 \label{S:eq:prop3}
\end{equation}
the second entry taken over the three choices of the singleton $k$, $\{i,j,k\}=\{1,2,3\}$. The optimum is three separate homodynes, or one homodyne and Bell homodyne on the other two modes.

\textit{Proof.} The records are the real parts of the outputs, with covariance $C=\operatorname{Re}(\mathcal U\mathcal N\mathcal U^\dagger)$ and response $D=\operatorname{Re}(\mathcal U\mathcal M\mathcal U^\dagger)$ for $\mathcal N=\operatorname{diag}(\nu_i)$ and $\mathcal M=\operatorname{diag}(\mu_i)$, and $F=\tfrac12\Tr[(C^{-1}D)^2]$. Two gauges cost nothing: a real orthogonal matrix on the left recombines the records, and diagonal phases on the right leave $C$ and $D$ unchanged. The phase of the first column makes its real and imaginary parts orthogonal, an orthogonal matrix aligns them with two axes, and the remaining two columns span the complement of the first; after their phases are removed, one mixing angle and one relative phase $\delta$ are left. Every network is therefore equivalent to
\begin{equation}
 \mathcal U=\begin{pmatrix} c_1 & ic_2s_1 & is_2s_1\\ is_1 & c_2c_1 & s_2c_1\\ 0 & -s_2e^{i\delta} & c_2e^{i\delta}\end{pmatrix},\qquad c_1^2+s_1^2=c_2^2+s_2^2=1,
 \label{S:eq:canon3}
\end{equation}
which gives
\begin{equation}
 C=\begin{pmatrix}A&0&x\\0&B&y\\x&y&Z\end{pmatrix},\qquad
 D=\begin{pmatrix}A_m&0&rx\\0&B_m&ry\\rx&ry&Z_m\end{pmatrix},
 \label{S:eq:CD3}
\end{equation}
with $A=c_1^2\nu_1+s_1^2(c_2^2\nu_2+s_2^2\nu_3)$, $B=s_1^2\nu_1+c_1^2(c_2^2\nu_2+s_2^2\nu_3)$, $Z=s_2^2\nu_2+c_2^2\nu_3$, $x=c_2s_2s_1(\nu_3-\nu_2)\sin\delta$, $y=c_2s_2c_1(\nu_3-\nu_2)\cos\delta$, the same expressions with $\mu$ in place of $\nu$ for $A_m$, $B_m$ and $Z_m$, and $r=(\mu_3-\mu_2)/(\nu_3-\nu_2)$. Write $\tau=\sin^2\delta$, $X=c_2^2s_2^2s_1^2(\nu_3-\nu_2)^2$ and $Y=c_2^2s_2^2c_1^2(\nu_3-\nu_2)^2$, so that $x^2=X\tau$ and $y^2=Y(1-\tau)$. If $\nu_2=\nu_3$, then $x=y=0$, $C$ is diagonal, $F$ is affine in $\tau$, and its maximum over the phase is at $\tau=0$ or $\tau=1$.

The phase sits at an end in general. The Fisher information is even in $x$ and in $y$, so with $\det C=d_0+\kappa\tau$, $d_0=A(BZ-Y)$ and $\kappa=AY-BX$, the numerator $N(\tau)=\Tr[(\operatorname{adj}(C)D)^2]$ is a polynomial in $\tau$. Its cubic terms cancel, and its quadratic coefficient is $n_2=2r^2(AY-BX)^2+(A_mY-B_mX)^2\ge0$. For $\kappa\ne0$, division by $d^2$ gives $2F=n_2/\kappa^2+(n_1/\kappa-2n_2d_0/\kappa^2)/d+\gamma/d^2$ with $\gamma=n_0-n_1d_0/\kappa+n_2d_0^2/\kappa^2$, and $\kappa^2\gamma$ is the square of
\begin{equation}
 2A^2BYZr-A^2BYZ_m-A^2B_mYZ-2AB^2XZr+AB^2XZ_m+AB_mXY+A_mB^2XZ-A_mBXY,
 \label{S:eq:gammasq}
\end{equation}
so $\gamma\ge0$; we checked the three polynomial identities of this paragraph by computer algebra. The Fisher information is therefore a convex quadratic in $1/d$, and $1/d$ is monotone in $\tau$ because $d$ is affine and positive on $[0,1]$, so the maximum over $\delta$ is at $\tau=0$ or $\tau=1$. For $\kappa=0$ the denominator is constant and $N$ itself is convex in $\tau$, with the same conclusion.

At either end, column phases bring the network to a real frame with one imaginary row, $\widetilde{\mathcal U}=(in^{T};V)$, where $n\in\mathbb R^3$ is a unit vector and $V$ is real with orthonormal rows and $Vn=0$: one record is the momentum combination $n\cdot p$ and the other two are position combinations. Positions and momenta of a thermal product are uncorrelated, so $F=\tfrac12(n^{T}\mathcal Mn/n^{T}\mathcal Nn)^2+\tfrac12\Tr[((V\mathcal NV^{T})^{-1}V\mathcal MV^{T})^2]$. Put $g_i=n_i^2$ and $p_i=(g_i/\nu_i)/\sum_j(g_j/\nu_j)$, which runs over the whole simplex as $n$ runs over the sphere. The identity $\det(VRV^{T})=\det R\;n^{T}R^{-1}n$ for positive diagonal $R$, applied to $R=\mathcal N+z\mathcal M$ and expanded in $z$, together with $\Tr X^2=(\Tr X)^2-2\det X$ for the $2\times2$ matrix $X=(V\mathcal NV^{T})^{-1}V\mathcal MV^{T}$, gives
\begin{equation}
 F(p)=\frac12\Big(\frac{a\cdot p}{b\cdot p}\Big)^{\!2}+\frac12\big(L-\lambda\cdot p\big)^2-\pi\cdot p,
 \label{S:eq:simplex3}
\end{equation}
with $b_i=\nu_i^2$, $a_i=\nu_i\mu_i$, $L=\sum_i\lambda_i$ and $\pi_i=\prod_{j\ne i}\lambda_j$. For an interior $p$ take $h\ne0$ with $\mathbf 1\cdot h=0$ and $b\cdot h=0$. Along the chord $p+uh$ the denominator $b\cdot p$ is fixed, the two squared terms are squares of affine functions of $u$ and the last term is affine, so $F$ is convex along the chord and its maximum lies on the boundary of the simplex, where some $p_k=0$, that is $n_k=0$. There the two real rows can be taken as $e_k$ and the unit vector orthogonal to $n$ in the plane of modes $i$ and $j$: mode $k$ is read by its own homodyne, and modes $i$ and $j$ by the beam-splitter family of Eq.~\eqref{S:eq:passive} with $t=n_i^2$. Proposition~2 puts the maximum of that family at $t=1/2$ or at an end, so the face value is the larger of $\tfrac12\lambda_k^2+[(\mu_i+\mu_j)/(\nu_i+\nu_j)]^2$ and $\tfrac12\sum_i\lambda_i^2$. Maximizing over the three faces gives Eq.~\eqref{S:eq:prop3}; coincident widths and vanishing responses follow by continuity. $\square$

\textit{Corollary (two modes and one vacuum ancilla).} Take mode 3 to be a vacuum ancilla, $\nu_3=1$ and $\mu_3=0$. A Bell pair of mode $i$ with the ancilla is the heterodyne detection of mode $i$, so Eq.~\eqref{S:eq:prop3} says that over every passive network of the two modes and one vacuum ancilla followed by homodynes, the supremum is the largest of $\FI_{\mathrm{Bell}}$, two homodynes, and a homodyne on one mode with heterodyne on the other. None of these exceeds $\max\{\FI_{\mathrm{Bell}},\FI_{\mathrm{local}}^{\max}\}$. Bell homodyne is therefore the optimum of that class wherever it beats the separate-readout optimum, for every width and response, and by Proposition~1 the class never exceeds $(4-2\sqrt2)\FI_{\mathrm{local}}^{\max}$. For three modes with ancillas among them the same reading gives the bound of Proposition~1 at $N'=3$: the optimum is a pairing, each Bell block of two real modes obeys Proposition~1, and each block that contains the ancilla is a single-mode reading.

\textit{Framework for the $N$-mode bound.---} The class has a compact geometric form. Let $N'$ count the modes together with any vacuum ancillas, let $L\subset\mathbb R^{2N'}$ be the Lagrangian subspace of the $N'$ commuting quadratures that the homodynes record, and let $A$ hold an orthonormal basis of $L$ in its rows, so that the records are $y=Ax$ with covariance $C=AVA^{T}$ and response $\partial C=ADA^{T}$, $D=\bigoplus_i\mu_iI_2$. With $B=C^{-1/2}AV^{1/2}$, which obeys $BB^{T}=I$, the Fisher information $\tfrac12\Tr[(C^{-1}\partial C)^2]$ becomes $\tfrac12\Tr[(B\Lambda B^{T})^2]$ with $\Lambda=V^{-1/2}DV^{-1/2}=\bigoplus_i\lambda_iI_2$ and $\lambda_i=\mu_i/\nu_i$. Since $\Pi=B^{T}B$ is the orthogonal projector onto $\tilde L=V^{1/2}L$,
\begin{equation}
 F=\tfrac12\Tr\!\left[(\Pi\Lambda\Pi)^2\right],
 \qquad \Pi\tilde\Omega\Pi=0,\qquad \tilde\Omega=\bigoplus_i\nu_i^{-1}J,
 \label{S:eq:compression}
\end{equation}
where the second relation is the Lagrangian condition $u^{T}\Omega v=0$ on $L$ rewritten for $\tilde L$, because $V^{-1/2}\Omega V^{-1/2}=\tilde\Omega$. The Fisher information is the energy of the compression of the block-scalar operator $\Lambda$ to an $N'$-dimensional subspace that is isotropic for a symplectic form weighted by the inverse widths, and every passive network with homodynes is one such subspace.

Write $Q_i=\Pi E_i\Pi$ for the compression to $\tilde L$ of the projector $E_i$ on the plane of mode $i$. Each $Q_i$ is positive with rank at most two, its eigenvalues $\alpha_i\ge\beta_i$ lie in $[0,1]$, $w_i=\alpha_i+\beta_i$ is the occupation of the plane, and $\sum_iQ_i=I$ on $\tilde L$. With the Gram matrix $Q_{ij}=\Tr(Q_iQ_j)\ge0$,
\begin{equation}
 F=\tfrac12\sum_{ij}\lambda_i\lambda_jQ_{ij}
 =\tfrac12\sum_i w_i\lambda_i^2-\tfrac12\sum_{i<j}Q_{ij}(\lambda_i-\lambda_j)^2 ,
 \label{S:eq:gramF}
\end{equation}
the second form following from $\sum_jQ_{ij}=w_i$. Separate homodynes are $Q_i=e_ie_i^{T}$; a Bell pair is $Q_i=\nu_i I_2/(\nu_i+\nu_j)$ and $Q_j=\nu_jI_2/(\nu_i+\nu_j)$ on one shared two-plane, so the warm mode occupies more than one unit and the cold one less. Equation~\eqref{S:eq:gramF} says where a gain can come from: occupation moves from narrow modes to wide ones, and every overlap between modes of different response ratio is paid for by $(\lambda_i-\lambda_j)^2$.

Four exact consequences of $\sum_iQ_i=I$ and of isotropy constrain the occupations and overlaps. Tracing $\sum_iQ_i=I$ gives
\begin{equation}
 \sum_i(\alpha_i+\beta_i)=N',\qquad
 \sum_{j\ne i}Q_{ij}=\alpha_i(1-\alpha_i)+\beta_i(1-\beta_i),
 \label{S:eq:E01}
\end{equation}
the second after multiplying by $Q_i$: a plane occupied by less than two units overlaps other planes by exactly the deficit. For the isotropy, let $\hat u_i,\hat v_i$ be the eigenvectors of $Q_i$ and $\Pi_i$ their span. The compression $\Pi J_i\Pi$ of the symplectic generator of mode $i$ is $\sqrt{\alpha_i\beta_i}\,(\hat u_i\hat v_i^{T}-\hat v_i\hat u_i^{T})$, an oriented area, and projecting $\Pi\tilde\Omega\Pi=0$ onto $\Pi_i$ gives the area balance $\sqrt{\alpha_i\beta_i}/\nu_i=-\sum_{j\ne i}\sqrt{\alpha_j\beta_j}\det(G_{ij})/\nu_j$, with $G_{ij}$ the $2\times2$ matrix of inner products between the eigenvectors of $Q_i$ and $Q_j$. The compression $R_j=G_{ij}\operatorname{diag}(\alpha_j,\beta_j)G_{ij}^{T}$ of $Q_j$ to $\Pi_i$ has $\det R_j=\alpha_j\beta_j\det(G_{ij})^2$ and $\sum_{j\ne i}R_j=\operatorname{diag}(1-\alpha_i,1-\beta_i)$, so the balance reads $\sqrt{\alpha_i\beta_i}/\nu_i\le\sum_{j\ne i}\sqrt{\det R_j}/\nu_j$. Two inequalities follow. Since $|\det M|\le\tfrac12\|M\|_F^2$ for any $2\times2$ matrix and $\|\operatorname{diag}(\alpha_i,\beta_i)^{1/2}G_{ij}\operatorname{diag}(\alpha_j,\beta_j)^{1/2}\|_F^2=Q_{ij}$,
\begin{equation}
 \frac{2\alpha_i\beta_i}{\nu_i}\le\sum_{j\ne i}\frac{Q_{ij}}{\nu_j};
 \label{S:eq:E2}
\end{equation}
and since $\sqrt{\det}$ is concave on positive $2\times2$ matrices,
\begin{equation}
 \nu_{i}'\sqrt{\alpha_i\beta_i}\le\nu_i\sqrt{(1-\alpha_i)(1-\beta_i)},\qquad \nu_i'=\min_{j\ne i}\nu_j .
 \label{S:eq:E3}
\end{equation}
Equation~\eqref{S:eq:E3} alone bounds the occupation by $w_i\le2\nu_i/(\nu_i+\nu_i')$, the heterodyne value, and both inequalities are equalities for a Bell pair.

The universal bound for $N'$ modes is a statement about $Q$. With $G=\operatorname{diag}(g_i)$, $g_i=\nu_i^2/\Gamma(\nu_i)$, the inequality $F\le(4-2\sqrt2)\,\FI_{\mathrm{local}}^{\max}$ for every response vector $\lambda$ is
\begin{equation}
 Q\preceq 2(4-2\sqrt2)\,G .
 \label{S:eq:matrixbound}
\end{equation}
We maximized $F/\FI_{\mathrm{local}}^{\max}$ over all $(\alpha_i,\beta_i,\nu_i,\lambda_i,Q_{ij})$ that satisfy Eqs.~\eqref{S:eq:E01}--\eqref{S:eq:E3}, with $\nu_i\ge1$ and no further geometry, by sequential quadratic programming from $40$ to $80$ random starts. At $N'=3$ and $N'=4$ the supremum is $4-2\sqrt2$ to six digits, reached only at copies of the two-mode supremum, one mode at $\nu_c$ paired with a vacuum mode; without Eq.~\eqref{S:eq:E3} the relaxed supremum at $N'=4$ is $1.1722$, so the determinant inequality is the one that closes it there. At $N'=5$ the four scalar conditions leave a margin of $4\times10^{-5}$: the supremum $1.17161$ is reached by two Bell-like pairs in which one mode of a pair has unequal occupations $(0.353,0.241)$ while its partner has $(0.707,0.707)$. Such a pair violates the operator condition $Q_i+Q_j\preceq I$, because the compression of the first mode to the partner's plane would have to stay below $0.293$, and the scalar conditions record only the traces of $\sum_iQ_i=I$. Keeping $\sum_iQ_i=I$ as an operator identity, with $Q_i=Y_i^{T}Y_i$ for an orthonormal $2N'\times N'$ matrix $Y$, and imposing only Eqs.~\eqref{S:eq:E2}--\eqref{S:eq:E3} in place of isotropy, the supremum returns to $4-2\sqrt2$ to six digits at $N'=3$, $4$ and $5$, from $24$ to $30$ random starts each. Proposition~3 settles $N'=3$, ancillas included. For $N'\ge4$, Proposition~1 rests on one algebraic lemma, Eq.~\eqref{S:eq:matrixbound} under $\sum_iQ_i=I$ and Eqs.~\eqref{S:eq:E2}--\eqref{S:eq:E3}. Because $2(4-2\sqrt2)G-Q$ has nonpositive off-diagonal entries, that lemma is equivalent to the existence of a positive vector $v$ with $[2(4-2\sqrt2)G-Q]\,v\ge0$; at the two-mode supremum $v\propto(\nu_c,1)$, the occupation vector of the Bell pair.

\section{Pure-seed reduction}

\begin{figure*}[!tp]
 \includegraphics{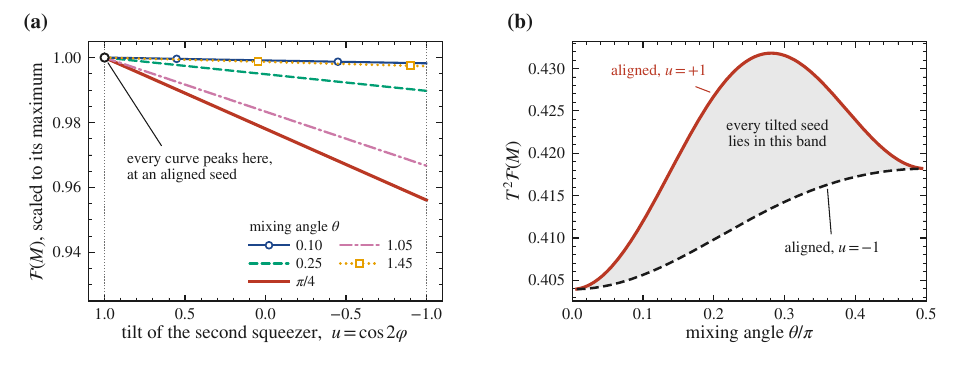}
 \caption{\label{S:fig:tilt}
 Tilting the second squeezer never helps, at the maximal-gain pair with $a=4$ and $b=0.5$.
 \textbf{(a)} Fisher information along the tilt $u=\cos2\phi$, each curve scaled to its own
 maximum, for five mixing angles. Every curve is monotonic and peaks at an aligned seed, which
 is what the sign of $L_1$ forces: the derivative can change sign only once, and only from
 negative to positive, so no interior tilt can be a maximum. \textbf{(b)} The two aligned seeds
 against the mixing angle. Every tilted seed lies in the band between them, so the optimum over
 the whole family is always aligned and the search reduces to $M=X\oplus X^{-1}$.}
\end{figure*}
\label{S:sec:pureseed}

End Matter D fixes every pure two-mode seed in Bloch--Messiah form, one beam splitter of mixing angle $\theta$ acting on two squeezed vacua of strengths $a$ and $b$. The axes of the second squeezer are tilted by an angle $\phi$ in its quadrature plane. End Matter D states that for $x_2>x_1$ the numerator of $d\FI/du$ is a straight line in $u=\cos2\phi$ with nonnegative slope, so that tilting one axis against the other only lowers the Fisher information. We prove that statement.

Two reductions End Matter D compresses are worth writing out. A common input phase commutes through the beam splitter, so only the relative orientation of the two squeezing axes survives, which is why one angle is enough. A fixed displacement of the seed translates the outcome distribution by a temperature-independent vector and cannot change a Fisher information. Only signs are used below, so we suppress the factor $T^2$ and write $\FI$ for $T^2\FI$ throughout this section.

The tilt enters through $u=\cos2\phi$, and the second squeezing and the mixing angle through
\begin{equation}
h=\frac{b+b^{-1}}2\ge1,\qquad \tau=\cos^2\theta.
\end{equation}
Inverting $V+M$ and taking the trace, we find
\begin{equation}
\FI(u)=\frac{\Psi(u)}{2\,\Phi(u)^2},\qquad
\partial_u\FI=\frac{L_0+L_1u}{2\,\Phi(u)^3},
\label{S:eq:orientation}
\end{equation}
with $\Psi$ a quadratic in $u$ and $\Phi$ linear in $u$.

Order the modes so that $x_2>x_1$ and write
\begin{equation}
\nu_2=n,\quad \nu_1=n+p,\qquad
\lambda_2=\ell,\quad\lambda_1=\ell+q.
\end{equation}
Both $\nu$ and $\lambda$ decrease in $x$, so $p\ge0$ and $q\ge0$, and $n\ge1$ because a thermal width is at least the vacuum width. Reducing the numerator, we find
\begin{equation}
L_1=-\frac{2\tau^2(1-\tau)^2(a^2-1)^2(h^2-1)p^2}{a^3}\,\mathcal Q,
\quad
\mathcal Q=(a-1)^2U+aW,
\end{equation}
using $a+a^{-1}-2=(a-1)^2/a$. The reduction splits $\mathcal Q$ into two polynomials in $n,p,q,\ell,\tau$ and $h$, one multiplying $(a-1)^2\ge0$ and one multiplying $a>0$. Each of the two is nonpositive, and that fixes the sign of $L_1$. The signs follow from
\begin{equation}
\partial_hU=-2nq(n+p)(2\ell np+\ell p^2+3n^2q+4npq+p^2q)\le0.
\end{equation}
That derivative carries neither $h$ nor $\tau$, so $U$ is affine in each of them separately. Since $h\ge1$, the polynomial $U$ therefore takes its largest value at $h=1$, and there it is a first-degree polynomial in $\tau$, so on $0\le\tau\le1$ it is largest at one of the two ends. Those two values are
\begin{align}
U_{\tau=0}&=-nK(\ell p+3n^2q+5npq+3nq+2p^2q+3pq),\\
U_{\tau=1}&=-(n+p)K(\ell p+3n^2q+npq+3nq+pq),
\end{align}
where $K=\ell p+n^2q+npq+nq+pq\ge0$, because $n\ge1$ and $\ell$, $p$ and $q$ are all nonnegative, $\ell=\lambda_2>0$ being a ratio of two positive thermal quantities. Both endpoint values are nonpositive, so $U\le0$ everywhere on the range.

Similarly, $\partial_hW=-2KK_2\le0$, with
\begin{equation}
K_2=\ell p[n+p(1-\tau)]+q[3n^3+4n^2p+3n^2+np^2+4np+p^2+\tau p(n-1)(n+p)].
\end{equation}
That derivative again carries no $h$, so $W$ is affine in $h$ and its $h=1$ value decides its sign. At $h=1$, $W=-(A_0\ell^2+B_0\ell q+C_0q^2)$, where
\begin{align}
A_0&=p^2(n^4+2n^3p+n^2p^2+4n+2p+3),\\
B_0&=-2p(n+1)^2(n+p)(n^2+np-2n-2p-3),\\
C_0&=(n+1)^3(n+p)^2(3n+2p+3).
\end{align}
The bracket is a quadratic form in $\ell$ and $q$. Such a form is nonnegative for all real $\ell$ and $q$ when its two square coefficients are nonnegative and four times their product is at least the square of the cross coefficient. Here $A_0$ and $C_0$ are sums of positive terms, so both are nonnegative, and
\begin{equation}
4A_0C_0-B_0^2=8n^2p^2(n+1)^3(n+p)^2(n+p+1)^3\ge0
\end{equation}
is the third condition. The bracket is therefore nonnegative and $W\le0$ at $h=1$, and $\partial_hW\le0$ carries that to every $h\ge1$. With $U\le0$ as well, $\mathcal Q\le0$ and $L_1\ge0$, which is the claim of End Matter D.

The function $\Phi(u)$ is a determinant of $V+M$, so it is positive for every admissible seed. Its cube keeps one sign on $-1\le u\le1$. The sign of $\partial_u\FI$ is therefore the sign of $L_0+L_1u$ alone, so the sign of $\partial_u\FI$ turns from negative to positive at most once and Eq.~\eqref{S:eq:orientation} has no interior maximum in $u$. The endpoints $u=\pm1$ exchange $b$ and $b^{-1}$ and reduce every pure seed to $M=X\oplus X^{-1}$ in $(q_1,q_2,p_1,p_2)$ ordering. We call such a seed aligned, because its two squeezing axes are parallel.

\section{Proof at the reference pair}
\label{S:sec:certpoint}

End Matter D reports $742$ strictly positive Bernstein coefficients and $8$ structural zeros at the reference pair. The polynomial behind that count is the numerator of the shortfall $\FI_{\mathrm{Bell}}-\FI_X$ of an aligned seed $X$, taken over a positive denominator, of degree four in each of the three compactified seed variables defined below. Quadratures are ordered $(q_1,q_2,p_1,p_2)$ throughout this section, so that the aligned seed is block diagonal.

For $X=\begin{psmallmatrix}\xi&\chi\\\chi&\zeta\end{psmallmatrix}$ and $\delta=\det X=\xi\zeta-\chi^2$,
\begin{equation}
2T^2\FI_X=\frac{N_1}{D_1^2}+\frac{N_2}{D_2^2},
\end{equation}
where
\begin{align}
D_1&=\nu_1\nu_2+\nu_1\zeta+\nu_2\xi+\delta,\\
N_1&=(\nu_2+\zeta)^2\mu_1^2+(\nu_1+\xi)^2\mu_2^2+2\chi^2\mu_1\mu_2,\\
D_2&=\nu_1\nu_2\delta+\nu_1\xi+\nu_2\zeta+1,\\
N_2&=(\nu_2\delta+\xi)^2\mu_1^2+(\nu_1\delta+\zeta)^2\mu_2^2+2\chi^2\mu_1\mu_2.
\end{align}
Here $D_1$ is the determinant of the $q$ block of $V+M$, and $D_2$ is $\delta$ times the determinant of its $p$ block. As a check, $\xi=\zeta=\delta=1$ and $\chi=0$ returns the double-heterodyne value $\mu_1^2/(\nu_1+1)^2+\mu_2^2/(\nu_2+1)^2$.

The two squeezing strengths $a$ and $b$ of the Bloch--Messiah form run to infinity, so we compress each into a unit interval by $\alpha=a/(1+a)$ and $\beta=b/(1+b)$, and we keep $\tau=\cos^2\theta$ for the mixing angle. The vacuum seed $a=b=1$ sits at $\alpha=\beta=1/2$. Infinite squeezing sits at the ends $\alpha=0,1$ and $\beta=0,1$, so the limits Bell homodyne needs lie on the faces of the cube. All finite seeds occupy the open cube $(0,1)^3$. Double heterodyne is the line $(1/2,1/2,\tau)$, and the two Bell limits are the corners $(1,0,1/2)$ and $(0,1,1/2)$. Write $P(\alpha,\beta,\tau)$ for the numerator of the shortfall $\FI_{\mathrm{Bell}}-\FI_X$ once the positive denominator built from $D_1$ and $D_2$ has been cleared, evaluated at $x_1=\operatorname{arccoth}(1+\sqrt2)$ and $x_2=3/2$. Cutting at $\alpha=1/2$ and at $\beta=1/2$ makes four boxes. In two of them $\alpha$ and $\beta$ lie on the same side of $1/2$. We call the other two the cross boxes, and each of those we cut again at $\tau=1/2$. Six boxes are left in all.

We prove that $P$ is nonnegative on the whole cube and strictly positive in its interior. A polynomial written in the Bernstein basis of a box is a combination of basis functions that are nonnegative on that box. If every coefficient of that combination is positive, the polynomial is positive on the whole box, so a list of signs is a proof. We expand $P$ this way on each subbox, in interval arithmetic at $120$ decimal digits. Every coefficient then comes with an interval guaranteed to contain its true value. Only the sign of that interval enters the argument. Table~\ref{S:tab:certificate} gives the count on each piece.

\begin{table}[ht]
\caption{\label{S:tab:certificate}Bernstein coefficients of $P$ on the six subboxes of the seed cube, at the reference pair $x_1=x_c$, $x_2=3/2$. The eight zero coefficients sit at the two balanced infinite-squeezing corners, two of them on each of the four boxes cut from a cross box, and at those corners $P$ and $\partial_\tau P$ vanish together. Every other coefficient is strictly positive.}
\begin{ruledtabular}
\begin{tabular}{lcc}
Region & Positive & Zeros\\
\hline
$\alpha,\beta\le1/2$ & 125 & 0\\
$\alpha,\beta\ge1/2$ & 125 & 0\\
Each of the four boxes cut from a cross box & 123 & 2\\
\end{tabular}
\end{ruledtabular}
\end{table}

The three rows sum to the $742$ positive coefficients and $8$ zeros that End Matter D reports. Every coefficient other than those eight is enclosed in an interval whose lower endpoint is positive. Every finite seed is therefore strictly suboptimal, and equality holds only at the two balanced Bell limits. The maximal-gain pair is a different pair, and the rectangle proof of Sec.~\ref{S:sec:region} covers it inside a tile.

\section{Proof on two rectangles}

\begin{figure*}[!tp]
 \includegraphics{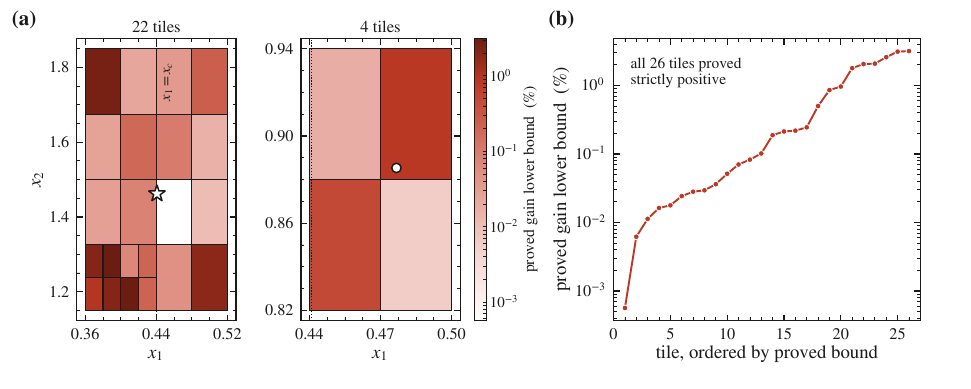}
 \caption{\label{S:fig:tiles}
 The two rectangles of thermal pairs on which Bell optimality is proved. \textbf{(a)} The tiles of each rectangle, colored by the proved lower bound on the gain in percent of the separate-readout value. The left rectangle carries $22$ tiles and surrounds the maximal-gain pair (star); the right one carries $4$ tiles and contains the coupled-resonator operating point at $g/\omega_0=0.3$ (circle). Tiles are finer along $x_1=x_c$ (dotted), where the separate-readout benchmark changes branch. Mode exchange maps each rectangle to a mirror region below the diagonal $x_1=x_2$, which inherits the same proof. \textbf{(b)} The proved lower bound on the gain, tile by tile, sorted. Every tile is proved both Bell-optimal over all Gaussian seeds and strictly inside the Bell phase, the weakest bound being $5.6\times10^{-6}$ of the separate-readout value, on the tile that straddles the crossover $x_c$.}
\end{figure*}
\label{S:sec:region}

End Matter D proves Bell optimality at the single thermal pair $x_1=x_c$, $x_2=3/2$. We carry that proof to whole rectangles in $(x_1,x_2)$. Each $x_i$ is a frequency divided by a temperature, so one rectangle holds many frequency ratios and, for each ratio, an interval of temperatures.

The obstruction is that a Bernstein proof needs coefficients that are numbers. At a fixed pair that is what Sec.~\ref{S:sec:certpoint} supplies: clearing the positive denominator leaves the polynomial $P(\alpha,\beta,\tau)$, in the two compactified squeezings and the mixing angle, with fixed exact coefficients. Over a rectangle each coefficient is instead a function of $x_1$ and $x_2$, and the expansion has no number to work with. The obvious repair encloses each coefficient in an interval covering its values on the rectangle, and it destroys the proof. The coefficients of $P$ are sums of large terms that almost cancel, and $P$ is small and positive because the cancellations are exact. Independent intervals forget which term cancels against which, so every Bernstein coefficient comes out as an interval wider than the spread it must cover, and each contains zero. The precision is lost in the arithmetic, and a smaller box suffers the same loss.

We keep the cancellations symbolic. Cut a rectangle into small rectangles and call one of them a tile. On a tile centered at $(\bar x_1,\bar x_2)$ with half-width $\Delta_i$ in $x_i$, write $\epsilon_i=x_i-\bar x_i$ and expand to second order,
\begin{equation}
 \nu_i=\nu_i(\bar x_i)+\nu_i'(\bar x_i)\,\epsilon_i
 +\tfrac12\nu_i''(\bar x_i)\,\epsilon_i^{2}+\rho_i ,
 \label{S:eq:taylormodel}
\end{equation}
with a prime an ordinary derivative in $x_i$; the onset law used logarithmic ones. The same expansion applies to $\mu_1$ and $\mu_2$, giving four remainders, $\rho_1$ and $\rho_2$ for the widths and $\rho_3$ and $\rho_4$ for the responses. The Lagrange form of the second-order remainder evaluates the second derivative at some point $\vartheta$ lying between $\bar x_i$ and $x_i$. Subtracting the second-order term we kept therefore leaves $\tfrac12[\nu_i''(\vartheta)-\nu_i''(\bar x_i)]\epsilon_i^{2}$ for a width. The same expression in $\mu_i$ bounds a response. The two points are separated by at most $\Delta_i$, so each remainder is bounded at order $\Delta_i^{3}$.

We treat each remainder as an extra variable of the polynomial, and call on its bound only at the end. The Taylor polynomials of Eq.~\eqref{S:eq:taylormodel} keep exact coefficients, so every cancellation among the monomials of $P$ survives symbolically. Terms quadratic and higher in the remainders reach order $\Delta_i^{6}$, and we bound those by a single number and subtract it before expanding. The result is exactly linear in the four remainders, with an exactly known coefficient on each. The proof reduces to one statement, that a single polynomial is positive on a nine-dimensional box of three seed variables, two thermal increments and four remainders. We expand it in the Bernstein basis of that box in interval arithmetic, at $100$ decimal digits, where a single thermal pair needed $120$. The extra variables hold the thermal variation exactly, so $100$ digits already decide every sign. Where an enclosure leaves a sign undecided the box is bisected, and the subdivision terminates with every coefficient of definite sign.

Two rectangles carry this proof. The first covers $x_1\in[0.36,0.52]$ and $x_2\in[1.15,1.85]$, cut into $22$ tiles, taken finer where the separate-readout benchmark changes branch. The second covers $x_1\in[0.44,0.50]$ and $x_2\in[0.82,0.94]$, cut into $4$ tiles of equal size. Write $\FI_{\mathrm{global}}^{\mathrm G}$ for the largest Fisher information any Gaussian measurement on the two modes reaches. Every tile carries both halves of the chain $\FI_{\mathrm{global}}^{\mathrm G}=\FI_{\mathrm{Bell}}>\FI_{\mathrm{local}}^{\max}$ that the Letter proves at a single pair. At each thermal pair inside a tile, every finite aligned seed has strictly smaller Fisher information than Bell homodyne. Tilting one squeezing axis lowers the Fisher information by Sec.~\ref{S:sec:pureseed}, and Sec.~\ref{S:sec:mixed} excludes mixed seeds, so the aligned family decides the whole question. The same expansion applied to the gain gives a positive lower bound on it over the tile. Its pairs therefore sit strictly inside the Bell phase, the set of pairs on which Bell homodyne beats that benchmark outright.

The gap $\FI_{\mathrm{Bell}}-\FI(M)$ closes only where the seed is the Bell measurement itself, at the two corners of the seed box with both squeezings infinite and equal. Those corners are fixed in the seed variables, so they stay exactly zero whatever values the remainders take.

Every tile of the first rectangle comes with a proved lower bound on the gain no smaller than $5.6\times10^{-6}$, and every tile of the second no smaller than $6.2\times10^{-5}$. Each bound is a fraction of the separate-readout benchmark, and each holds across a whole tile, so it is a floor under the gain on that tile. The weakest of those bounds belongs to the tile $x_1\in[0.44,0.48]$, $x_2\in[1.325,1.5]$, whose $x_1$ interval straddles the crossover $x_c$, so the separate-readout benchmark changes branch inside it. That tile holds the maximal-gain pair, where the gain reaches $12.699\%$. The second rectangle holds the coupled-resonator operating point: at $g/\omega_0=0.3$ the normal modes at $\omega_0\pm g$ stand in the ratio $1.3$ to $0.7$.

The same construction applies to any rectangle we cover with tiles, and it decides a tile when the gain on that tile stays away from zero. On the boundary of the Bell phase the gain falls to zero, so no tile straddling that boundary can be proved strictly inside the phase. The Bell phase extends past both rectangles, and these two are the rectangles we prove.

Bell homodyne is therefore the Gaussian optimum on both rectangles and at the reference pair, and Sec.~\ref{S:sec:ceiling} proves it for every passive network of the two modes followed by homodynes. The search over pure seeds in Sec.~\ref{S:sec:ceiling} found nothing above $\max\{\FI_{\mathrm{local}}^{\max},\FI_{\mathrm{Bell}}\}$ at four hundred random pairs of widths and responses.

\section{Mixed seeds}
\label{S:sec:mixed}

The Letter restricts the search to pure seeds~\cite{Weedbrook2012}. We prove that no mixed Gaussian seed beats the best pure one, so the restriction is exact. Williamson decomposition gives a mixed seed
\begin{equation}
M=S\left[\bigoplus_j\sigma_jI_2\right]S^T,\qquad \sigma_j\ge1,
\end{equation}
with $\sigma_j$ the symplectic eigenvalues of the seed. Define the pure seed $M_0=SS^T$. Here $S$ is symplectic, which means it is the quadrature action of some network of beam splitters and squeezers. Any such $S$ has determinant one, so $\det M_0=1$ and $M_0+i\Omega\ge0$. The seed $M_0$ therefore adds the least noise a Gaussian detector can add. Then
\begin{equation}
M=M_0+N,\qquad
N=S\left[\bigoplus_j(\sigma_j-1)I_2\right]S^T\ge0.
\end{equation}
The outcome distribution for $M$ is the outcome for $M_0$ convolved with a parameter-independent Gaussian of covariance $N$. That convolution is classical post-processing of the outcomes, and the classical Fisher information cannot increase under it, so mixed seeds cannot exceed the pure optimum.

\section{Coupled resonators}
\label{S:sec:resonators}

The Letter shows that two coupled identical resonators realize an unequal-frequency thermal pair, and that the Bell receiver for their normal modes is two bare homodyne detectors~\cite{Miyanaga2021,Bothner2021,Lu2023}. We prove both claims from the diagonalization. That setting is the relative phase of the two local oscillators~\cite{OckeloenKorppi2016,MercierDeLepinay2021,Kotler2021}. The temperature comes with it, because the coupling fixes the frequency ratio and the temperature alone places the pair on the $x$ axis.

Let each resonator have frequency $\omega_0$, and let a coupling $g$ that swaps excitations between them join the two,
\begin{equation}
 H=\hbar\omega_0\left(a_1^\dagger a_1+a_2^\dagger a_2\right)+\hbar g\left(a_1^\dagger a_2+a_2^\dagger a_1\right).
 \label{S:eq:resH}
\end{equation}
Substituting $a_1=(a_++a_-)/\sqrt2$ and $a_2=(a_+-a_-)/\sqrt2$ cancels every cross term and leaves $H=\hbar(\omega_0+g)a_+^\dagger a_++\hbar(\omega_0-g)a_-^\dagger a_-$. The normal modes $a_\pm=(a_1\pm a_2)/\sqrt2$ therefore diagonalize Eq.~\eqref{S:eq:resH} at frequencies $\omega_0\pm g$, both positive for $0<g<\omega_0$. They obey the same commutation relations as the bare modes, so the Gibbs weight $e^{-H/k_{\mathrm B}T}$ factorizes across them, and their amplitude map is a balanced beam splitter, passive and therefore symplectic. The equilibrium covariance in the normal-mode quadratures is accordingly $\nu_+I_2\oplus\nu_-I_2$, with $\nu_\pm=\coth x_\pm$, $\mu_\pm=x_\pm\csch^2x_\pm$ and
\begin{equation}
 x_\pm=\frac{\hbar(\omega_0\pm g)}{2k_{\mathrm B}T}=x_0\left(1\pm\frac{g}{\omega_0}\right),
 \qquad
 x_0=\frac{\hbar\omega_0}{2k_{\mathrm B}T},
 \label{S:eq:resx}
\end{equation}
which is the thermal pair of Eq.~\eqref{S:eq:thermal} at frequency ratio $(\omega_0+g)/(\omega_0-g)$. Throughout this section the gain is the fraction by which the Bell Fisher information exceeds the best reading of the two normal modes with one detector each. That benchmark is the sum of Eq.~\eqref{S:eq:local} over the pair $x_\pm$.

The splitting between the two normal modes is $s=\ln[(\omega_0+g)/(\omega_0-g)]$, exactly, for every $g$ in $(0,\omega_0)$. The onset law expands about the logarithmic midpoint of the pair, here $\sqrt{x_+x_-}=x_0\sqrt{1-(g/\omega_0)^2}$, which equals $x_0$ up to a correction of order $(g/\omega_0)^2$. It gives the gain as a coefficient times $s^2$, and that coefficient is largest, $0.13024$, when the midpoint sits at $x=0.66678$. At weak coupling the best temperature drives the midpoint there, and the gain approaches $0.13024\,s^2$.

Composing the Bell beam splitter with the normal-mode transformation leaves only an exchange of the two resonator labels. We show that the identity holds for the recorded quadratures as well as for the mode amplitudes. Both mixings are real orthogonal matrices acting on the mode amplitudes. Such a matrix acts on the two $q$ quadratures by itself, acts on the two $p$ quadratures by itself, and never turns a $q$ into a $p$. Writing $q_\pm$, $p_\pm$ for the normal-mode quadratures and $q_i$, $p_i$ for the bare ones, this gives $q_\pm=(q_1\pm q_2)/\sqrt2$ and $p_\pm=(p_1\pm p_2)/\sqrt2$. The receiver takes the $q$ difference at its first output and the $p$ sum at its second, and those are
\begin{align}
 \frac{q_+-q_-}{\sqrt2}&=\frac{(q_1+q_2)-(q_1-q_2)}{2}=q_2,
 \label{S:eq:quadA}\\
 \frac{p_++p_-}{\sqrt2}&=\frac{(p_1+p_2)+(p_1-p_2)}{2}=p_1 .
 \label{S:eq:quadB}
\end{align}
The subscript $\pm$ labels a normal mode here, whereas in the main text it labelled a difference and a sum. The two records sit on different resonators, so $q_2$ and $p_1$ commute and one shot fixes both of them sharply. Our Bell splitter is $\tfrac1{\sqrt2}\begin{psmallmatrix}1&-1\\1&1\end{psmallmatrix}$; its transpose delivers $q_1$ and $p_2$, the same statement with the labels exchanged.

A detector on one bare resonator sees nothing of the mismatch. Tracing out the other leaves a thermal state of variance $\bar\nu=(\nu_++\nu_-)/2$ and response $\bar\mu=(\mu_++\mu_-)/2$, the same two numbers for both resonators. The mismatch survives only in the cross-covariance $(\nu_+-\nu_-)/2$ of the two resonators, which a joint reading collects.

At fixed coupling the temperature is the only free parameter, and it slides both $x_\pm$ along the $x$ axis together. Maximizing the gain over it leaves one number per coupling (Table~\ref{S:tab:resonators}). The weakest coupling there is the onset law at work: $s=\ln(11/9)=0.2007$ gives $0.13024\,s^2=0.52\%$, the tabulated value, and the maximizing temperature places the midpoint at $0.6649$ against the peak at $0.66678$. Accuracy falls as the splitting grows. Two rows of the table land inside the rectangles proved in Sec.~\ref{S:sec:region}. The row at $g/\omega_0=0.3$ falls in the second rectangle. The row at $g/\omega_0=0.5368$ falls in the first, and it reproduces the maximal-gain pair of the Letter. On both we prove in interval arithmetic that every finite Gaussian seed does strictly worse than Bell homodyne, so Bell homodyne is the Gaussian optimum there. The two normal modes are separated in frequency by $2g$, so a measurement on the field radiated out of the resonators needs a detection bandwidth above $2g$ to resolve them~\cite{Clerk2010,Eichler2011,Menzel2010}.

\begin{table}[b]
\caption{\label{S:tab:resonators}Two identical resonators of frequency $\omega_0$ joined by a coupling $g$ that swaps excitations between them. Each row gives the temperature at which the gain of Bell homodyne over the best separate readout of the normal modes is maximal, and the thermal pair that temperature produces through Eq.~\eqref{S:eq:resx}.}
\begin{ruledtabular}
\begin{tabular}{ccccc}
$g/\omega_0$ & $k_{\mathrm B}T/\hbar\omega_0$ & $x_-$ & $x_+$ & gain\\
\hline
$0.1$ & $0.7482$ & $0.6015$ & $0.7351$ & $0.52\%$\\
$0.2$ & $0.7430$ & $0.5383$ & $0.8075$ & $2.12\%$\\
$0.3$ & $0.7342$ & $0.4767$ & $0.8853$ & $4.87\%$\\
$0.5368$ & $0.5255$ & $0.4407$ & $1.462$ & $12.699\%$\\
\end{tabular}
\end{ruledtabular}
\end{table}

Over a range of couplings the best temperature is fixed in closed form, and the crossover of Eq.~\eqref{S:eq:local} fixes it. The optimal temperature falls sharply at the strongest coupling of Table~\ref{S:tab:resonators}. For every coupling in $0.35916\le g/\omega_0\le0.60720$ the maximizing temperature satisfies $x_-=x_c$ exactly, that is
\begin{equation}
 \frac{k_{\mathrm B}T}{\hbar\omega_0}=\frac{1-g/\omega_0}{2x_c},
 \label{S:eq:pin}
\end{equation}
which holds the lower normal mode on the crossover; we prove this below. The last temperature in the table is that closed form at $g/\omega_0=0.5368$. The gain has a corner at that temperature. Rounding $k_{\mathrm B}T/\hbar\omega_0$ to two decimal places there already costs $0.16$ percentage points of gain.

The corner in the benchmark is what pins that temperature. The separate-readout benchmark of the normal modes is the larger of the homodyne and heterodyne values, so as a function of $x_-$ it bends upward at $\nu_c$. It sits in the denominator of the gain, and a reciprocal turns an upward bend into a downward one, so the gain bends downward there. Write $\gamma=g/\omega_0$, $r=(1+\gamma)/(1-\gamma)$ for the normal-mode ratio and $\varphi(t)=G(t,rt)$ for the gain as a function of $t=x_-$, so that the temperature enters through $t$ alone. The two one-sided slopes at the corner, $\sigma_-=\varphi'(x_c^-)$ from the warm side and $\sigma_+=\varphi'(x_c^+)$ from the cold side, are explicit functions of $\gamma$. Each decreases through zero once, $\sigma_+$ at $\gamma_{\mathrm{lo}}=0.359153$ and $\sigma_-$ at $\gamma_{\mathrm{hi}}=0.607205$, normal-mode ratios $2.1209$ and $4.0917$; the roots are enclosed by bisection in interval arithmetic and the decrease is established within $0.01$ of each root. Between the roots the slopes straddle zero, $\sigma_->0$ and $\sigma_+<0$, so the gain rises into the corner and falls away from it. That the corner is the global maximum in $t$ follows in three steps, verified in interval arithmetic as in Sec.~\ref{S:sec:cap}. On the warm strip $x_c-0.03\le t<x_c$ the gain is concave in $t$ for every $\gamma$ in the interval, and on the cold strip $x_c<t\le x_c+0.03$ it is concave for $\gamma\le0.55$ and falling for $\gamma\ge0.55$; a concave function whose slope at the corner points toward the corner lies below its corner value, so both strips lie below $\varphi(x_c)$. Away from the strips, on $0.1\le t\le x_c-0.03$ and on $x_c+0.03\le t\le3$, the inequality $\varphi(t)<\varphi(x_c)$ holds directly, on $2.9\times10^{5}$ boxes. For $t<0.1$ both normal modes lie below the crossover and the gain is negative by the argument of Sec.~\ref{S:sec:cap}, and for $t\ge3$ the gain is below $\coth^23-1<1\%$ while the corner gain exceeds $6\%$ throughout the interval. Outside the interval the two slopes share a sign, and the temperature then improves the gain by moving off the corner. For $0.05\le\gamma<\gamma_{\mathrm{lo}}$ both slopes are positive, so the optimum settles at $x_->x_c$, where the three weaker rows of Table~\ref{S:tab:resonators} sit; for $\gamma_{\mathrm{hi}}<\gamma\le0.95$ both are negative, which pushes the optimum across to $x_-<x_c$. The thermal cap of $12.699\%$ is fixed by the same corner, with two free parameters in place of the one here.

The relative phase of the two local oscillators is the setting that delivers the advantage. Write $\lambda_\pm=\mu_\pm/\nu_\pm$, $\nu_\Delta=(\nu_--\nu_+)/2$ and $\mu_\Delta=(\mu_--\mu_+)/2$; both differences are positive, because $\nu$ and $\mu$ decrease in $x$ while $x_+>x_-$. Homodyning resonator $i$ at local-oscillator phase $\varphi_i$ records $Y_i=q_i\cos\varphi_i+p_i\sin\varphi_i$. Each bare quadrature is a balanced combination of the two normal-mode quadratures, so both records have variance $\bar\nu$. Their covariance is $-\nu_\Delta(\cos\varphi_1\cos\varphi_2+\sin\varphi_1\sin\varphi_2)=-\nu_\Delta\cos\Delta\varphi$, with $\Delta\varphi=\varphi_1-\varphi_2$. Differentiating the covariance matrix with respect to temperature gives the same shape, $\bar\mu$ on the diagonal and $-\mu_\Delta\cos\Delta\varphi$ off it. Both matrices have equal diagonal entries, so both are diagonal in the sum and difference basis of the two records, with eigenvalues the diagonal entry plus and minus the off-diagonal one. The Fisher information therefore splits into two independent readings,
\begin{equation}
 T^2\FI(\Delta\varphi)=\frac12\left[\frac{\bar\mu-\mu_\Delta\cos\Delta\varphi}{\bar\nu-\nu_\Delta\cos\Delta\varphi}\right]^2
 +\frac12\left[\frac{\bar\mu+\mu_\Delta\cos\Delta\varphi}{\bar\nu+\nu_\Delta\cos\Delta\varphi}\right]^2 .
 \label{S:eq:dial}
\end{equation}
The two signs there index the record combinations, not the normal modes. At $\Delta\varphi=0$ the records are an orthogonal recombination of one quadrature from each normal mode, and Eq.~\eqref{S:eq:dial} returns $\tfrac12(\lambda_+^2+\lambda_-^2)$. Every row of Table~\ref{S:tab:resonators} has $x_\pm\ge x_c$, so homodyne is the better single-mode reading and that value is the full separate-readout optimum of the normal modes. At $\Delta\varphi=90^\circ$ the off-diagonal entries vanish and Eq.~\eqref{S:eq:dial} returns the Bell value $[(\mu_++\mu_-)/(\nu_++\nu_-)]^2$. Sending $\cos\Delta\varphi$ to its negative exchanges the two terms, so the gain depends on the phase through $\cos^2\Delta\varphi$ alone. Writing $2t-1=\cos\Delta\varphi$ turns Eq.~\eqref{S:eq:dial} into Eq.~\eqref{S:eq:passive} at transmissivity $t$, with the lower normal mode in the role of the first mode there. That family has no interior maximum, so the best phase setting is one of the two ends, and the $90^\circ$ end is the larger one in every row of Table~\ref{S:tab:resonators}. Turning one local-oscillator phase is the only change to the receiver.

\section{Beyond the rotating-wave approximation}
\label{S:sec:rwa}

Section~\ref{S:sec:resonators} joins the two resonators by a coupling $g$ that swaps excitations between them, Eq.~\eqref{S:eq:resH}. That coupling is the rotating-wave form of a bilinear coupling between the two resonator coordinates. The rotating-wave form keeps only the terms that conserve the total number of excitations. The full coupling also contains, with the same weight, terms that create or destroy one excitation in each resonator at once. Dropping them is justified only while $g$ is small against the resonator frequency $\omega_0$. The Letter uses couplings up to $g/\omega_0=0.54$, where their effect is large: the coordinate coupling that this exchange strength stands for, $\varkappa=2g=1.07\,\omega_0$, lies past the instability derived below. We therefore treat the full bilinear coupling exactly. We find the same gain function at a corrected frequency ratio, and one fixed unbalanced beam splitter as its receiver.

The exact treatment starts from the coordinate coupling itself,
\begin{equation}
 H=\frac{\hbar\omega_0}{2}\sum_{i=1,2}\left(q_i^2+p_i^2\right)+\hbar\varkappa\,q_1q_2,
 \label{S:eq:rwaH}
\end{equation}
with the dimensionless quadratures of Sec.~\ref{S:sec:conventions} and a coupling strength $\varkappa$ in frequency units. Writing each $q_i$ in mode amplitudes and keeping only the excitation-conserving terms returns Eq.~\eqref{S:eq:resH} with $g=\varkappa/2$. The normal coordinates $q_\pm=(q_1\pm q_2)/\sqrt2$ and $p_\pm=(p_1\pm p_2)/\sqrt2$ of Sec.~\ref{S:sec:resonators} diagonalize Eq.~\eqref{S:eq:rwaH} exactly, because $2q_1q_2=q_+^2-q_-^2$. What remains is $H=\sum_\pm\tfrac{\hbar}{2}\left[(\omega_0\pm\varkappa)q_\pm^2+\omega_0p_\pm^2\right]$, two oscillators of unequal stiffness and common mass. Their frequencies are
\begin{equation}
 \omega_\pm=\omega_0\sqrt{1\pm\varkappa/\omega_0},
 \label{S:eq:rwaomega}
\end{equation}
both real for $\varkappa<\omega_0$, whereas the rotating-wave form gives $\omega_0\pm g$. The canonical quadratures of each normal mode are the rescaled pair
\begin{equation}
 Q_\pm=\varsigma_\pm q_\pm,\qquad P_\pm=p_\pm/\varsigma_\pm,\qquad \varsigma_\pm=\sqrt{\omega_\pm/\omega_0},
 \label{S:eq:rwaQP}
\end{equation}
which obey $[Q_\pm,P_\pm]=i$ and bring the Hamiltonian to $\sum_\pm\tfrac{\hbar\omega_\pm}{2}(Q_\pm^2+P_\pm^2)$. The rescaling is a squeezing of each normal mode by the fixed factor $\varsigma_\pm$, which the coupling alone sets~\cite{Weedbrook2012}. The Gibbs state is therefore a product of two thermal states, one per normal mode. In the canonical quadratures its covariance is $\nu_+I_2\oplus\nu_-I_2$ with $\nu_\pm=\coth x_\pm$, $\mu_\pm=x_\pm\csch^2x_\pm$ and $x_\pm=\hbar\omega_\pm/(2k_{\mathrm B}T)$, the thermal pair of Eq.~\eqref{S:eq:thermal}. Squeezing a normal mode maps the Gaussian measurements of that mode onto themselves. The separate-readout benchmark of the normal modes is therefore the sum of Eq.~\eqref{S:eq:local} over $x_\pm$, as in Sec.~\ref{S:sec:resonators}. The Bell value in the canonical quadratures is $[(\mu_++\mu_-)/(\nu_++\nu_-)]^2$, also as before. We conclude that the gain over the best separate readout of the normal modes is the gain function of the Letter. It is evaluated at the exact frequency ratio
\begin{equation}
 \frac{\omega_+}{\omega_-}=\sqrt{\frac{1+\varkappa/\omega_0}{1-\varkappa/\omega_0}}
 \label{S:eq:rwaratio}
\end{equation}
in place of the rotating-wave ratio $(1+g/\omega_0)/(1-g/\omega_0)$. The two agree to second order in the coupling. The exact ratio then grows faster, because the lower normal mode, the soft mode from here on, softens as $\sqrt{1-\varkappa/\omega_0}$.

That softening changes the receiver and leaves the gain function alone, so we now derive the receiver in the laboratory frame. The Bell quadratures of the normal modes are $Q_+-Q_-$ and $P_++P_-$, and Eq.~\eqref{S:eq:rwaQP} writes them in bare quadratures as
\begin{align}
 Q_+-Q_-&=\frac{(\varsigma_+-\varsigma_-)\,q_1+(\varsigma_++\varsigma_-)\,q_2}{\sqrt2},
 \label{S:eq:rwaQ}\\
 P_++P_-&=\frac{(1/\varsigma_++1/\varsigma_-)\,p_1+(1/\varsigma_+-1/\varsigma_-)\,p_2}{\sqrt2}.
 \label{S:eq:rwaP}
\end{align}
They commute, since $[Q_+,P_+]$ and $[Q_-,P_-]$ cancel in $[Q_+-Q_-,P_++P_-]$. In the bare quadratures, the two coefficient vectors are orthogonal. Their inner product $(\varsigma_+-\varsigma_-)(1/\varsigma_++1/\varsigma_-)+(\varsigma_++\varsigma_-)(1/\varsigma_+-1/\varsigma_-)$ vanishes, the two products contributing $\pm(\varsigma_+/\varsigma_{-}-\varsigma_{-}/\varsigma_+)$. A beam splitter acts on $(q_1,q_2)$ and on $(p_1,p_2)$ by one and the same orthogonal matrix. Two orthogonal coefficient vectors are therefore exactly what its two ports deliver. Take the unbalanced splitter
\begin{equation}
 \frac{1}{\sqrt{2(\varsigma_+^2+\varsigma_-^2)}}
 \begin{pmatrix}\varsigma_+-\varsigma_- & \varsigma_++\varsigma_-\\ -(\varsigma_++\varsigma_-) & \varsigma_+-\varsigma_-\end{pmatrix},
 \label{S:eq:rwaBS}
\end{equation}
whose rows are orthogonal and of unit length. Its first row applied to $(q_1,q_2)$ is $Q_+-Q_-$ up to a constant factor. Its second row applied to $(p_1,p_2)$ is $P_++P_-$ up to a constant factor. Multiplying the coefficients of Eq.~\eqref{S:eq:rwaP} by $-\varsigma_+\varsigma_-$ reproduces that row up to its normalization. A constant factor on a homodyne record changes no Fisher information. A $q$ homodyne on the first output and a $p$ homodyne on the second therefore measure the pair in one shot. The splitter mixes the resonators with amplitude ratio $(\sqrt{\omega_+}-\sqrt{\omega_-}):(\sqrt{\omega_+}+\sqrt{\omega_-})$. The share of resonator 1 on the first port is therefore
\begin{equation}
 \frac{(\sqrt{\omega_+}-\sqrt{\omega_-})^2}{2(\omega_++\omega_-)},
 \label{S:eq:rwashare}
\end{equation}
and equals $7.8\%$ at the maximal-gain ratio $3.318$. The coupling alone fixes that share, and it vanishes in the rotating-wave limit $\varsigma_+=\varsigma_-$. There Eq.~\eqref{S:eq:rwaBS} reduces to an exchange of the two labels up to a sign, and the ports carry the bare quadratures $q_2$ and $p_1$ of Eqs.~\eqref{S:eq:quadA} and \eqref{S:eq:quadB}.

The plain bare homodynes $q_2$ and $p_1$ remain a valid Gaussian receiver at every coupling. By Eq.~\eqref{S:eq:rwaQP}, $q_2=(Q_+/\varsigma_+-Q_-/\varsigma_-)/\sqrt2$ and $p_1=(\varsigma_+P_++\varsigma_-P_-)/\sqrt2$. The thermal state has no correlation between any $Q$ and any $P$, so the two records are independent. Each is a centered Gaussian whose only temperature dependence is its variance. For a single record of variance $V$, Eq.~\eqref{S:eq:FI} gives $T^2\FI=\tfrac12(T\partial_T\ln V)^2$, one half of the squared logarithmic derivative of the variance. The two variances are $\tfrac12(\nu_+/\varsigma_+^2+\nu_-/\varsigma_-^2)$ and $\tfrac12(\varsigma_+^2\nu_++\varsigma_-^2\nu_-)$, with $T\partial_T\nu_\pm=\mu_\pm$, so
\begin{equation}
 T^2\FI_{\mathrm{bare}}=\frac12\left[\frac{\mu_+/\omega_++\mu_-/\omega_-}{\nu_+/\omega_++\nu_-/\omega_-}\right]^2
 +\frac12\left[\frac{\omega_+\mu_++\omega_-\mu_-}{\omega_+\nu_++\omega_-\nu_-}\right]^2 .
 \label{S:eq:rwabare}
\end{equation}
Each record weights the two normal modes by a power of their frequency, $\omega_\pm^{-1}$ in the first and $\omega_\pm$ in the second. Those weights are the squeezing factors $\varsigma_\pm^{\mp2}$, and the rotating-wave treatment sets them to one. Each term then returns half the Bell value, as Eqs.~\eqref{S:eq:quadA} and \eqref{S:eq:quadB} require. Table~\ref{S:tab:rwa} compares the two receivers at the temperature that holds the soft mode on the crossover, $x_-=x_c$. We fix the temperature by the crossover rule of Eq.~\eqref{S:eq:pin} with $\omega_-$ in place of $\omega_0-g$ at every coupling. Section~\ref{S:sec:resonators} proves it to be the maximizing temperature for ratios between $2.121$ and $4.091$, which covers the rows at $\varkappa/\omega_0=0.80$ and $0.8335$, and at the cap it gives $k_{\mathrm B}T/\hbar\omega_0=0.463$. The bare homodynes track the optimum at weak coupling and fall behind as the coupling grows, because the unequal weights in Eq.~\eqref{S:eq:rwabare} pull them away from the Bell combination. We check the optimality of the first gain column numerically. Take any coupling and temperature of Table~\ref{S:tab:rwa}, and maximize the Fisher information over all two-mode passive networks of the bare resonators followed by two homodynes. That maximum agrees with the tabulated optimal gain to four digits. The unbalanced splitter of Eq.~\eqref{S:eq:rwaBS} is therefore the best passive receiver, as the balanced one was in Sec.~\ref{S:sec:passivenet}. Three rows reach further. At $\varkappa/\omega_0=0.60$, $0.80$ and $0.8335$ the normal-mode pairs $(x_-,x_+)$ lie inside the rectangles proved in Sec.~\ref{S:sec:region}. A symplectic change of frame maps the Gaussian measurement class onto itself, so there the Bell value is the optimum over every Gaussian measurement of the bare resonators.

\begin{table}[b]
\caption{\label{S:tab:rwa}Two identical resonators of frequency $\omega_0$ coupled through their coordinates with strength $\varkappa$, Eq.~\eqref{S:eq:rwaH}, at the temperature that holds the soft normal mode on the crossover, $x_-=x_c$. The optimal gain is the gain of the Letter at the exact normal-mode ratio $\omega_+/\omega_-$, delivered by the splitter of Eq.~\eqref{S:eq:rwaBS}. The bare-homodyne gain is Eq.~\eqref{S:eq:rwabare}. Both are relative to the best separate readout of the normal modes and quoted in percent.}
\begin{ruledtabular}
\begin{tabular}{cccc}
$\varkappa/\omega_0$ & $\omega_+/\omega_-$ & optimal gain & bare-homodyne gain\\
\hline
$0.20$ & $1.225$ & $0.44\%$ & $0.43\%$\\
$0.40$ & $1.528$ & $2.12\%$ & $2.03\%$\\
$0.60$ & $2.000$ & $6.04\%$ & $5.38\%$\\
$0.80$ & $3.000$ & $12.31\%$ & $8.93\%$\\
$0.8335$ & $3.318$ & $12.70\%$ & $8.45\%$\\
$0.90$ & $4.359$ & $10.04\%$ & $4.21\%$\\
\end{tabular}
\end{ruledtabular}
\end{table}

\begin{figure*}[!tp]
 \includegraphics{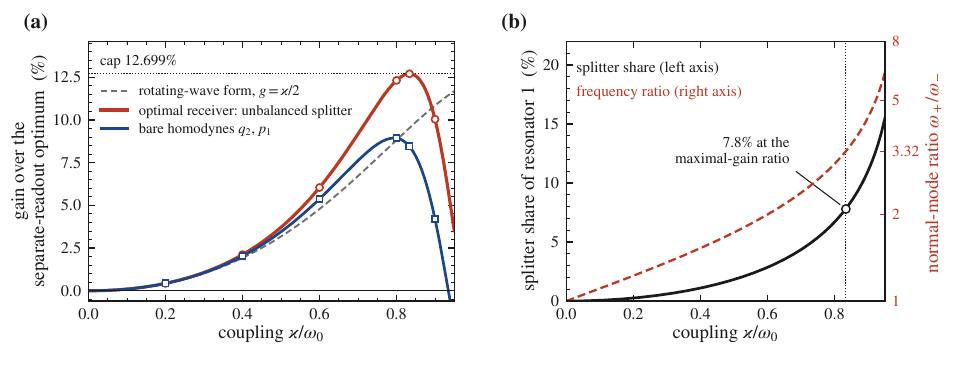}
 \caption{\label{S:fig:rwa}
 The exact bilinear coupling. \textbf{(a)} Gain over the best separate readout of the normal modes against the coupling $\varkappa/\omega_0$ from $0$ to $0.95$. The temperature holds the soft mode on the crossover, $x_-=x_c$. Red is the optimal receiver, the gain of the Letter at the ratio of Eq.~\eqref{S:eq:rwaratio}, delivered by the splitter of Eq.~\eqref{S:eq:rwaBS}. Blue is the plain bare homodynes $q_2$ and $p_1$ of Eq.~\eqref{S:eq:rwabare}. The six couplings of Table~\ref{S:tab:rwa} are marked. Grey dashed is the exchange model of Sec.~\ref{S:sec:resonators} at $g=\varkappa/2$ under the same temperature rule, which is why it lies below the values of Table~\ref{S:tab:resonators} at weak coupling, and the dotted line is the thermal cap of $12.699\%$. \textbf{(b)} The share of resonator 1 on the first port of the splitter, Eq.~\eqref{S:eq:rwashare}, against the coupling (left axis). The right axis carries the normal-mode ratio $\omega_+/\omega_-$ of Eq.~\eqref{S:eq:rwaratio}. The cap ratio $3.318$ is marked at $\varkappa/\omega_0=0.8335$, where the share is $7.8\%$.}
\end{figure*}

The two models differ most visibly in the coupling that reaches the cap. The cap ratio $3.318$ is reached at $\varkappa/\omega_0=0.8335$, where the soft mode sits at $\omega_-=0.408\,\omega_0$. The coupling drives that mode toward zero frequency, and $\varkappa\to\omega_0$ is the instability where it arrives there. The exchange model needs $g/\omega_0=0.5368$ for the same ratio. If that exchange coupling were the rotating-wave form of Eq.~\eqref{S:eq:rwaH}, the underlying coupling would be $\varkappa=2g=1.07\,\omega_0$, past the instability. The rotating-wave approximation therefore overstates the coupling that a coordinate-coupled experiment needs to reach the cap. A momentum coupling $\hbar\varkappa\,p_1p_2$ has the same spectrum and gives every result above with $q$ and $p$ exchanged, so only the ratio $\omega_+/\omega_-$ and the temperature enter the gain. On any platform the measured splitting of the two normal modes therefore fixes that ratio, and the gain follows from the curve of the Letter. Figure~\ref{S:fig:rwa} draws the two receivers and the splitter share for couplings up to $0.95\,\omega_0$, and the exchange model at $g=\varkappa/2$ lies below the exact curve at every coupling.

\section{Loss, squeezing and finite samples}
\label{S:sec:robustness}

The Letter quotes two efficiency thresholds, a squeezing scale and a Fisher-information ratio, and its Fig.~1(c) shows a temperature window~\cite{Clerk2010,Scigliuzzo2020}. Each is a statement about an apparatus that is not ideal, and we derive them from one detector model. In that model an inefficient homodyne and an inefficient heterodyne add different amounts of vacuum noise, so loss moves the homodyne-heterodyne crossover as well as the two Fisher informations.

Every homodyne in either receiver has the same efficiency $\eta$. An efficiency $\eta$ scales the quadrature amplitude by $\sqrt\eta$ and adds $1-\eta$ units of vacuum noise. Dividing the record by $\sqrt\eta$ restores the signal and leaves the added noise at $(1-\eta)/\eta$ units. A heterodyne reads both quadratures at once. It splits the mode on a balanced splitter and homodynes each output, and vacuum enters the splitter's empty port. That vacuum contributes one unit of noise on its own. The split also halves the signal variance reaching each detector, while each detector still adds its $1-\eta$. Dividing by $\eta/2$ restores the signal and leaves $1+2(1-\eta)/\eta=(2-\eta)/\eta$ units. At $\eta=1$ only the empty-port unit survives, the Arthurs--Kelly unit that the uncertainty principle adds for reading two conjugate quadratures in one shot, so both rules return the ideal receivers. The Bell value, two homodynes on one balanced splitter, becomes
\begin{equation}
T^2\FI_{\mathrm{Bell}}(\eta)=\left[\frac{\mu_1+\mu_2}
{\nu_1+\nu_2+2(1-\eta)/\eta}\right]^2,
\end{equation}
and the separate-readout benchmark becomes
\begin{equation}
T^2\FI_{\mathrm{local}}^{\max}(\eta)=\sum_{i=1,2}\mu_i^2\max\left\{
\frac{1}{2\left[\nu_i+\frac{1-\eta}{\eta}\right]^2},
\frac{1}{\left[\nu_i+\frac{2-\eta}{\eta}\right]^2}\right\},
\end{equation}
which is Eq.~\eqref{S:eq:local} with each branch carrying its own detector noise. At the maximal-gain pair we find the two curves crossing at $\eta=57.95\%$, and Bell homodyne exceeding the ideal lossless benchmark at $\eta=90.20\%$.

A receiver assembled from squeezed light is imperfect in a second way, and the Letter quotes the squeezing it has to reach. Cap both squeezings of the aligned family of Sec.~\ref{S:sec:pureseed} at a common bound and maximize the gain over the seeds that survive. We quote that bound in decibels of quadrature squeezing. The number is ten times the base-ten logarithm of the ratio between the vacuum variance and the smaller variance of the seed, so the vacuum seed sits at $0$ dB. At the maximal-gain pair the best seed within a $0$ dB budget is double heterodyne, whose gain is negative. The best seed first beats the separate-readout benchmark at $5.4$ dB; it recovers half the Bell gain at $10.0$ dB and reaches nine tenths of that gain only at $17.8$ dB. A beam splitter followed by two homodyne detectors reaches the infinite-squeezing limit outright, which is why the coupled resonators of Sec.~\ref{S:sec:resonators} need no squeezed light at all.

Running the same comparison against temperature gives the tolerance of a fixed pair, and we quote two pairs. The maximal-gain pair $x_1=x_c$, $x_2=1.4621888$ keeps a positive gain over $0.769\,T_0<T<1.538\,T_0$, with $T_0$ the temperature that puts mode~1 exactly on the crossover, and those are the two ends of the window in Fig.~1(c) of the Letter. The reference pair $x_1=x_c$, $x_2=3/2$ is slightly wider, at $0.777\,T_0<T<1.561\,T_0$. Both ends of either window are zeros of the exact gain at fixed frequencies. The splitting of either pair is far outside the small-$s$ range where the onset law applies. We locate the four ends by evaluating the Bell value of Eq.~\eqref{S:eq:twoaverages} against the envelope of Eq.~\eqref{S:eq:local} directly.

A ratio of Fisher informations becomes a ratio of errors only after enough shots, and we state the condition. With $n$ independent repetitions, the maximum-likelihood estimate of $T$ from either receiver settles on the true temperature as $n$ grows, and its spread about that value becomes Gaussian. Its mean-square error approaches the inverse Fisher information divided by $n$. At the maximal-gain pair the Bell receiver's limiting mean-square error falls below the separate-readout one by the $12.699\%$ of Sec.~\ref{S:sec:cap}, and at every other pair by less. At finite $n$ the estimate acquires a bias of order $1/n$, which enters the mean-square error at order $1/n^2$. The two receivers therefore separate at the Fisher-information ratio once $n$ is large enough for that correction to be negligible.

\section{Pairing a frequency bank}
\label{S:sec:pairing}

The Letter works with two modes. We extend the same receiver to a bank of many modes at one temperature, once the modes are matched into disjoint pairs~\cite{Proctor2018,Guo2020}. A matched pair of modes is read either by Bell homodyne or by separate readout, whichever is larger, and a mode left over is read by its own single-mode optimum.

Write $\FI_i^{\mathrm{loc}}$ for the single-mode optimum of mode $i$, given by Eq.~\eqref{S:eq:local}, and define the weight of a pair as the Bell gain on it,
\begin{equation}
w_{ij}=\max\{0,\FI_{\mathrm{Bell}}^{(ij)}-\FI_i^{\mathrm{loc}}-\FI_j^{\mathrm{loc}}\}.
\end{equation}
A matching $\mathcal M$ is a way of splitting the bank into disjoint pairs, with each mode in at most one pair. The receivers it names act on disjoint sets of modes, so they can all run at once, and their Fisher informations add,
\begin{equation}
\FI_{\mathrm{pair}}=\sum_i\FI_i^{\mathrm{loc}}+\max_{\mathcal M}\sum_{(i,j)\in\mathcal M}w_{ij},
\end{equation}
where the maximum runs over every such matching.

We evaluate this protocol on a ten-mode logarithmic bank spanning $x=0.15$ to $2.5$ at a design temperature $T_{\mathrm d}$. It gives a $3.75\%$ gain at $T_{\mathrm d}$, and between $1.7\%$ and $5.7\%$ over $0.55<T/T_{\mathrm d}<1.8$. The maximum is over the disjoint-pair class, one two-mode receiver to a pair.

\begin{figure*}[!tp]
 \includegraphics{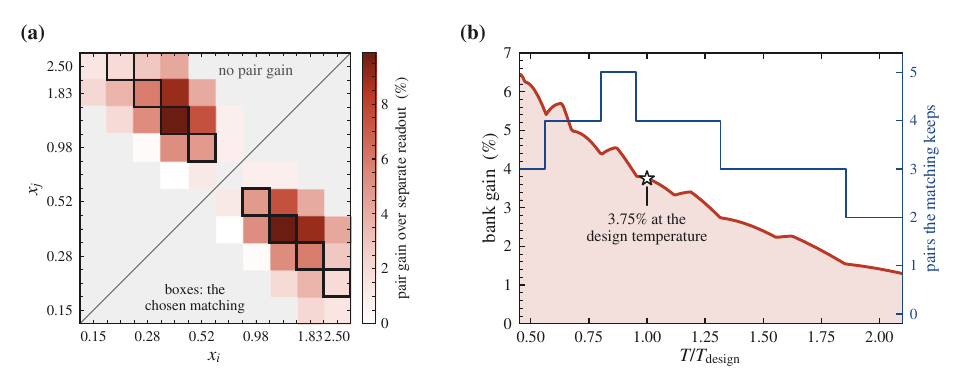}
 \caption{\label{S:fig:pairing}
 A ten-mode bank whose frequencies are spaced logarithmically from $x=0.15$ to $x=2.5$ at the
 design temperature. \textbf{(a)} Relative gain of a balanced Bell measurement on modes $i$ and
 $j$ over their separate readout. Grey marks the pairs that gain nothing, so a
 collective reading is worth making only for the pairs that straddle the crossover, and the best
 single pair gains $9.9\%$. Black boxes are the pairs the maximum-weight matching selects.
 \textbf{(b)} Gain of the whole bank against temperature, together with the number of pairs the
 matching keeps. The matching sheds pairs as the bank heats and every mode drifts onto the same
 branch, and the gain falls with it.}
\end{figure*}

\section{Amplifier as receiver}
\label{S:sec:amplifier}

The Letter's Bell receiver records the pair $q_-$ and $p_+$ of Sec.~\ref{S:sec:passivenet} behind a balanced beam splitter. We prove that a phase-preserving amplifier at large gain performs that measurement on its own signal and idler bands. Write $a_{\mathrm s}$ and $a_{\mathrm i}$ for the annihilation operators of the two bands and $\mathcal G$ for the power gain. Two-mode squeezing gives the amplified signal~\cite{Caves1982,Clerk2010}
\begin{equation}
 b_{\mathrm s}=\sqrt{\mathcal G}\,a_{\mathrm s}+\sqrt{\mathcal G-1}\,a_{\mathrm i}^\dagger .
 \label{S:eq:tms}
\end{equation}
In-phase and quadrature (IQ) demodulation of the amplified signal is a heterodyne of $b_{\mathrm s}$~\cite{Eichler2011}. It reads, up to a factor $\sqrt2$, the real and imaginary parts of $b_{\mathrm s}+a_{\mathrm v}^\dagger$, with $a_{\mathrm v}$ the vacuum mode of its empty port. With $q=(a+a^\dagger)/\sqrt2$ and $p=(a-a^\dagger)/(i\sqrt2)$ for every mode, the in-phase record $Y_{\mathrm I}$ and the quadrature record $Y_{\mathrm Q}$ are
\begin{equation}
 Y_{\mathrm I}=\sqrt{\mathcal G}\,q_{\mathrm s}+\sqrt{\mathcal G-1}\,q_{\mathrm i}+q_{\mathrm v},\qquad
 Y_{\mathrm Q}=\sqrt{\mathcal G}\,p_{\mathrm s}-\sqrt{\mathcal G-1}\,p_{\mathrm i}-p_{\mathrm v}.
 \label{S:eq:iq}
\end{equation}
They commute, since $i\mathcal G-i(\mathcal G-1)-i=0$. Divided by the common factor $\sqrt{2\mathcal G}$, they are $(q_{\mathrm s}+q_{\mathrm i})/\sqrt2$ and $(p_{\mathrm s}-p_{\mathrm i})/\sqrt2$, up to an idler weight $\sqrt{1-1/\mathcal G}$ and a vacuum term of size $1/\sqrt{2\mathcal G}$. That pair commutes on its own, since $[q_{\mathrm s},p_{\mathrm s}]-[q_{\mathrm i},p_{\mathrm i}]=i-i=0$. A phase rotation of the idler by $\pi$ sends $q_{\mathrm i}$ and $p_{\mathrm i}$ to their negatives. The pair is then $q_-$ and $p_+$ of Sec.~\ref{S:sec:passivenet}, with the signal as mode 1. A thermal state is invariant under phase rotations, so both pairs have the same statistics. At large gain the readout is therefore the Bell measurement of signal and idler.

That measurement reads two thermal bands at once. Let $\nu_{\mathrm s},\mu_{\mathrm s}$ and $\nu_{\mathrm i},\mu_{\mathrm i}$ be their widths and responses. Each record of Eq.~\eqref{S:eq:iq} then has width $\mathcal G\nu_{\mathrm s}+(\mathcal G-1)\nu_{\mathrm i}+1$ and response $\mathcal G\mu_{\mathrm s}+(\mathcal G-1)\mu_{\mathrm i}$. A thermal product state has no correlation between $q$ and $p$, so the two records are independent. Each carries half the squared ratio of response to width, the homodyne entry of Eq.~\eqref{S:eq:local}, and we derive
\begin{equation}
 T^2\FI=\left[\frac{\mathcal G\mu_{\mathrm s}+(\mathcal G-1)\mu_{\mathrm i}}{\mathcal G\nu_{\mathrm s}+(\mathcal G-1)\nu_{\mathrm i}+1}\right]^2.
 \label{S:eq:ampFI}
\end{equation}
With a vacuum idler, $\nu_{\mathrm i}=1$ and $\mu_{\mathrm i}=0$, this is $\mu_{\mathrm s}^2/(\nu_{\mathrm s}+1)^2$ at every gain, the heterodyne branch of Eq.~\eqref{S:eq:local}. The idler vacuum is Caves's half quantum of added noise~\cite{Caves1982,Clerk2010}. With the idler band on the same thermal source at the same temperature, the large-gain records divided by $\sqrt{2\mathcal G}$ have the Bell width $(\nu_{\mathrm s}+\nu_{\mathrm i})/2$ and response $(\mu_{\mathrm s}+\mu_{\mathrm i})/2$. Equation~\eqref{S:eq:ampFI} is then the Bell value of Eq.~\eqref{S:eq:twoaverages}. The receiver needs neither a beam splitter nor a frequency converter, and it has no local-oscillator phase to set, because $Y_{\mathrm I}$ and $Y_{\mathrm Q}$ are already the right pair. The two bands differ in frequency by construction. A four-wave-mixing traveling-wave amplifier pumped at $\omega_{\mathrm p}$ pairs the signal frequency $\omega_{\mathrm s}$ with $\omega_{\mathrm i}=2\omega_{\mathrm p}-\omega_{\mathrm s}$~\cite{Macklin2015}, and a three-wave-mixing amplifier with $\omega_{\mathrm i}=\omega_{\mathrm p}-\omega_{\mathrm s}$~\cite{Bergeal2010,RoyDevoret2016}. A broadband matched load at temperature $T$ on the input line, or one on each input of a two-port device, feeds both bands.

A real line attenuates both bands and a real amplifier adds noise. We derive how both enter the loss model of Sec.~\ref{S:sec:robustness}. Let $\eta_0$ be the power transmission from the load to the amplifier, the same in both bands. Let $n_{\mathrm{ex}}$ be the noise the amplifier adds from its own internal degrees of freedom, beyond what the idler port delivers, in quanta after division by the gain. A transmission $\eta_0$ replaces the lost fraction of a band by vacuum, so each width becomes $\eta_0\nu+1-\eta_0$ and each response $\eta_0\mu$. One quantum of added noise is two units of width, because the vacuum's half quantum is one unit. Referred to the input and divided by $\sqrt{\mathcal G}$, the large-gain Bell record therefore has width $\eta_0(\nu_{\mathrm s}+\nu_{\mathrm i})+2(1-\eta_0)+2n_{\mathrm{ex}}$, twice the width of the $q_-$ record of Sec.~\ref{S:sec:robustness}, and response $\eta_0(\mu_{\mathrm s}+\mu_{\mathrm i})$. Dividing both by $\eta_0$ leaves the ideal response and the width $\nu_{\mathrm s}+\nu_{\mathrm i}+2(1-\eta_0+n_{\mathrm{ex}})/\eta_0$. The Bell value of Sec.~\ref{S:sec:robustness} at homodyne efficiency $\eta$ has the denominator $\nu_1+\nu_2+2(1-\eta)/\eta$, so the two agree exactly when
\begin{equation}
 \eta=\frac{\eta_0}{1+n_{\mathrm{ex}}}.
 \label{S:eq:etamap}
\end{equation}
The vacuum-idler record maps the same way onto the heterodyne width $\nu_{\mathrm s}+(2-\eta)/\eta$, so one efficiency describes the Bell setting and the heterodyne setting. A second amplifier of noise $n_{\mathrm H}$ quanta, referred to its own input, adds $n_{\mathrm H}/\mathcal G$ to $n_{\mathrm{ex}}$. At $\mathcal G=100$ ($20$ dB) with $n_{\mathrm H}=10$ this alone gives $n_{\mathrm{ex}}=0.1$ and $\eta=0.91$. The finite-gain record makes this exact. Let the second amplifier, with everything after it, replace the vacuum unit of Eq.~\eqref{S:eq:ampFI} by $2n_{\mathrm H}$. Dividing numerator and denominator by $\mathcal G$,
\begin{equation}
 T^2\FI=\left[\frac{\mu_{\mathrm s}+(1-1/\mathcal G)\,\mu_{\mathrm i}}{\nu_{\mathrm s}+(1-1/\mathcal G)\,\nu_{\mathrm i}+2n_{\mathrm H}/\mathcal G}\right]^2,
 \label{S:eq:ampfinite}
\end{equation}
which is the Bell value at $\eta=1/(1+n_{\mathrm H}/\mathcal G)$ except for the weight $1-1/\mathcal G$ on the idler band, a correction of relative order $1/\mathcal G$. For the $4$ and $10$ GHz pair of Table~\ref{S:tab:amplifier} with $n_{\mathrm H}=10$, the record reaches $0.901$ of the ideal Bell value at $\mathcal G=100$, where the idler weight accounts for $0.2\%$, and $0.989$ at $\mathcal G=1000$.

The homodyne setting maps differently. Pumped degenerately, the amplifier gives one signal quadrature the amplitude gain $\sqrt{\mathcal G}+\sqrt{\mathcal G-1}$, close to $2\sqrt{\mathcal G}$, so the second amplifier's noise referred to the input is $n_{\mathrm H}/(4\mathcal G)$ quanta, a quarter of the phase-preserving value. The homodyne benchmark therefore runs at the efficiency $\eta_{\mathrm h}=\eta_0/(1+n_{\mathrm H}/2\mathcal G)$ while the Bell record runs at $\eta_0/(1+n_{\mathrm H}/\mathcal G)$, and a fair comparison on one chain uses both. At $\mathcal G=100$ with $n_{\mathrm H}=10$ and no line loss, the $4$ and $10$ GHz pair then gains $4.0\%$ against $9.1\%$ at one common efficiency, and with $\eta_0=0.8$ it gains $1.1\%$ against $6.5\%$. At $\mathcal G=1000$ the two mappings differ by half a percent of gain, $9.4\%$ against $9.9\%$ without line loss and $7.0\%$ against $7.6\%$ at $\eta_0=0.8$. Table~\ref{S:tab:amplifier} follows the Letter's convention of one common efficiency, and line loss, common to every setting, dominates that efficiency once the first stage has $30$ dB of gain or a second parametric stage.

That loss model and the Letter's temperature rule select the pair. The rule puts the lower band on the crossover, $x_{\mathrm s}=x_c$ with $x_{\mathrm s}=\hbar\omega_{\mathrm s}/(2k_{\mathrm B}T)$, so the signal frequency fixes the temperature. Since $\hbar\omega/k_{\mathrm B}$ is $48$ mK per GHz, the rule reads $T=54.5$ mK per GHz of signal frequency. Table~\ref{S:tab:amplifier} lists four pairs around a $7$ GHz pump. The pair at $4$ and $10$ GHz is the best of the four at $\eta=0.8$ and $0.65$, gaining $10.0\%$ ideally, $7.7\%$ at $\eta=0.8$ and $5.0\%$ at $\eta=0.65$. Its ratio of $2.5$ lies between $2.121$ and $4.091$, where Sec.~\ref{S:sec:resonators} proves the rule to be the maximizing temperature. The $3$ and $11$ GHz pair sits closer to the cap, at $12.3\%$, and falls to $0.8\%$ at $\eta=0.65$. Its idler band is cold, with width $1.08$ and response $0.28$. It adds little response to the Bell sum, while the detector noise $2(1-\eta)/\eta$ enters the Bell width in full. A closer partner brings more response, so the $4$ and $10$ GHz pair holds up once the chain is lossy.

\begin{table}[!htb]
\caption{\label{S:tab:amplifier}Four-wave-mixing amplifier pairs, $\omega_{\mathrm s}+\omega_{\mathrm i}=2\omega_{\mathrm p}$, at the temperature that puts the signal band on the crossover. The ratio is $\omega_{\mathrm i}/\omega_{\mathrm s}$. Gains of Bell homodyne over the separate-readout benchmark are in percent, ideal and at three homodyne efficiencies $\eta$ of Sec.~\ref{S:sec:robustness}, one common efficiency for every setting.}
\begin{ruledtabular}
\begin{tabular}{ccccccccc}
$\omega_{\mathrm s}/2\pi$ & $\omega_{\mathrm i}/2\pi$ & $\omega_{\mathrm p}/2\pi$ & $T$ & ratio & ideal & $\eta=0.8$ & $\eta=0.65$ & $\eta=0.5$\\
\hline
$4$ GHz   & $10$ GHz  & $7$ GHz & $218$ mK & $2.50$ & $10.0$ & $7.7$ & $5.0$ & $1.3$\\
$4.5$ GHz & $9.5$ GHz & $7$ GHz & $245$ mK & $2.11$ & $7.0$  & $5.9$ & $4.5$ & $2.2$\\
$3$ GHz   & $11$ GHz  & $7$ GHz & $163$ mK & $3.67$ & $12.3$ & $6.3$ & $0.8$ & $-5.7$\\
$5$ GHz   & $9$ GHz   & $7$ GHz & $272$ mK & $1.80$ & $4.3$  & $3.9$ & $3.2$ & $2.0$\\
\end{tabular}
\end{ruledtabular}
\end{table}

Both separate-readout benchmarks run on the same chain. Diverting the idler band through a filter into a cold load makes the record the heterodyne of the signal band, by Eq.~\eqref{S:eq:ampFI}. Pumping degenerately, with the idler band on the signal band, makes the amplification phase-sensitive~\cite{RoyDevoret2016}. One signal quadrature is then amplified and the other squeezed, so at large gain the record is a homodyne of the amplified one. A width estimated from $N$ samples has a relative error of $\sqrt{2/N}$, so twenty thousand samples fix it to one percent and resolve a five percent difference in Fisher information. The whole comparison of separate readout with Bell homodyne therefore runs on one readout line, with the pump and the idler load as its two controls.

The pump setting also pairs a whole bank. Section~\ref{S:sec:pairing} reads a bank of modes at one temperature through disjoint pairs. A pump at $\omega_{\mathrm p}$ reads every band $\omega$ together with its mirror band $2\omega_{\mathrm p}-\omega$, or $\omega_{\mathrm p}-\omega$ for three-wave mixing. That is a disjoint pairing of the bank, selected by one setting. A real thermal field is a bank of modes at one temperature and many frequencies, so the mismatch the Letter needs is the generic case.

\section{Data availability}
\label{S:sec:availability}

The data that support the findings of this work are available within the
Letter and this Supplemental Material; no additional data were generated. Every number in the Letter and in this Supplement follows from the equations given, and the Bernstein coefficient counts in Table~\ref{S:tab:certificate} and in Table~I of the Letter, together with the box counts of Secs.~\ref{S:sec:cap} and \ref{S:sec:resonators}, are the complete output of the proofs.

\end{document}